\documentclass[a4paper,8pt]{article}   	
\pdfoutput=1
\usepackage{jheppub}
\usepackage{amsfonts}
\usepackage{gensymb} 
\usepackage{pdfpages} 
\usepackage{mathtools}
\usepackage{wrapfig}
\usepackage[T1]{fontenc}
\usepackage{enumitem}
\makeatother 
\usepackage{braket}
\usepackage{slashed}
\allowdisplaybreaks 

\DeclareMathOperator{\arccosh}{arccosh}

\usepackage{wrapfig}
\usepackage{floatrow}
\usepackage{caption}

\newcommand{\ms}{\overline{\textrm{MS}}}

\newcommand{\Vtree}{V^{(0)}}
\newcommand{\Vone}{V^{(1)}}
\newcommand{\f}{\varphi}
\newcommand{\g}{\,\mathrm{GeV}}
\newcommand{\tzero}{t_*^{(0)}}

\title{Single-scale Renormalisation Group Improvement\\ of Multi-scale Effective Potentials}
\author[a,c]{Leonardo Chataignier,}
\author[a]{Tomislav Prokopec,}
\author[b]{Michael G. Schmidt}
\author[a,d]{and Bogumi\l a~\'Swie\.zewska}

\emailAdd{lcmr@thp.uni-koeln.de}
\emailAdd{t.prokopec@uu.nl}
\emailAdd{M.G.Schmidt@thphys.uni-heidelberg.de}
\emailAdd{b.swiezewska@uu.nl}

\affiliation[a]{Institute for Theoretical Physics, Spinoza Institute \& EMME$\Phi$, Utrecht University, \\Princetonplein 5, 3584 CC Utrecht, The Netherlands}
\affiliation[b]{Institute for Theoretical Physics, Universit\"{a}t Heidelberg, \\Philosophenweg 16, D-69120 Heidelberg, Germany}
\affiliation[c]{Institute for Theoretical Physics, University of Cologne, \\Z\"{u}lpicher Stra\ss e 77, 50937 K\"{o}ln, Germany}
\affiliation[d]{Faculty of Physics, University of Warsaw, \\Pasteura 5, 02-093 Warsaw, Poland}

\abstract{
We present a new method for renormalisation group improvement of the effective potential of a~quantum field theory with an arbitrary number of scalar fields.  The method amounts to solving the~renormalisation group equation for the effective potential with the boundary conditions chosen on the hypersurface where quantum corrections vanish. This hypersurface is defined through a suitable choice of a field-dependent value for the renormalisation scale. The method can be applied to any order in perturbation theory and it is a generalisation of the standard procedure valid for the one-field case. In our method, however, the choice of the renormalisation scale does not eliminate individual logarithmic terms but rather the entire loop corrections to the effective potential.  It allows us to evaluate the improved effective potential for arbitrary values of the scalar fields using the tree-level potential with running coupling constants as long as they remain perturbative. This opens the~possibility of studying  various applications which require an analysis of multi-field effective potentials across different energy scales. In particular, the issue of stability of the scalar potential can be easily studied beyond tree level.
}
\arxivnumber{1801.05258}

\begin{document}
\maketitle

\section{Introduction}

Despite the absence of direct evidence for physics beyond the Standard Model (SM) of Particle Physics, there are still unresolved puzzles that need to be considered. For example, dark matter, matter-antimatter asymmetry in the Universe, neutrino masses and the strong CP problem call for an explanation. Many of these issues can be addressed within extensions of the SM with extra scalar fields. It is thus of great interest to understand such models in detail and, in particular, to determine what the effects of  quantum corrections on the dynamics are.

A basic tool to study  quantum corrections of a given quantum field theory is the effective action. In the approximation of zero external momentum (also called local potential approximation or LPA in short) it reduces to the effective potential which  can be used to determine the true vacuum of the theory and to study its stability, as well as to compute loop corrections to couplings (as third or fourth derivatives of the effective potential) and masses (second derivatives). If extended scalar sectors are considered, the effective potential  typically exhibits different mass scales, which are the different eigenvalues of the field-dependent Hessian matrix of the tree-level potential. In this case, we refer to the potential as a multi-scale effective potential. If a model with only one scalar field is considered,  the effective potential exhibits a single mass eigenvalue, in which case it  is referred to as a~single-scale potential. Even if a single scalar field is present, many mass scales can be introduced by the~existence of different particle species coupling to the scalar field. In such a~case a single-scalar-field potential can be considered to be a~multi-scale potential. An example of such a~setting is the SM.

Multi-scale potentials are considerably more complicated than their single-scale counterparts. The~reason, apart from the higher dimensionality of the configuration space, is that the loop corrections introduce terms that are proportional to powers of logarithms of the ratios of the mass eigenvalues to the renormalisation scale $\mu$. The values of the masses can be vastly different, depending on the values of couplings and the region of the field-space we study. If the mass scales are very different, it is not possible to choose a value of the renormalisation scale such that all the ratios are of the same order and all the logarithmic terms do not grow large. In this case the large logarithms, which appear at each loop order,  invalidate the perturbative expansion of the effective potential.

S. Coleman and E. Weinberg (CW) analysed the problem of large logarithms in the case of massless $\phi^4$-theory~\cite{Coleman:1973jx}. In this single-scale case, large logarithms appear when the field-dependent mass is very different from the renormalisation scale. CW showed that one may use the renormalisation group (RG) to resum the large logarithms into a running coupling constant. Integrating the $\beta$-function corresponds to summing a power series in the logarithmic term to a closed-form expression contained in the running coupling. In this way, the effective potential is improved: even if there are large logarithmic terms, perturbation theory is to be trusted as long as the running coupling is perturbative and is, therefore, a~valid expansion parameter. B. Kastening extended this method to  massive $\phi^4$-theory \cite{Kastening}. In both cases, knowledge of the renormalisation group functions to one-loop order is sufficient to resum the~highest powers of logarithmic terms, referred to as leading logarithms, to a closed-form expression.

Unfortunately, the situation is more complicated for models with extended scalar sectors which exhibit more than one mass eigenvalue. Various logarithms that appear in the usual expression for the effective potential cannot be resummed with a single renormalisation scale, i.e., with the usual renormalisation group techniques. For this reason, multi-scale methods were first employed in \cite{Einhorn} and developed in subsequent works \cite{Ford5,Bando2,Ford,Ford3,Ford4,Steele3}. Such methods introduce several renormalisation scales and, therefore, one needs to work with partial renormalisation group equations. An explicit application to O(N)-symmetric $\phi^4$-theory can be found in \cite{Ford4}.

Since multi-scale methods introduce several technical complications, it would be desirable to devise a framework in which the usual renormalisation group equation would suffice to improve the multi-scale effective potential, at least in a certain region of the parameter space of a given model. In \cite{Bando2,Casas}, it was shown that one can use the standard renormalisation group equation to improve a multi-scale effective potential if one uses the decoupling theorem in mass-independent renormalisation schemes to reduce the number of different mass eigenvalues present in the expression for the effective potential below a~given energy threshold. To achieve this, one has to modify the expression for the effective potential by introducing  step functions which realise the decoupling of high-mass scales. Remarkably, it was shown in \cite{Casas} within the context of the Higgs--Yukawa model that this method is equivalent to the multi-scale method used in \cite{Ford3,Ford4}. The equivalence of the two methods can be understood in terms of resummation, i.e., both resum the same class of logarithmic terms.

The main point of the present paper is a presentation of a new method to RG improve multi-scale effective potentials using the traditional, unmodified RG equation containing a single renormalisation scale. We achieve this by solving the RG equation with suitably chosen boundary conditions, generalising the method used in the one-field case.  The method here developed is conceptually simple and can be implemented numerically in a rather straightforward way. Moreover, it allows us to analyse the issue of stability of the RG improved potential and to cast the stability conditions in the simple form of the~tree-level conditions evaluated at a large scale.

It will become clear that this method is not equivalent to the multi-scale methods because it does not resum the leading logarithms as they are defined in the literature. Nevertheless, it allows us to resum important logarithmic contributions to truncations of the effective potential at a given loop order. In particular, we prove that, if one can determine a single dominant logarithm in a certain region of parameter space, resummation of its powers will yield the same result as obtained with the use of our method. Thus, it provides a good approximation to the full effective potential as long as the running coupling constants remain perturbative.

The paper is structured as follows. In section~\ref{sec:pre-RG}, we review the simple case of massless $\phi^4$-theory and introduce some preliminary notions that are useful for the method presented in the following section. In~section~\ref{chap:RG}, we introduce our method of RG improvement of the multi-scale effective potential at one-loop level. In the following section we discuss validity of the method and show that it reproduces the result of a resummation of powers of a dominant logarithm, if such a logarithm can be identified.  Section~\ref{sec:applications} is devoted to the presentation of several applications of our method, in particular the study of the stability of the improved effective potential, which includes higher-loop order contributions. Moreover, we study there some of the general issues mentioned in the preceding part of the paper within the context of concrete models. Finally we conclude and give an outlook to further applications in section~\ref{sec:conclusions}. For convenience and clarity, we focus on the~one-loop improvement of a general theory in the main body of the paper. The method presented in section~\ref{chap:RG} is generalised to an arbitrary number of loops in appendix~\ref{chap:O-N-phi4}, where we show that the full effective potential (to all loop orders) can be written in the tree-level form.

Throughout the text, we use dimensional regularisation and the $\ms$ renormalisation scheme. We adopt units in which $c = 1$, but we  keep factors of $\hbar$ explicit. The $n$-loop term in the loop expansion of the effective potential is denoted by $\hbar^n V^{(n)}$.

\section{Massless $\phi^4$-theory \label{sec:pre-RG}}

We begin by considering massless $\phi^4$-theory to illustrate our method of RG-improvement in a simple setting. After dimensional regularisation, the effective potential for this model reads~\cite{Coleman:1973jx,Quiros}\footnote{The potential in ref.~\cite{Coleman:1973jx} has slightly different form due to a difference in the renormalisation condition for the self-coupling~$\lambda$.}
\begin{equation}\label{eq:phi4-one-loop-eff-pot}
V(\mu;\lambda,\phi) = \frac{1}{4}\lambda\phi^4+\frac{9\hbar\lambda^2\phi^4}{64\pi^2}\left[\log\frac{3\lambda\phi^2}{\mu^2}-\frac{3}{2}\right]+\mathcal{O}(\hbar^2) \ ,
\end{equation}
where $\phi$ is the background field. The full effective potential cannot depend on the arbitrary renormalisation scale $\mu$. Nevertheless, its truncations to a given loop order depend on $\mu$. For this reason, we choose to include $\mu$ as a separate argument of $V$. In general, the one-loop approximation~(\ref{eq:phi4-one-loop-eff-pot}) is reliable if the $\mathcal{O}(\hbar^2)$ terms in the effective potential are not large. Otherwise, perturbation theory ceases to be valid. Perturbativity is guaranteed if the couplings are small parameters and there are no large logarithmic contributions in the loop corrections.

The $n$-th loop order effective potential contains products of the coupling and logarithm with powers given by, at most, $\lambda^{n+1}\left[\hbar\log\frac{3\lambda\phi^2}{\mu^2}\right]^n$. Such terms are referred to as the leading logarithms of massless $\phi^4$-theory. The loop expansion of the effective potential is thus a power series in $\hbar\lambda$ and in $\hbar\lambda\log\frac{3\lambda\phi^2}{\mu^2}$ and truncations at a given 
loop order provide useful approximations if both parameters are small.

If the logarithm is large, one may use the renormalisation group equation to resum the logarithmic terms and regain perturbativity~\cite{Coleman:1973jx}. The idea behind the RG equation is the freedom of choosing the renormalisation scale $\mu$.  In general, for each coupling parameter and field with (finite) field normalisation $Z$, one defines the renormalisation group (RG) functions
\begin{equation}\label{eq:ele-pediu}
\begin{aligned}
\beta\equiv\beta(\lambda)&:=\mu\frac{\mathrm{d}\lambda}{\mathrm{d}\mu}\ , \\
\gamma\equiv\gamma(\lambda)&:=\mu\frac{\mathrm{d}\log Z}{\mathrm{d}\mu} \ ,
\end{aligned}\end{equation}
which are referred to as $\beta$-function and anomalous dimension, respectively. The independence of the effective potential on the choice of $\mu$ can be expressed as the RG (Callan-Symanzik) equation (see also section~\ref{sec:RG-improvement})
\begin{equation}\label{eq:RG-phi4}
\mu \frac{\mathrm{d}}{\mathrm{d}\mu} V (\mu;\lambda,\phi)= \left[\mu \frac{\partial}{\partial\mu} + \beta(\lambda) \frac{\partial}{\partial\lambda}-\frac{1}{2}\gamma(\lambda) \phi\frac{\partial}{\partial\phi}\right]V(\mu;\lambda,\phi)=0.
\end{equation}
It may be regarded as a partial differential equation (PDE) in the parameter space spanned by $(\mu;\lambda,\phi)$. We can thus solve the RG equation for the effective potential using the method of characteristics (cf.\;e.g.\;ref.~\cite{FritzJohn}). In short, the method consists of solving the Cauchy problem for eq.~(\ref{eq:RG-phi4}) by specifying the boundary conditions on a regular and noncharacteristic hypersurface, i.e., a hypersurface which is not tangent to the flow of the RG equation, and then transmitting this information to the full parameter space along characteristic curves. The running renormalisation scale, couplings and fields are simply the coordinates of points along a characteristic curve.

We choose the boundary to be the hypersurface where the one-loop correction to the effective potential vanishes and we  refer to it as the tree-level hypersurface. It is given by the equation (cf. eq.~(\ref{eq:phi4-one-loop-eff-pot}))
\begin{equation}\label{eq:boundary-phi4}
\log\frac{3\lambda\phi^2}{\mu^2}-\frac{3}{2}=0 \ .
\end{equation}
We can thus parametrise the tree-level hypersurface by coordinates $\xi=(\lambda_*, \phi_*)$ and express $\mu$ at this surface as a function of $\lambda_*$ and $\phi_*$, $\mu \equiv \mu_*(\lambda_*,\phi_*)$. We require that on this surface the solution of eq.~\eqref{eq:RG-phi4} has the tree-level form.

The characteristic equations and boundary conditions then read
\begin{equation}\label{eq:char-eqs-phi4}
\begin{aligned}
\frac{\mathrm{d}}{\mathrm{d}t'}\bar{\mu}(t',\xi) = \bar{\mu}(t',\xi)\ ,& \quad \bar{\mu}(t'=0,\xi)=\mu_*(\lambda_*,\phi_*) \ ,\\
\frac{\mathrm{d}}{\mathrm{d}t'}\bar{\lambda}(t',\xi) = \beta(\bar{\lambda})\ ,&\quad \bar{\lambda}(t'=0,\xi)=\lambda_{*}\ ,\\
\frac{\mathrm{d}}{\mathrm{d}t'}\bar{\phi}(t',\xi) = -\frac{1}{2}\gamma(\bar{\lambda})\bar{\phi}(t',\xi)\ ,&\quad \bar{\phi}(t'=0,\xi)=\phi_{*}\ ,\\
\frac{\mathrm{d}}{\mathrm{d}t'}V(t', \xi) = 0\ ,& \quad V(t'=0,\xi)=\Vtree(\lambda_*, \phi_{*})=\frac{1}{4}\lambda_*\phi_*^4\ .
\end{aligned}
\end{equation}
By integrating the above system, one finds a characteristic curve family given by $\bar{\mu},\bar{\lambda},\bar{\phi}$. A \emph{particular} choice of boundary conditions $\xi=(\lambda_*, \phi_*)$ corresponds to a \emph{particular} curve in the family, such that different curves are labelled by $\xi$. Each curve is parametrised by $t'$, which corresponds to the displacement along the curve. 

The last of eqs.~\eqref{eq:char-eqs-phi4} implies that the effective potential is constant along characteristic curves. Any choice of characteristic parameter $t'$ (any choice of $\bar{\mu}$) is equally valid, as long as the Cauchy problem is well defined. If we know how the couplings, masses and fields run along characteristic curves, the~scale independence of the effective potential provides a way to RG-improve it by incorporating the effect of the running of its parameters, such that the contributions from radiative corrections are minimised by a suitable choice of scale. Running along a characteristic curve can thus improve the validity of the perturbative expansion of the effective potential at a given loop order.

\begin{figure}[ht]
\center
\includegraphics[width=.55\textwidth]{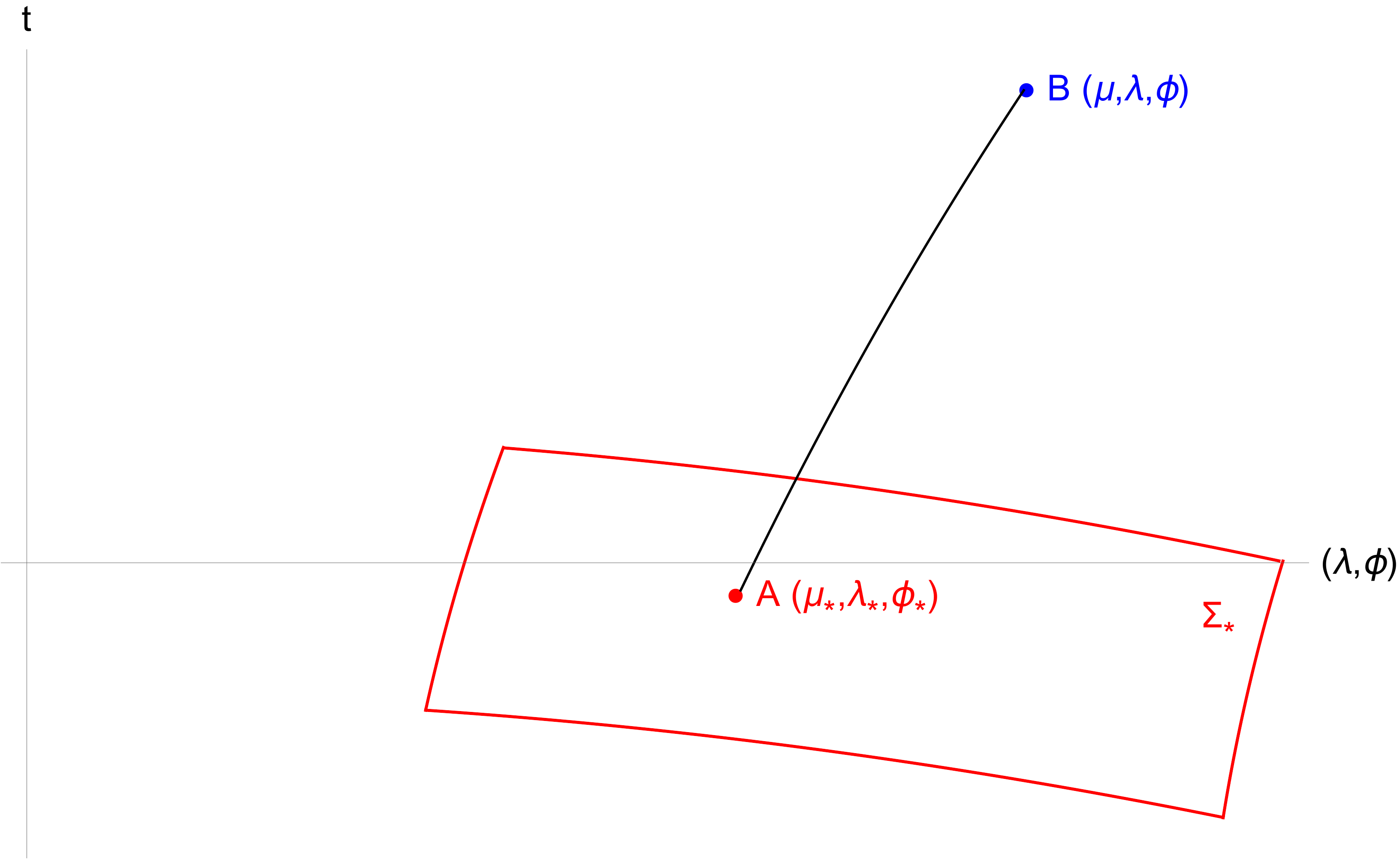}
\caption{The tree-level hypersurface is symbolically represented by $\Sigma_*$. If we start from point $A$ and travel a distance $t'_*$ along the characteristic curve to reach point $B$, then we return to A by starting from $B$ and by travelling the same distance in the opposite direction.\label{fig:inversion}}
\end{figure}

For massless $\phi^4$-theory, the one-loop anomalous dimension of the scalar field vanishes, which implies  that, at one-loop order, the field variable $\phi$ is independent of the choice of the renormalisation scale~\cite{Coleman:1973jx, Sher} and, therefore, it is constant along a characteristic curve. We may thus write $\bar{\phi}(t',\xi) = \phi_*$. Since this is valid in the entire parameter space, we  drop the subscript star for the field variable. We then find the~solution
\begin{equation}
V(t', \xi) = \frac{1}{4}\lambda_*\phi^4 \ ,
\end{equation}
which is the RG improved potential. It is desirable to express it in terms of the general variables $(\mu;\lambda,\phi)$ of the parameter space, rather than the characteristic curve parameters $(t',\lambda_*,\phi_*)$. This can be achieved simply by noting that if\footnote{The evolution of the coupling parameter along a characteristic curve does not depend on the boundary value $\phi_*$ due to eqs.~(\ref{eq:char-eqs-phi4}). For this reason, we omit $\phi_*$ from the arguments of $\bar{\lambda}$. Nonetheless, both $t_*$ and $\mu_*$ depend on $\phi$ through eq.~\eqref{eq:t*-phi4}.}
\begin{equation}
\lambda=\bar{\lambda}(t'_*,\lambda_{*}) \ 
\end{equation}
then
\begin{equation}\label{eq:inversion}
\bar{\lambda}(t_*=-t'_*,\lambda)=\lambda_{*} \ .
\end{equation}
The above equations\footnote{Eq.~(\ref{eq:inversion}) corresponds to an inversion of the characteristic curve family \cite{FritzJohn}.}  correspond to the fact that, if we start from a given point $A$ in the parameter space and travel a distance $t'_*$ along a characteristic curve to reach point $B$, then we can return to point $A$ by starting from $B$ and by travelling the distance $t'_*$ in the opposite direction. This is illustrated in figure~\ref{fig:inversion}.

After changing the variables, we find the RG improved potential in the following form
\begin{equation}\label{eq:Vimproved-phi4}
V(\mu;\lambda,\phi) = \frac{1}{4}\bar{\lambda}(t_*,\lambda)\phi^4+ \mathcal{O}(\hbar^2)\ ,
\end{equation}
where $t_* =-t_*'= \log\frac{\mu_*}{\mu}$ (cf. first line of eq.~\eqref{eq:char-eqs-phi4}) and $\mu_*$ can be found by inserting eq.~\eqref{eq:inversion} into eq.~\eqref{eq:boundary-phi4},
\begin{equation}\label{eq:t*-phi4}
\log\frac{3\bar{\lambda}(t_*,\lambda)\phi^2}{\mu_*^2}-\frac{3}{2}=0 \ .
\end{equation}
The method that led to the solution in eq.~\eqref{eq:Vimproved-phi4} is equivalent to choosing the renormalisation scale such that the one-loop correction to the effective potential vanishes. Indeed, since the effective potential does not change along characteristic curves, we obtain
\begin{align}
V(\mu; \lambda,\phi)&=\Vtree(\lambda,\phi)+\hbar\Vone(\mu, \lambda,\phi)+ \mathcal{O}(\hbar^2)\nonumber\\*
&=\Vtree(\bar{\lambda}(t,\lambda),\phi)+\hbar\Vone(\bar{\mu}(t,\mu),\bar{\lambda}(t,\lambda),\phi)+ \mathcal{O}(\hbar^2).\label{eq:t-invariance}
\end{align}
By choosing $t=t_*$ (or $\mu = \mu_*$) such that $\Vone(\bar{\mu}(t_*,\mu),\bar{\lambda}(t_*,\lambda),\phi)=0$, we obtain
\begin{equation}\label{eq:Vimproved-1}
V(\mu, \lambda,\phi)=\Vtree(\bar{\lambda}(t_*,\lambda),\phi) +\mathcal{O}(\hbar^2) \ ,
\end{equation}
which is the solution in eq.~\eqref{eq:Vimproved-phi4}, up to two-loop RG improved terms, which we do not consider here (see appendix~\ref{chap:O-N-phi4}). The solution is valid as long as the running coupling evaluated at $t_*$ remains perturbative.

An explicit solution of the above one-loop RG improvement is obtained by integrating the $\beta$-function, which reads~\cite{Coleman:1973jx}
\begin{align}
\beta &=\frac{9\hbar\lambda^2}{8\pi^2}+\mathcal{O}(\hbar^2)\equiv\hbar\beta^{(1)}+\mathcal{O}(\hbar^2) \ . \label{eq:phi4-beta}
\end{align}
Integrating the first of eqs.~(\ref{eq:ele-pediu}) and using eq.~(\ref{eq:phi4-beta}), we obtain the one-loop running coupling
\begin{align}\label{eq:running-coupling-phi4}
\bar{\lambda}(t,\lambda) &= \frac{\lambda}{1-\frac{9\hbar\lambda}{8\pi^2}t} \ .
\end{align}
With $t=\log \frac{\bar{\mu}(t,\mu)}{\mu}$ we can interpret $\bar{\lambda}(t,\lambda)$ as the value of the coupling parameter when the renormalisation scale is chosen to be $\bar{\mu}$, which is related to the corresponding value at the arbitrary reference scale $\mu$ by a resummation of terms of all orders in $\hbar\lambda\log\frac{\bar{\mu}}{\mu}$. This observation is the key to regaining perturbativity in the case of large logarithms. Indeed, the closed form in eq.~(\ref{eq:running-coupling-phi4}) guarantees that, as long as $|\bar{\lambda}(t,\lambda)| < 4\pi$, the logarithm in the denominator can grow large without invalidating a perturbative expansion in terms of the running parameter $\bar{\lambda}(t,\lambda)$. Therefore, if the potential evaluated at the scale $\mu$ exhibits large logarithmic terms, then we should evaluate the potential at the scale $\bar{\mu}$, where these logarithms have been resummed into $\bar{\lambda}(t,\lambda)$.

The RG improved potential incorporates this resummation, since we can now rewrite eq.~(\ref{eq:Vimproved-phi4}) as follows,
\begin{equation}\label{eq:Vimproved-phi4-2}
V(\mu;\lambda,\phi) = \frac{1}{4}\bar{\lambda}(t_*,\lambda)\phi^4=\frac{1}{4}\frac{\lambda}{1-\frac{9\hbar\lambda}{16\pi^2}\log\frac{\mu_*^2}{\mu^2}} \phi^4.
\end{equation}
The logarithm in the denominator reads\footnote{In the original work of CW~\cite{Coleman:1973jx}, the resummation of leading logarithms was accomplished by choosing $\bar{\mu}$ such that the effective potential at this scale contains no logarithmic terms. This can be accomplished by setting 
$\bar{\mu}^2 = 3\bar{\lambda}(\bar{\mu},\lambda)\phi^2 = 3\lambda\phi^2 + \mathcal{O}(\hbar).$
It is clear that this choice differs from that in eq.~(\ref{eq:our-choice}) by a multiplicative constant. Thus, both are equally suited to resum the leading logarithms \emph{with respect to} the scale $\mu$. (CW did not work in the $\ms$ scheme which leads to slight differences in their expressions.)}
\begin{equation}\label{eq:our-choice}
\log\frac{\mu_*^2}{\mu^2}=\log\frac{3\lambda_*\phi^2}{\mu^2}-\frac{3}{2} = \log\frac{3\lambda\phi^2}{\mu^2}-\frac{3}{2}+\mathcal{O}(\hbar) \ ,
\end{equation}
which implies the RG improved potential in eq.~\eqref{eq:Vimproved-phi4-2} automatically includes a resummation of the leading logarithms, i.e., 
\begin{equation}\label{eq:Vimproved-phi4-3}
V(\mu;\lambda,\phi) =  \frac{1}{4}\phi^4\sum_{n = 0}^{\infty}\lambda^{n+1}\left[\frac{9\hbar}{16\pi^2}\log\frac{\mu_*^2}{\mu^2}\right]^n = \frac{1}{4}\phi^4\sum_{n = 0}^{\infty}\lambda^{n+1}\left[\frac{9\hbar}{16\pi^2}\log\frac{3\lambda\phi^2}{\mu^2}\right]^n + \cdots \ ,
\end{equation}
where the ellipses hide terms of subleading-logarithmic order (see below). Thus, the one-loop RG running of the coupling parameter is sufficient to resum into a closed form the highest powers of the mass logarithm that appear at each loop-level.

It is important to note that the RG-improved potential has a double interpretation. To see this, we combine eqs.~(\ref{eq:Vimproved-phi4-2}) and~(\ref{eq:Vimproved-phi4-3}) to obtain
\begin{equation}\label{eq:Vimproved-phi4-4}
V(\mu;\lambda,\phi) =  \frac{1}{4}\bar{\lambda}(t_*,\lambda)\phi^4 = \frac{1}{4}\phi^4\sum_{n = 0}^{\infty}\lambda^{n+1}\left[\frac{9\hbar}{16\pi^2}\log\frac{3\lambda\phi^2}{\mu^2}\right]^n + \cdots \ .
\end{equation}
On the left hand side, the potential is the lowest order effective potential evaluated at the scale $\mu_*$, which is a solution to eq.~(\ref{eq:t*-phi4}). On the right hand side, it is the potential evaluated at the scale $\mu$ for which terms of all orders in $\hbar$ have been included. It is for this reason that eq.~(\ref{eq:Vimproved-phi4-4}) deserves to be referred to as the one-loop RG-improved potential.

In general, one may define terms of the kind
\begin{equation}\label{eq:n-to-leading-logs-phi4}
\lambda^{n+k+1}\hbar^{n+k}\left[\log\frac{3\lambda\phi^2}{\mu^2}\right]^k \ , \ k = 0,1,2,... \ ,
\end{equation}
to be of $n$-th-to-leading logarithmic order \emph{with respect to} the renormalisation scale $\mu$. In the same way that the tree-level potential evaluated at a suitable field-dependent scale corresponds to a resummation of all leading logarithms, the one-loop term $V^{(1)}$ evaluated at that scale  contains the resummation of subleading logarithms. The issue of resummation of subleading terms is also addressed in this paper in the context of multi-scale potentials.

In what follows, we generalise the method introduced herein to multi-scale effective potentials. Evidently, it is not possible to choose the scale $\bar{\mu}$  to cancel  different mass logarithms individually, which would be a na\"{i}ve generalisation of the CW approach. Nonetheless, we will argue that it is possible to improve the effective potential by solving the RG equation with the boundary given at the hypersurface of vanishing quantum corrections. To put it differently, the appropriate generalisation to the multi-scale case consists of choosing the \emph{single} renormalisation scale to be $\bar{\mu} = \mu_*$, which is the scale at which $V^{(1)} = 0$. For simplicity in the following section we focus on the improvement procedure at one-loop order. The method presented in the following section is generalised to an arbitrary loop order in appendix~\ref{chap:O-N-phi4}.

\section{The Tree-Level Hypersurface \label{chap:RG}}

We  analyse a theory with $N_{\phi}$ scalar fields and $N_{\lambda}$ couplings in addition to vector and fermionic fields. We denote the couplings (possibly including mass terms) by $\lambda = (\lambda_1, ..., \lambda_{N_{\lambda}})$ and the classical scalar fields by $\phi = (\phi_1,...,\phi_{N_{\phi}})$. There are $N_m$ field-dependent mass eigenvalues, denoted by $m = (m_1, ..., m_{N_m})$. Note that each eigenvalue is a function of the couplings as well as the fields. In this notation, the parameter space is spanned by $(\mu;\lambda,\phi)$, i.e., by the renormalisation scale, the couplings and scalar fields of the theory. We  denote the effective potential with a tilde, i.e., $\tilde{V}(\mu;\lambda,\phi)$, for reasons that will become apparent in what follows.

\subsection{The One-Loop Effective Potential and the Pivot Logarithm}

When multiple scalars or vector fields and fermions are present, one must generalise eq.~(\ref{eq:phi4-one-loop-eff-pot}) to account for  different physical degrees of freedom of each particle species. The renormalised effective potential thus reads
\begin{align}
\tilde{V}(\mu;\lambda,\phi) &= \tilde{V}^{(0)}(\lambda,\phi)+\hbar \tilde{V}^{(1)}(\mu,\lambda,\phi) +\mathcal{O}\left(\hbar^2\right) , \nonumber\\*
\tilde{V}^{(1)}(\mu,\lambda,\phi) &= \frac{1}{64\pi^2}\sum_{a}n_am_a^4(\lambda,\phi)\left[\log\frac{m_a^2(\lambda,\phi)}{\mu^2}-\chi_a \right],\label{eq:general-one-loop-term}
\end{align}
where the index $a$ runs over all particle species and the tree-level field-dependent mass eigenvalues are denoted by $m_a(\lambda,\phi)$. The $n_a$ factor accounts for the number of degrees of freedom a given particle carries and is given for a particle of spin $s_a$ by the formula
$$
n_a=(-1)^{2s_a} Q_a N_a (2s_a+1),
$$
where $Q_a=1,2$ for uncharged and charged particles, respectively, and $N_a=1,3$ for uncoloured and coloured particles. Moreover, $\chi_a=\frac{5}{6}$ for vector bosons and $\chi_a=\frac{3}{2}$ for scalars and fermions (when dimensional regularisation and $\ms$ scheme are used).

A convenient way of writing the one-loop term is obtained by rewriting each mass logarithm as
\begin{equation}\label{eq:mass-logs-rho}
\log\frac{m_a^2(\lambda,\phi)}{\mu^2} = \log\frac{m_a^2(\lambda,\phi)}{\mathcal{M}^2}+\log\frac{\mathcal{M}^2}{\mu^2} \ ,
\end{equation}
where $\mathcal{M}$ is any non-vanishing function with mass dimension equal to one. It will be referred to as the pivot scale.  For example, if there are $N_{\phi}$ scalar fields, we can choose $\mathcal{M}$ to be the radial variable in the scalar-field configuration space
\begin{equation}
\mathcal{M}^2 = \rho^2 := \sum_{j=1}^{N_{\phi}}\phi_{j}^2\ .\label{eq:radial-pivot}
\end{equation}
Any other choice of pivot scale is equally valid. We can then rewrite eq.~(\ref{eq:general-one-loop-term}) for an arbitrary pivot scale as
\begin{equation}\label{eq:one-loop-AB}
\tilde{V}^{(1)}(\mu,\lambda,\phi) = \mathbb{A}+\mathbb{B}\log\frac{\mathcal{M}^2}{\mu^2} \ ,
\end{equation}
where we defined the functions
\begin{equation}\label{eq:AB-fcts}
\begin{aligned}
\mathbb{A} &= \frac{1}{64\pi^2}\sum_{a}n_a m_a^4(\lambda,\phi)\left[\log\frac{m_a^2(\lambda,\phi)}{\mathcal{M}^2}-\chi_a \right] \ , \\
\mathbb{B} &=\frac{1}{64\pi^2}\sum_{a}n_a m_a^4(\lambda,\phi) \ .
\end{aligned}
\end{equation}
Since $\mathcal{M}$ is arbitrary, we note that different choices $\mathcal{M}$ and $\mathcal{M}'$ are related by the transformations
\begin{equation}\label{eq:rho-transform}
\begin{aligned}
\mathbb{A} &= \frac{1}{64\pi^2}\sum_{a}n_a m_a^4(\lambda,\phi)\left[\log\frac{m_a^2(\lambda,\phi)}{\mathcal{M}^{\prime2}}-\chi_a\right]+ \frac{1}{64\pi^2}\sum_{a}n_am_a^4(\lambda,\phi)\log\frac{\mathcal{M}^{\prime2}}{\mathcal{M}^2} \\&\equiv \mathbb{A}'+\mathbb{B}'\log\frac{\mathcal{M}^{\prime2}}{\mathcal{M}^2} \ , \\
\mathbb{B} &\equiv \mathbb{B}' \ .
\end{aligned}
\end{equation}
The one-loop term~(\ref{eq:one-loop-AB}) is invariant under the above transformations. Using expression~(\ref{eq:one-loop-AB}) is advantageous since the explicit dependence on the renormalisation scale is contained only in the logarithm $\log\frac{\mathcal{M}^2}{\mu^2}$ if $\mathcal{M}$ is independent of $\mu$. We will refer to this logarithm as the pivot logarithm. 

\subsection{Renormalisation Group Improvement\label{sec:RG-improvement}}

The generalisation of eqs.~(\ref{eq:ele-pediu}) gives the $\beta$-functions and anomalous dimensions
\begin{equation}\label{eq:beta-gamma}
\begin{aligned}
\beta_i \equiv \beta_i(\lambda) &= \mu\frac{\mathrm{d}}{\mathrm{d}\mu}\lambda_i(\mu) = \sum_{l=1}^{\infty}\hbar^l\beta_i^{(l)} \ ,\\
\gamma_j \equiv \gamma_j(\lambda) &= \mu\frac{\mathrm{d}}{\mathrm{d}\mu}\log Z_j(\mu) = \sum_{l=1}^{\infty}\hbar^l \gamma_j^{(l)} \ , 
\end{aligned}
\end{equation}
where $i = 1,...,N_{\lambda}, \ j = 1,...,N_{\phi}$ and $Z_j$ is the normalisation of the field $\phi_j$. The RG equation for the effective potential then reads
\begin{equation}\notag 
\mu\frac{\mathrm{d}\tilde{V}}{\mathrm{d}\mu} \equiv \left(\mu\frac{\partial}{\partial \mu}+\sum_{i=1}^{N_{\lambda}} \beta_i \frac{\partial}{\partial \lambda_i}-\frac{1}{2}\sum_{a=1}^{N_{\phi}}\gamma_j\phi_j\frac{\partial}{\partial\phi_j}\right)\tilde{V}(\mu; \lambda, \phi) = \left(\mu\frac{\partial}{\partial \mu}+\sum_{i=1}^{N_{\lambda}} \beta_i \frac{\partial}{\partial \lambda_i}\right)\tilde{V}(\mu; \lambda, 0) \ .
\end{equation}
Note that the left hand side of the above equation is only zero if the vacuum energy $\tilde{V}(\mu;\lambda,0)$ is scale invariant\footnote{This was not taken into account in eq.~(\ref{eq:RG-phi4}) because the vacuum energy is set to zero in massless $\phi^4$-theory.}. It is always possible to perform a field-independent shift of the potential such that the scale dependence of the vacuum energy is cancelled~\cite{Ford5, Ford}. We define
\begin{equation}\label{eq:tilde-pot-no-vacuum}
\begin{aligned}
V(\mu;\lambda,\phi) &:= \tilde{V}(\mu;\lambda,\phi)+\delta\Lambda(\mu,\lambda)  \ , \\
\Lambda(\mu,\lambda) &:=\tilde{V}(\mu;\lambda,0)+\delta\Lambda(\mu,\lambda) \ ,
\end{aligned}
\end{equation}
and demand\footnote{Recall that we include the mass parameters into the set of couplings $\lambda$, that is why $\Lambda$ depends on $\lambda$.}
\begin{equation}\notag
\left(\mu\frac{\partial}{\partial \mu}+\sum_{i=1}^{N_{\lambda}} \beta_i \frac{\partial}{\partial \lambda_i}\right)\Lambda(\mu,\lambda) = 0 \ .
\end{equation}
In this way, the potential $V(\mu;\lambda,\phi)$ satisfies the equation
\begin{equation}\label{eq:RG-potential2}
\mu\frac{\mathrm{d}V}{\mathrm{d}\mu}(\mu;\lambda,\phi) = \left(\mu\frac{\partial}{\partial \mu}+\sum_{i=1}^{N_{\lambda}} \beta_i \frac{\partial}{\partial \lambda_i}-\frac{1}{2}\sum_{a=1}^{N_{\phi}}\gamma_j\phi_j\frac{\partial}{\partial\phi_j}\right)V(\mu; \lambda, \phi) = 0 \ ,
\end{equation}
which can be interpreted as a partial differential equation on a domain of the parameter space spanned by $(\mu;\lambda,\phi)$. 

 Furthermore, it is useful to define the $l$-th loop order derivative as
\begin{equation}\label{eq:d-loops}
\mathrm{d}^{(l)} := \sum_{i = 1}^{N_{\lambda}}\beta_i^{(l)}\frac{\partial}{\partial\lambda_i}-\frac{1}{2}\sum_{a = 1}^{N_{\phi}}\gamma_j^{(l)}\phi_j\frac{\partial}{\partial\phi_j} \ .
\end{equation}
As an example,  the one-loop truncation of the RG equation~(\ref{eq:RG-potential2}) reads
\begin{equation}
0 = \mu\frac{\mathrm{d}V}{\mathrm{d}\mu}(\mu;\lambda,\phi) = \hbar\mu\frac{\partial V^{(1)}}{\partial\mu}+\hbar\mathrm{d}^{(1)}V^{(0)}+\mathcal{O}(\hbar^2) = -2\hbar\mathbb{B}+\hbar\mathrm{d}^{(1)}V^{(0)}+\mathcal{O}(\hbar^2)\label{eq:d1V0} \ ,
\end{equation}
where we used eq.~(\ref{eq:one-loop-AB}).

\subsection{The Hypersurface of Vanishing One-Loop Corrections}\label{sec:hyper-one-loop}

In this section we generalise  the method of one-loop RG improvement described in section~\ref{sec:pre-RG} to the multi-scale case. In appendix~\ref{chap:O-N-phi4} the method is generalised to any loop order. As already indicated in section~\ref{sec:pre-RG}, the procedure is to solve the Cauchy problem for the RG equation~\eqref{eq:RG-potential2} with the method of characteristics. The boundary condition is given on a hypersurface where the one-loop correction to the effective potential vanishes. As mentioned earlier, we refer to this hypersurface as the tree-level hypersurface.

The choice of boundary as the tree-level hypersurface is motivated by the fact that, while it is not possible to suppress all logarithms individually with a \emph{single} renormalisation scale, one can suppress the~quantum corrections altogether, when evaluating the potential on this hypersurface. 

The Cauchy problem admits solutions only if the boundary hypersurface is regular, i.e., with a non-vanishing normal vector, and noncharacteristic. A hypersurface defined by the equation $\Sigma(\mu,\lambda,\phi) = 0$ is said to be characteristic if the gradient $\nabla\Sigma$ is orthogonal to the flow of the RG equation (i.e.\;the surface is tangent to the flow). If it is not orthogonal, then the hypersurface is noncharacteristic and we can use the RG flow to evolve the boundary data to the whole of parameter space. We postpone the discussion of whether or not the tree-level hypersurface is noncharacteristic to section~\ref{sec:B0}, where we address this issue in detail.

As before, the tree-level boundary is given by the equation
\begin{equation}\label{eq:boundary}
\hbar\Vone(\mu;\lambda,\phi)=0 \ ,
\end{equation}
which can be used to eliminate one of the parameters as a function of the others. In this way, if we use eq.~\eqref{eq:boundary} to write $\mu$ in terms of $\lambda$ and $\phi$, we can parametrise the boundary by $\xi=(\lambda_*,\phi_*)$ and write $\mu = \mu_*(\lambda_*,\phi_*)$. The characteristic curve family and boundary conditions are then given by the analog of the one-field eq.~\eqref{eq:char-eqs-phi4},
\begin{equation}\label{eq:RG-sys}
\begin{aligned}
\frac{\mathrm{d}}{\mathrm{d}t'}\bar{\mu}(t',\xi) = \bar{\mu}(t', \xi) \ ,& \quad \bar{\mu}(t'=0,\xi)=\mu_* \ , \\
\frac{\mathrm{d}}{\mathrm{d}t'}\bar{\lambda}_i(t',\xi) = \beta_i(\bar{\lambda}) \ ,&\quad \bar{\lambda}_i(t'=0,\xi)=\lambda_{i*} \ ,\\
\frac{\mathrm{d}}{\mathrm{d}t'}\bar{\phi}_j(t',\xi) = -\frac{1}{2}\gamma_j(\bar{\lambda})\bar{\phi}_j(t',\xi) \ ,&\quad \bar{\phi}_j(t'=0,\xi)=\phi_{j*}\ , \\
\frac{\mathrm{d}}{\mathrm{d}t'}V(t',\xi) = 0 \ ,& \quad V(t'=0,\xi)=\Vtree(\lambda_*, \phi_*).
\end{aligned}
\end{equation}
After integration, to obtain the solution in the $(\mu, \lambda, \phi)$ space we must invert the characteristic curve family, i.e., we express $(\mu_*;\lambda_*,\phi_{*})$ in terms of $(\mu;\lambda,\phi)$. We can proceed in full analogy with the case of $\phi^4$-theory. The value of $t_*$ which measures the displacement along the characteristic curve from a given point in the parameter space to the boundary is given by the defining equation of the boundary
\begin{equation}\label{eq:define-implicit-t*}
\hbar\Vone(\bar{\mu}(t_*,\mu,\lambda,\phi),\bar{\lambda}(t_*,\mu,\lambda,\phi),\bar{\phi}(t_*,\mu,\lambda,\phi))=0 \ .
\end{equation}
The general solution for the one-loop RG-improved effective potential at an arbitrary point $(\mu; \lambda, \phi)$ of parameter space reads
\begin{equation}
V(\mu; \lambda, \phi)=\Vtree(\bar{\lambda}(t_*,\mu,\lambda,\phi),\bar{\phi}(t_*,\mu,\lambda,\phi))+\mathcal{O}(\hbar^2)\label{eq:V-improved}
\end{equation}
and is valid in the regime where the running couplings evaluated at $t_*$ remain perturbative. As before, the improvement procedure can be interpreted as follows: we exploit the invariance of the effective potential along a characteristic curve to choose a characteristic displacement $t=t_*$ such that the one-loop correction vanishes. Equation~\eqref{eq:V-improved} represents an RG-improved potential as it contains terms of all orders in $\hbar$, which appear as a result of the integration of the characteristic equations for the running couplings and fields.

The solution to eq.~\eqref{eq:define-implicit-t*} can be found by using eq.~(\ref{eq:one-loop-AB}). Indeed, we can write the implicit solution\footnote{This equation holds for $\mathbb{A}, \mathbb{B}\neq0$ and $\mathcal{M}\neq\mu_*$. For the case of  $\mathbb{A}=0$ we need also $\mathbb{B}=0$ or  $\mathcal{M}=\mu_*$. We devote section~\ref{sec:B0} to the discussion of the $\mathbb{B}=0$ case.}
\begin{equation}\label{eq:mu-kills-CW}
\bar{\mu}(t_*) = \mu_* = \mathcal{M}\exp\left\{\frac{1}{2}\frac{\mathbb{A}\left(\bar{\lambda}(t_*),\bar{\phi}(t_*),\mathcal{M}\right)}{\mathbb{B}\left(\bar{\lambda}(t_*),\bar{\phi}(t_*)\right)}\right\} \ ,
\end{equation}
where for notational simplicity we denote $\bar{\mu}(t_*,\mu,\phi,\lambda)\equiv \bar{\mu}(t_*)$, $\bar{\lambda}(t_*,\mu,\phi,\lambda)\equiv\bar{\lambda}(t_*)$ and $\bar{\phi}(t_*,\mu,\phi,\lambda)\equiv\bar{\phi}(t_*)$ to keep only the $t_*$ dependence explicit. The implicit solution of eq.~\eqref{eq:mu-kills-CW} corresponds to the characteristic displacement

\begin{equation}\label{eq:t*-first}
t_* = \frac{1}{2}\log\frac{\mu_*^2}{\mu^2} = \frac{1}{2}\log\frac{\mathcal{M}^2}{\mu^2}+\frac{1}{2}\frac{\mathbb{A}\left(\bar{\lambda}(t_*),\bar{\phi}(t_*), \mathcal{M}\right)}{\mathbb{B}\left(\bar{\lambda}(t_*),\bar{\phi(t_*)}\right)} \ , 
\end{equation}
which can also be written as
\begin{equation}\label{eq:V1/2Brho4}
t_* = \frac{V^{(1)}\left(\mu,\bar{\lambda}(t_*),\bar{\phi}(t_*)\right)}{2\mathbb{B}\left(\bar{\lambda}(t_*),\bar{\phi}(t_*)\right)} \ .
\end{equation}
Evidently, eq.~(\ref{eq:V1/2Brho4}) is independent of $\mathcal{M}$ and one easily verifies that eq.~(\ref{eq:mu-kills-CW}) remains unchanged under a redefinition of the pivot scale, given in eqs.~(\ref{eq:rho-transform}).
Moreover, by expanding the running parameters to lowest order in $\hbar$, we obtain
\begin{equation}
t_*= t_*^{(0)}+ \mathcal{O}(\hbar)= \frac{1}{2}\log\frac{\mathcal{M}^2}{\mu^2}+\frac{1}{2}\frac{\mathbb{A}\left(\lambda,\phi, \mathcal{M}\right)}{\mathbb{B}\left(\lambda,\phi\right)}+\mathcal{O}(\hbar) =\frac{V^{(1)}(\mu; \lambda, \phi)}{2\mathbb{B}(\lambda, \phi)}+\mathcal{O}(\hbar) \ .\label{eq:tstar-0}
\end{equation}
Thus, to lowest order in $\hbar$, we recover the one-loop approximation of the effective potential
\begin{align*}
V(\mu;\lambda, \phi) &= V^{(0)}(\lambda_*,\phi_*) = V^{(0)}(\lambda,\phi)- \hbar\left[\mathrm{d}^{(1)}V^{(0)}\right]t_*^{(0)}+\mathcal{O}(\hbar^2) \\*
&= V^{(0)}(\lambda,\phi)+\hbar\mathbb{B}\left(\lambda,\phi\right)\left[\log\frac{\mathcal{M}^2}{\mu^2}+\frac{\mathbb{A}\left(\lambda,\phi,{\mathcal{M}}\right)}{\mathbb{B}\left(\lambda,\phi\right)}\right]+\mathcal{O}(\hbar^2) \\*
& = V^{(0)}(\lambda,\phi)+\hbar V^{(1)}(\mu,\lambda,\phi)+\mathcal{O}(\hbar^2) \ .
\end{align*}
Had we retained terms with higher powers of $\hbar$, which originate from the integration of the one-loop $\beta$-functions and anomalous dimensions, we would obtain a solution $V(\mu(t_*);\lambda(t_*),\phi(t_*))$ which formally contains terms from higher-loop orders. This represents an RG improvement of the potential and is a direct generalisation of the case of massless $\phi^4$-theory analysed in section~\ref{sec:pre-RG}.

We thus conclude that by solving the RG equation~(\ref{eq:RG-potential2}) with the method of characteristics, we can use the standard RG techniques to improve a multi-scale effective potential, as long as we can identify a regular and noncharacteristic tree-level hypersurface. It is worth emphasising that the resulting  RG-improved potential is the tree-level potential evaluated at some field-dependent scale. This means that there is no explicit scale dependence, and in particular no explicit logarithmic contributions are present. This clearly indicates that the validity range of the improved potential is vastly larger than that of the perturbative one-loop potential. The improved potential is valid as long as the running coupling constants evaluated at $t_*$ remain perturbative.

Up to now we have presented the method of RG-improvement of the effective potential at one-loop level.  In appendix~\ref{chap:O-N-phi4}, the method is generalised to higher-loop orders. The principles of the method are the same beyond the one-loop order --- we evaluate the potential at a hypersurface where loop corrections vanish. This results in an RG-improved potential in the form of the tree-level potential evaluated at a field dependent scale $t_*$, which can be computed in perturbation theory. Therefore, the difference between the one- and two-loop improved potentials can only originate from the running of the coupling constants and fields. Thus as long as the $t_*$ scale does not receive large corrections and we are far from Landau poles, the one-loop approximation is accurate. An explicit evaluation of the accuracy of the one-loop approximation is presented in section~\ref{chap:O-N-phi4-2} and in appendix~\ref{app:ON} for the O(N)-symmetric $\phi^4$ theory. The next section is devoted to a detailed analysis of validity and accuracy of the method.

Finally, let us comment on the issue of gauge dependence. Evidently, the fixed-loop approximations to the effective potential depend on the choice of gauge (see e.g.~ref.~\cite{Andreassen:prl, Andreassen, Konstandin}). Nonetheless, our method of improvement is valid in any gauge since we choose the renormalisation scale such that the loop corrections vanish, which can be done for any choice of gauge.

\section{Validity and Accuracy of the Method\label{sec:applicability}}

\subsection{Validity\label{sec:B0}}

The method of RG improvement described in the present paper is only valid when a unique solution of the implicit equation  for $t_*$, eq.~\eqref{eq:define-implicit-t*}, exists. The possibility to write down an explicit approximate solution $t_*^{(0)}$ given in eq.~\eqref{eq:tstar-0}, suggests that the full solution should also exist. Nevertheless, this issue is rather subtle since to obtain the result for $t_*^{(0)}$ one neglects the running of the couplings and fields, only focusing on the explicit $\mu$ dependence. The two effects might, however, balance leading to no solution, or on the contrary, they can act oppositely leading to double solutions. A general discussion of the uniqueness and existence of the solution is beyond the scope of this paper. Nonetheless, in section~\ref{sec:applications} we discuss these issues within the context of specific models.

As we have mentioned in section~\ref{sec:hyper-one-loop}, the tree-level hypersurface can serve as a boundary for the~Cauchy problem only if it is regular and noncharacteristic. In this section we examine the properties of the tree-level hypersurface and discuss the limitations they impose on our method. Below we focus on the $t_*^{(0)}$ approximation (cf.\  eq.~\eqref{eq:tstar-0}). 

The tree-level hypersurface can be defined by eq.~\eqref{eq:boundary} up to $\mathcal{O}(\hbar^2)$ terms. It is noncharacteristic if its gradient is not orthogonal to the flow of the RG equation, which is given by the following condition
\begin{equation}
0 \neq\hbar\mu\frac{\partial V^{(1)}}{\partial\mu}+\hbar\sum_{i = 1}^{N_{\lambda}}\beta_i\frac{\partial V^{(1)}}{\partial\lambda_i}-\frac{\hbar}{2}\sum_{a = 1}^{N_{\phi}}\gamma_j\phi_j\frac{\partial V^{(1)}}{\partial\phi_j} = \hbar\mu\frac{\partial V^{(1)}}{\partial\mu}+\mathcal{O}(\hbar^2) = -2\hbar\mathbb{B}+\mathcal{O}(\hbar^2) \ .
\end{equation}
The $\mathcal{O}(\hbar^2)$ terms correspond to the running of the couplings and the fields and these are exactly the terms that we ignore when we use the $t_*^{(0)}$ approximation of eq.~\eqref{eq:tstar-0}. We thus see that in the $t_*^{(0)}$ approach the tree-level hypersurface is noncharacteristic if $\mathbb{B}\neq0$. This is in agreement with  eqs.~\eqref{eq:t*-first}--\eqref{eq:tstar-0}, which are not defined for $\mathbb{B} = 0$. 

If we repeat the above calculation for the hypersurface $\hbar\mathbb{B}\equiv\hbar\mathbb{B}(\lambda,\phi)=0$ itself, we obtain
\begin{equation}\label{eq:B-char}
\hbar\mu\frac{\partial \mathbb{B}}{\partial\mu}+\hbar\sum_{i = 1}^{N_{\lambda}}\beta_i\frac{\partial\mathbb{B}}{\partial\lambda_i}-\frac{\hbar}{2}\sum_{a = 1}^{N_{\phi}}\gamma_j\phi_j\frac{\partial\mathbb{B}}{\partial\phi_j} = \hbar\mu\frac{\partial\mathbb{B}}{\partial\mu}+\mathcal{O}(\hbar^2)= \mathcal{O}(\hbar^2),
\end{equation}
since there is no explicit $\mu$ dependence in $\mathbb{B}$. Thus, the $\hbar\mathbb{B}=0$ hypersurface is characteristic up to the terms that we neglect in the $t_*^{(0)}$ approximation. This means that the RG flow cannot transform a point that is contained in this hypersurface to a point which is  outside it. This hypersurface is then isolated from the rest of the parameter space and divides it  into two regions that cannot be connected by the RG flow. Even if it has some nontrivial intersection with the $\hbar\Vone=0$ hypersurface, the flow cannot start from a point at the $\hbar\mathbb{B}=0$ hypersurface and arrive at another point at the $\hbar\Vone=0$ boudary because the latter hypersurface is also characteristic when $\mathbb{B}=0$. Thus the method ceases to be valid at the subspace given by $\hbar\mathbb{B}=0$ which is of codimension one in the whole parameter space spanned by $(\mu; \lambda, \phi)$.

One should note, however, that the one-loop correction to the effective potential given by eq.~\eqref{eq:one-loop-AB} indicates that $\hbar V^{(1)}$ has no explicit $\mu$-dependence if $\mathbb{B}$ vanishes. In this case the one-loop effective potential
\begin{equation}\label{eq:one-loop-B-zero}
V(\mu;\lambda,\phi) = V^{(0)}(\lambda,\phi)+\hbar\mathbb{A}(\lambda,\phi,\mathcal{M})
\end{equation}
has no explicit $\mu$-dependence and is thus an approximate solution to the RG equation, up to the $\mathcal{O}(\hbar^2)$ terms that have been neglected. It is important to note that there is no inconsistency here since the neglected terms would contribute to the two-loop RG-improved potential which we do not consider here (see appendix~\ref{chap:O-N-phi4}).

Therefore, if we \emph{define} the one-loop RG-improved potential to be the simple sum in eq.~\eqref{eq:one-loop-B-zero} for $\mathbb{B}=0$, and to be given by eq.~\eqref{eq:V-improved} in the rest of the parameter space, then the method presented here, within the $\tzero$ approximation, covers the entire parameter space. In section~\ref{sec:SU2cSM} we analyse the effective potential for the SU(2)cSM model and explicitly show how matching for the two solutions works.


\subsection{Resummation of Logarithmic Terms\label{sec:pert-struct}}
Closely related to the issue of accuracy of a fixed-loop-order improved effective potential is the issue of resummation of logarithmic terms. In section~\ref{sec:pre-RG}, in the case of massless $\phi^4$-theory, we concluded that
by running to the surface of vanishing one-loop correction we resummed the leading logarithms of the theory. However, when more scalar fields are present and different mass eigenvalues are featured in the effective potential, the choice~(\ref{eq:mu-kills-CW}) does not resum  all the leading logarithms as they are defined in the literature. In this section we  thoroughly examine the reason for this  and we  explain why this method is still able to resum the largest (dominant) logarithms in most of the parameter space, thereby providing a reliable approximation to the full effective potential.

The effective potential
can be computed in perturbation theory as the loop expansion
\begin{equation}\label{eq:eff}
V(\mu;\lambda,\phi) = \sum_{l = 0}^{\infty}\hbar^lV^{(l)}(\mu,\lambda,\phi) \ ,
\end{equation}
where $V^{(l)}(\mu,\lambda,\phi)$ is the $l$-th loop order contribution and, in particular, $V^{(0)}(\lambda,\phi)$ is the tree-level term. In general, the highest powers of the mass logarithms that appear in the renormalised $l$-th loop order term are of the form~\cite{Kastening2,Bando2,Ford}
\begin{equation}\label{eq:leading-logs-simple}
\begin{aligned}
&\hbar^l \prod_{a=1}^{N_m}\left(\log\frac{m_a^2(\lambda,\phi)}{\mu^2}\right)^{n_a} \ , \ \sum_{a = 1}^{N_m} n_a = l \ , \ n_a = 0, 1,... \ ,
\end{aligned}\end{equation}
and are referred to as the leading logarithms of the theory.
The appearance of explicit logarithmic terms can jeopardise the validity of perturbation theory if the mass logarithms are large. In particular, this implies that for large field values, which generally yield large logarithms, the loop expansion is not reliable. If one is interested in the large field behaviour of the effective potential, e.g. to study the vacuum stability of a given model, then one needs to enlarge the region of parameter space for which perturbation theory holds.

In the one-field case ($N_{\phi}= N_m = 1$), this is accomplished by resumming the (sub)leading logarithms by choosing $\mu$ to be the only mass eigenvalue. When $N_{m} > 1$, it is not possible to suppress all logarithms individually, since there is only one renormalisation scale. For this reason, one needs to employ multi-scale techniques in order to resum the leading logarithms, as they are defined in eq.~(\ref{eq:leading-logs-simple}), with correct coefficients. These techniques employ several arbitrary renormalisation scales. Work in this direction was initiated in \cite{Einhorn} and later pursued in \cite{Ford,Ford3,Ford4,Steele3}. In this paper we develop an alternative way of resumming logarithms in the effective potential. We now explain how this method resums large logarithms. 

We begin by noting that we can express the potential as an expansion in powers of a pivot logarithm, as was done\footnote{The term ``pivot logarithm'' was not used in~\cite{Kastening2,Ford}.} in~\cite{Kastening2, Ford}. Such an expansion is very instructive because it allows us to collect all the explicit dependence on the renormalisation scale in only one term. To achieve this, we use eq.~(\ref{eq:mass-logs-rho}) and we define $\log\frac{\mathcal{M}^2}{\mu^2}$ to be the pivot logarithm. Inserting eq.~(\ref{eq:mass-logs-rho}) into eq.~(\ref{eq:eff}) leads to the series
\begin{equation}\label{eq:powers-pivot-0}
V(\mu,\lambda,\phi) =\sum_{l=0}^{\infty}\hbar^l\sum_{n=0}^lw_{n}^{(l)}(\lambda,\phi)\left[\log\frac{\mathcal{M}^2}{\mu^2}\right]^n \ ,
\end{equation}
where $w^{(l)}_{n}\left(\lambda,\phi\right)$ may include logarithms of the ratios $\frac{m_a}{\mathcal{M}}$. Furthermore, we have $\mathbb{A} = w^{(1)}_0$ and $\mathbb{B} = w^{(1)}_1$ in eq.~(\ref{eq:one-loop-AB}).

The expansion in powers of a pivot logarithm is best suited when the pivot logarithm is the dominant one, i.e. when logarithms of the ratios  $\frac{m_a(\lambda, \phi)}{\mathcal{M}}$ are subleading and we have the relation
\begin{equation}\label{eq:dominant-pivot}
\left|\log\frac{\mathcal{M}^2}{\mu^2}\right| \gg \max_a\left\{\left|\log\frac{m_a^2(\lambda, \phi)}{\mathcal{M}^2}\right|\right\} \ .
\end{equation}
The advantage of working with a pivot-logarithm expansion of the effective potential becomes clear when we note that, due to eq.~(\ref{eq:mass-logs-rho}), all the mass logarithms $\log\frac{m_a(\lambda,\phi)^2}{\mu^2}$ are approximately equal when eq.~\eqref{eq:dominant-pivot} holds and we recover the one-field case. Thus the \emph{single} scale renormalisation group is sufficient to resum large pivot logarithms. It is clear that in certain regions of the parameter space a multi-scale problem can be reduced to a single-scale one by choosing a dominant pivot logarithm as described above.  However, the pivot might be different in different regions of the parameter space. In section~\ref{sec:resummation-t*}, we discuss how the method of the tree-level hypersuface allows us to resum large pivot logarithms without the need to explicitly compute a pivot scale $\mathcal{M}$ such that eq.~\eqref{eq:dominant-pivot} is satisfied.

To better understand eq.~(\ref{eq:dominant-pivot}), let us examine simple examples. The case in which $N_m = 1$ is trivial, since we may choose the pivot scale to be the only mass eigenvalue of the theory, i.e., $\mathcal{M} = m(\lambda,\phi)$ and eq.~\eqref{eq:dominant-pivot} is trivially satisfied. In the case with $N_m = 2$, and at a point in the parameter space where $\mu < m_1(\lambda,\phi) \leq m_2(\lambda,\phi)$ it is  sufficient to choose $\mathcal{M} \gg m_2(\lambda,\phi)$ for eq.~(\ref{eq:dominant-pivot}) to be satisfied, since $\left|\log\frac{\mathcal{M}}{\mu}\right| = \log\frac{\mathcal{M}}{\mu} \gg \log\frac{\mathcal{M}}{m_1} = \left|\log\frac{m_1}{\mathcal{M}}\right| = \max\left\{\left|\log\frac{m_1}{\mathcal{M}}\right|,\left|\log\frac{m_2}{\mathcal{M}}\right|\right\}$.

Moreover, a concrete example
is obtained by defining the pivot scale to be the radius in the scalar-field configuration space and taking the large-field limit, i.e. $\mathcal{M}=\rho \to \infty$ at fixed values of the ratios (angles) $\frac{\phi_j}{\rho}$. In this limit, all mass logarithms are equal to the radial logarithm. For very large values of the~fields along a particular direction in parameter space we thus conclude that, as expected, the angular logarithms are subleading with respect to the pivot logarithm (except for infrared effects if some of the masses tend to zero). 

In general, to keep the logarithms on the right-hand-side of inequality~\eqref{eq:dominant-pivot} small, one should try to find the~pivot scale $\mathcal{M}$ as close as possible to the $m_a$. Possible definitions of the pivot logarithm are e.g.
$$
\left|\log\frac{\mathcal{M}^2}{\mu^2}\right|:= \max_a \left(\left|\log\frac{m_a^2}{\mu^2}\right|\right),
$$
or $\mathcal{M}$ can be chosen as $\mathcal{M}'$ that minimises the expression
$$
\max_a \left(\left|\log\frac{m_a^2}{\mathcal{M' }^2}\right|\right),
$$
which leads to the following definition of $\mathcal{M}$ 
$$
\mathcal{M}=\sqrt{\left(\min_{a} m_a\right)\left( \max_{a} m_a\right)}.
$$
In both cases the strong inequality~\eqref{eq:dominant-pivot} leading to pivot dominance is problematic for (exponentially) large mass ratios and can only be fulfilled when $\mu \ll \mathcal{M}, m_a$ or $\mu \gg \mathcal{M}, m_a$.

Both resummations in leading logarithms and in the leading powers of the pivot logarithm  yield the~same result if all the mass eigenvalues are degenerate ($N_m = 1$). If they are not, the resummation in leading powers of the pivot logarithm is  only reliable when condition~(\ref{eq:dominant-pivot}) is satisfied.

We may change the summation variables in eq.~(\ref{eq:powers-pivot-0}) to obtain
\begin{align}
V(\mu,\lambda,\phi) &=\sum_{l=0}^{\infty}\hbar^l\sum_{n=0}^lw_{n}^{(l)}(\lambda,\phi)\left[\log\frac{\mathcal{M}^2}{\mu^2}\right]^n =\sum_{l=0}^{\infty}\sum_{n=0}^{\infty}\hbar^{l+n}w_{n}^{(l+n)}(\lambda,\phi)\left[\log\frac{\mathcal{M}^2}{\mu^2}\right]^n  \nonumber\\
&=:\sum_{l=0}^{\infty}\hbar^l f_{l}(\hbar;\mu,\lambda,\phi) ,\label{eq:resum-pivot}
\end{align}
where we defined the functions
\begin{equation}\label{eq:radial-logs-leading}
f_l(\hbar;\mu,\lambda,\phi) = \sum_{n=0}^{\infty}\hbar^{n}w_{n}^{(n+l)}\left(\lambda,\phi\right)\left[\log\frac{\mathcal{M}^2}{\mu^2}\right]^n \ ,
\end{equation}
which will be referred to as the $l$-th-to-leading functions. It is also convenient to define
\begin{equation}\label{eq:powers-wn}
w_{n}\left(\lambda,\phi\right) = \sum_{k=n}^{\infty}\hbar^kw_n^{(k)}\left(\lambda,\phi\right) =\mathcal{O}(\hbar^n) \ ,
\end{equation}
such that the potential can be written as
\begin{equation}\label{eq:radial-logs}
V(\mu,\lambda,\phi) = \sum_{n = 0}^{\infty}w_n(\lambda,\phi)\left[\log\frac{\mathcal{M}^2}{\mu^2}\right]^n \ .
\end{equation}
From the above considerations, we see that a practical alternative to the resummation of the $l$-th-to-leading logarithms is to work with the pivot logarithm, if it is the dominant one, and to resum the $l$-th-to-leading functions of eq.~(\ref{eq:radial-logs-leading}). This amounts to changing the set of relevant logarithms,
\begin{equation}\label{eq:sets-of-logs}
\mathbb{S}_1 = \left\{\log\frac{m_a^2}{\mu^2}, \ a = 1,..., N_m\right\} \leftrightarrow \mathbb{S}_2 = \left\{\log\frac{\mathcal{M}^2}{\mu^2}\ ,\ \log\frac{m_a^2}{\mathcal{M}^2}, \ a = 1,..., N_m\right\} \ .
\end{equation}
If one chooses to work with the set $\mathbb{S}_1$, multi-scale techniques are needed to resum all the logarithms with correct coefficients. On the other hand, if the pivot logarithm is the dominant one, we can work with the set $\mathbb{S}_2$ and use a single renormalisation scale to resum the powers of the pivot logarithm to a closed-form expression depending on the remaining logarithms. Expanding this result in powers of the pivot logarithm, one reproduces the correct coefficients of the elements of the set $\mathbb{S}_2$ in the effective potential.  We  prove this is true to leading order in the next section, whereas the general case is left for appendix~\ref{chap:O-N-phi4}.

Evidently, the field-dependent mass eigenvalues  vary for different regions of parameter space. Thus, to use the pivot logarithm resummation one would have to make different choices of the pivot scale $\mathcal{M}$ for different domains, such that eq.~(\ref{eq:dominant-pivot}) is satisfied. This shortcoming of the pivot logarithm method will be overcome in section~\ref{sec:resummation-t*}. There we show that, whenever a pivot logarithm can be identified and resummed, by running to the tree-level hypersurface one can resum the same contributions. Therefore, if the tree-level hypersurface method is applied, there is no need to determine a pivot logarithm fulfilling eq.~\eqref{eq:dominant-pivot}.

\subsubsection{Resummation of Powers of the Pivot Logarithm\label{sec:pivot-log}}

Let us now examine the resummation of powers of the pivot logarithm. The derivatives of the coefficients $w_n$ can be treated in perturbation theory (cf. eq.~(\ref{eq:d-loops})),
\begin{equation}
\mu\frac{\mathrm{d}}{\mathrm{d}\mu}w_n(\lambda,\phi)\equiv\sum_{l =1}^{\infty} \hbar^l\left(\sum_{i=1}^{N_{\lambda}}\beta_i^{(l)} \frac{\partial}{\partial \lambda_i}-\frac{1}{2}\sum_{a=1}^{N_{\phi}}\gamma_j^{(l)}\phi_j\frac{\partial}{\partial\phi_j}\right)w_n(\lambda,\phi) \equiv\sum_{l =1}^{\infty} \hbar^l \mathrm{d}^{(l)}w_n(\lambda,\phi) \ . \label{eq:powers-d}
\end{equation}
Inserting eqs.~(\ref{eq:beta-gamma}), (\ref{eq:resum-pivot}) and (\ref{eq:powers-d}) into eq.~(\ref{eq:RG-potential2}) yields
\begin{equation}\notag
\sum_{s=n+1}^{\infty}\hbar^s\sum_{l=1}^{s-n}\mathrm{d}^{(l)}w_n^{(s-l)} = 2(n+1)\sum_{s=n+1}^{\infty}\hbar^s w_{n+1}^{(s)}-2\frac{n+1}{\mathcal{M}}\sum_{s=n+2}^{\infty}\hbar^s\sum_{l=1}^{s-n-1} w_{n+1}^{(s-l)}\mathrm{d}^{(l)}\mathcal{M} \ ,
\end{equation}
which leads to the two recursive relations
\begin{align}
\mathrm{d}^{(1)} w_n^{(n)} &= 2(n+1) w_{n+1}^{(n+1)} \ , \label{eq:recursive-wnn} \\*
\sum_{l=1}^{s-n}\mathrm{d}^{(l)} w_n^{(s-l)}&=2(n+1) w_{n+1}^{(s)}-2\frac{n+1}{\mathcal{M}}\sum_{l=1}^{s-n-1} w_{n+1}^{(s-l)}\mathrm{d}^{(l)}\mathcal{M} \ , \ s > n+1 \ . \label{eq:recursive-wns}
\end{align}
By solving eq.~(\ref{eq:recursive-wnn}), we obtain the relation
\begin{equation}\label{eq:recursive-w00}
\left[\mathrm{d}^{(1)}\right]^nV^{(0)}\equiv\left[\mathrm{d}^{(1)}\right]^n w_0^{(0)} = 2^n n! w_n^{(n)} \ .
\end{equation}
The leading function (cf. eq.~\eqref{eq:radial-logs-leading}) in the effective potential reads
\begin{equation}\label{eq:pivot-f0}
f_0(\hbar;\mu,\lambda,\phi) = \sum_{n=0}^{\infty}\hbar^nw_n^{(n)}\left(\lambda,\phi\right)\left[\log\frac{\mathcal{M}^2}{\mu^2}\right]^n \ .
\end{equation}
We can now use eq.~(\ref{eq:recursive-w00}) to write eq.~(\ref{eq:pivot-f0}) in closed form,
\begin{align}
f_0(\hbar;\mu,\lambda,\phi)&=\sum_{n=0}^{\infty}\hbar^n\frac{1}{2^nn!}\left\{\left[\mathrm{d}^{(1)}\right]^nV^{(0)}(\lambda,\phi)\right\}\left[\log\frac{\mathcal{M}^2}{\mu^2}\right]^n  \notag\\
&=\sum_{n=0}^{\infty}\left\{\frac{1}{n!}\left[\hbar\mathrm{d}^{(1)}\right]^nV^{(0)}(\lambda,\phi)\right\}\left[\frac{1}{2}\log\frac{\mathcal{M}^2}{\mu^2}\right]^n  \notag\\
&= V^{(0)}\left(\bar{\lambda}(t_{\mathcal{M}}),\bar{\phi}(t_{\mathcal{M}})\right) \ , \label{eq:leading-logs-pivot}
\end{align}
where the running parameters $\bar{\lambda}, \bar{\phi}$ are defined as before, with the $\beta$-functions and anomalous dimensions truncated to one-loop order. We also denote $t_{\mathcal{M}} =\frac{1}{2}\log\frac{\mathcal{M}^2}{\mu^2} $.

Thus, we can RG-improve the effective potential by resumming the leading powers of the pivot logarithm into a closed-form expression which is the tree-level term evaluated at the point $\left(\bar{\lambda}(t_{\mathcal{M}}),\bar{\phi}(t_{\mathcal{M}})\right)$. Resummation of subleading terms can be done with the aid of eq.~(\ref{eq:recursive-wns}). However, to find closed-form expressions for subleading terms, it is more convenient to use recursive relations for the $f_k$ functions instead of eqs.~(\ref{eq:recursive-wnn}) and~(\ref{eq:recursive-wns}), as was done in \cite{Kastening2}. This is reviewed in appendix~\ref{chap:O-N-phi4}.

To see that it is not possible to resum all the leading logarithms with the above method, it is sufficient to note that $f_0$ only resums the highest powers of the pivot logarithm, which have the coefficients $w^{(n)}_n$. These coefficients do not in general coincide with the coefficients of the leading logarithms of eq.~(\ref{eq:leading-logs-simple}). For example, we have
\begin{equation}\notag
w^{(1)}_1 = \mathbb{B} \ ,
\end{equation}
which is a sum of coefficients of leading logarithms (cf. eqs.~(\ref{eq:AB-fcts}) and~(\ref{eq:leading-logs-simple})). This remains true to higher loop orders. A concrete example is given in appendix~\ref{chap:O-N-phi4}. Moreover, the above results imply that, starting from the tree-level form
\begin{equation}\label{eq:tree-level-pivot-t}
V^{(0)}\left(\bar{\lambda}(t),\bar{\phi}(t)\right) \ ,
\end{equation}
there is no choice of $t$ (choice of pivot scale) for which a Taylor expansion of eq.~(\ref{eq:leading-logs-pivot}) would generate the leading logarithms of eq.~(\ref{eq:leading-logs-simple}). This follows from the fact that the coefficients $w_n^{(n)}$ of this Taylor expansion are not the coefficients of leading logarithms. 

Nevertheless, if the pivot logarithm is the dominant one, the above method correctly resums the (sub)leading functions of the effective potential as they are defined in eq.~(\ref{eq:radial-logs-leading}). In particular, a Taylor expansion of $f_0$ reproduces the correct coefficients of the leading powers of the pivot logarithm in the effective potential. For this result to hold it is paramount that the pivot logarithm is dominant.

Indeed, consider the case in which $N_m = 1$. The pivot scale can be chosen to be the only mass eigenvalue of the theory, $\mathcal{M}=m$, such that eq.~(\ref{eq:leading-logs-pivot}) resums leading logarithms (cf. section~\ref{sec:pre-RG}). 
This is not true if the theory contains different mass eigenvalues because in this case the coefficients $w_n^{(k)}$ with $k>n+1$ include logarithms of the ratios $\frac{m_a}{\mathcal{M}}$. If the theory contains many large logarithms, subleading terms in the expansion (\ref{eq:resum-pivot}) might become comparable to or greater than the leading function, invalidating the perturbative expansion.

Therefore,
eq.~(\ref{eq:leading-logs-pivot}) is a valid truncation of eq.~(\ref{eq:resum-pivot}) only in regions of parameter space in which the logarithms of the ratios $\frac{m_a}{\mathcal{M}}$ are 
smaller than or negligible in comparison to the pivot logarithm.
As we commented previously, 
the choice of $\mathcal{M}$ in eq.~(\ref{eq:dominant-pivot}) varies for different domains in parameter space and thus in principle it might be difficult or even impossible to choose the pivot scale. We will now demonstrate that with the tree-level hypersurface method it is possible to overcome this difficulty.

\subsection{The Tree-Level Hypersurface: Resummation\label{sec:resummation-t*}}
We start from the expression for the one-loop RG-improved effective potential given in eq.~\eqref{eq:V-improved}, with $t_*$ approximated as in eq.~\eqref{eq:tstar-0}. 
We note that $t_*^{(0)}$ is invariant under redefinitions of $\mathcal{M}$ due to eqs.~(\ref{eq:rho-transform}). Starting from any arbitrary definition of $\mathcal{M}$, we exploit this invariance to redefine it such that eq.~\eqref{eq:dominant-pivot} is fulfilled. We then note that eq.~\eqref{eq:V-improved} can be expanded in a Taylor series
$$
V^{(0)}(\bar{\lambda}(t_*,\lambda),\bar{\phi}(t_*,\phi)) = \sum_{n = 0}^{\infty}\hbar^n w^{(n)}_n(\lambda,\phi)(2 t_*)^n = \sum_{n = 0}^{\infty}\hbar^n w^{(n)}_n(\lambda,\phi)\left(\log\frac{\mathcal{M}^2}{\mu^2}\right)^n+\cdots \ ,
$$
where we used eq.~(\ref{eq:recursive-w00}) and~\eqref{eq:tstar-0}. The ellipses hide terms with positive powers of $\frac{\mathbb{A}}{\mathbb{B}}$.  Such terms are subleading if $\mathcal{M}$ fulfils eq.~(\ref{eq:dominant-pivot}). The first term in the above equation coincides with the leading function~(\ref{eq:pivot-f0}) of the pivot logarithm.
We conclude that eq.~\eqref{eq:V-improved} necessarily includes a~resummation of a dominant pivot logarithm \emph{with respect to} the scale $\mu$. It is not, however, necessary to identify which is this dominant logarithm in a given region of parameter space because of the invariance of $t_*$ given in eq.~(\ref{eq:V1/2Brho4}) (and approximately in eq.~\eqref{eq:tstar-0}) under redefinitions of the pivot scale. We have thus overcome the issues outlined in the end of section~\ref{sec:pivot-log}. 

To put it differently, the characteristic displacement $t_*$ given in eqs.~(\ref{eq:V1/2Brho4}) and~\eqref{eq:tstar-0} automatically resums the highest powers of dominant logarithms in any region of parameter space. Therefore, the $\mathcal{O}(\hbar^2)$ terms in eq.~\eqref{eq:V-improved} are necessarily subleading, i.e., they are not the largest logarithms appearing in the set $\mathbb{S}_2$ of relation~(\ref{eq:sets-of-logs}).  Thus, whenever a multiscale potential can be reduced to a single-pivot-scale one using inequality~\eqref{eq:dominant-pivot}, our method accounts for the contributions from this scale.

We have just shown that the RG improved potential given by formulas~\eqref{eq:V-improved} and \eqref{eq:V1/2Brho4} reproduces the resummation of the powers of the pivot logarithm in the case when a dominant scale can be identified. One should underline, however, that the tree-level hypersurface method of RG improvement is not limited to the cases where a dominant logarithm exists and resums large logarithmic corrections regardless of the existence of the pivot logarithm. To lowest order in perturbation theory, it improves the effective potential by resumming powers of the logarithms that appear in the one-loop correction. This is achieved by evaluating the tree-level potential at the field-dependent scale given in eqs.~(\ref{eq:V1/2Brho4})--\eqref{eq:tstar-0}, which generalises the result of the pivot logarithm resummation in eq.~\eqref{eq:leading-logs-pivot}, and is a reliable approximation in perturbation theory as long as the running couplings $\bar{\lambda}(t_*,\lambda)$ are small. In appendix~\ref{chap:O-N-phi4}, we present a more general way of computing the characteristic displacement in perturbation theory and we  give a formula to compute $t_*$ to any order in $\hbar$.

\section{Applications\label{sec:applications}}
In this section we demonstrate our method of RG improvement in varous settings. First we study the issue of stability of generic scalar potentials. Next we discuss this issue within the context of the~SU(2)cSM, an~extension of the conformal Standard Model with an~extra SU(2)$_X$ gauge group and a~new scalar doublet. Within this model we also discuss the issue of validity of the choice of the tree-level hypersurface as a~boundary hypersurface for the RG equation. Furthermore, we perform an analysis of the~Higgs--Yukawa model in order to compare our method with the approach of the decoupling method of ref.~\cite{Casas}. Finally, we present computations beyond one-loop order within the context of the massless O(N)-symmetric model.
\subsection{Stability of the RG-Improved Effective Potential\label{sec:stability}}
A fundamental requirement that a scalar potential has to fulfil in order to be physically acceptable is boundedness from below, also referred to as stability or positivity (in the large field limit) of the potential. If a potential is unbounded from below, no state of  lowest energy exists and thus the theory based on such a potential is unstable. An effective potential which can be expressed as a polynomial of the scalar fields is bounded from below if it is positive in the large-field limit in any direction of the scalar-field configuration space. For the tree-level potential, this requirement can be translated into \emph{copositivity} conditions  on the~matrix of quartic couplings, if $V^{(0)}$ is a biquadratic form~\cite{Kannike:2012}. Indeed, in this case in the large-field limit, any terms which are quadratic in the fields can be neglected and the tree-level potential has the~scale-invariant form
\begin{equation}\notag
V^{(0)} =  \frac{1}{4}\sum_{a,b=1}^{N_{\phi}} \lambda_{ab}\phi_a^2\phi_b^2 \ .
\end{equation}
It is bounded from below if the coupling matrix $\tilde{\lambda}$ satisfies the \emph{copositivity} condition
\begin{equation}\label{eq:copositive}
\eta^T\tilde{\lambda}\eta = \sum_{a,b}\lambda_{ab}\eta_a\eta_{b} > 0 \ ,
\end{equation}
where $\eta$ is a vector with non-negative components in the basis $\{\phi_1^2,...,\phi_{N_{\phi}}^2\}$. A matrix $\tilde{\lambda}$ which satisfies eq.~(\ref{eq:copositive}) is called copositive~\cite{Hadeler,Kannike:2012}. For example, for a model with two scalar fields eq.~\eqref{eq:copositive} is fulfilled if $\lambda_{11}, \lambda_{22}>0$ and $\mathrm{det}\tilde{\lambda}>0$. 2. Analytical tree-level stability conditions are still not known for generic multi-scalar potentials (for recent developments see ref.~\cite{Ivanov:2018} and references therein).

The study of stability of scalar potentials acquires additional complications beyond tree level.  The loop corrections involve logarithimic contributions which become divergent in the large field limit for a fixed value of the~renormalisation scale $\mu$. This shows that simple perturbative approximations to the effective potential are not suitable for the study of stability of the potential and RG improvement is essential~\cite{Sher}.

With the method of RG improvement presented in section~\ref{chap:RG}, one can write the one-loop improved potential in a tree-level form, where the running couplings and fields are evaluated at a suitable field-dependent scale $t_*$ (cf.\,eqs.~\eqref{eq:V-improved}, \eqref{eq:V1/2Brho4} and~\eqref{eq:t*-first}). This implies that the tree-level stability criteria are applicable to the improved potential as long as they are written in terms of the running couplings evaluated at $t_*$ in the large field limit.

Let us now compute the field dependent scale $t_*$ in the large-field limit. Due to the invariance of eq.~(\ref{eq:t*-first}) under redefinitions of the pivot scale $\mathcal{M}$, we may choose the pivot scale to be the radial variable in the scalar field configuration space, as in eq.~\eqref{eq:radial-pivot}, and rewrite eq.~\eqref{eq:tstar-0} as 
\begin{equation}\notag
t_* = \frac{1}{2}\log\frac{\rho^2}{\mu^2}+\frac{1}{2}\frac{\mathbb{A}\left(\lambda,\phi,\rho\right)}{\mathbb{B}\left(\lambda,\phi\right)} +\mathcal{O}(\hbar) =  \frac{1}{2}\log\frac{\rho^2}{\mu^2}+\frac{1}{2}\frac{\mathfrak{a}\left(\lambda,\frac{\phi}{\rho}\right)}{\mathfrak{b}\left(\lambda,\frac{\phi}{\rho}\right)}+\mathcal{O}(\hbar) \ ,
\end{equation}
where we defined $\mathfrak{a}:=\frac{\mathbb{A}}{\rho^4}$ and $\mathfrak{b}:=\frac{\mathbb{B}}{\rho^4}$. It is straightforward to see that both $\mathfrak{a}$ and $\mathfrak{b}$ only depend on the angular variables $\frac{\phi}{\rho}$. In the large field limit, the angles are kept fixed and we take $\rho\to\infty$, which leads to\footnote{In practice, there is always a cutoff to the model we study, it can either be a scale close to the Landau pole of one of the couplings or a scale at which effects of the high-energy  theory (e.g.\ quantum gravity) become significant. Thus in fact we cannot take the limit of $\rho\to\infty$. Nonetheless, we consider cases in which the cut-off is high enough to take $\rho$ such that $\frac{1}{2}\frac{\mathfrak{a}\left(\lambda,\frac{\phi}{\rho}\right)}{\mathfrak{b}\left(\lambda,\frac{\phi}{\rho}\right)}$ becomes negligible.}
\begin{equation}
t_* \stackrel{\rho\to\infty}{=}\frac{1}{2}\log\frac{\rho^2}{\mu^2} \ .
\end{equation}
Thus, $t_*$ is a monotonic function of the scalar fields for very large field values. This implies that we can assess the stability of the one-loop RG-improved effective potential by simply evaluating the tree-level stability conditions with the running couplings evaluated at some large scale. Evidently, the method is only valid if the scalar couplings do not become non-perturbative below that scale.

This tree-level criterion for the stability of RG-improved potentials was advocated in ref.~\cite{Sher} and is frequently used in the literature. However, to our best knowledge, the validity of such a criterion has never been rigorously proven. The method of the tree-level hypersurface presented herein can thus confirm that these stability criteria are justified. With the generalisation of the method presented in appendix~\ref{chap:O-N-phi4}, we conclude that the stability of the $n$-loop improved effective potential can be assessed by the tree-level stability conditions, where $n$ is the loop order at which the RG functions are truncated. 

\subsection{SU(2)cSM model\label{sec:SU2cSM}}
In this section we briefly discuss some properties of the RG-improved potential for a specific model in order to supplement the formal derivations of the preceeding sections with a concrete example. The model studied here is the conformal SM extended with a scalar field which is a singlet under the SM gauge group while a doublet under a new local SU(2)$_X$ symmetry. The SM fields are singlets under SU(2)$_X$. We refer to the model as SU(2)cSM for simplicity. It is an example of a classically conformal model in which all mass scales are generated through loop corrections. It has been studied in refs.~\cite{Hambye, Carone, Khoze, Karam2, Khoze-Plascencia} and is also a subject of a forthcoming paper~\cite{project2}. 

The tree-level potential for the model is given by 
\begin{equation}
\Vtree(h,\varphi)=\frac{1}{4}\left(\lambda_1 h^4 + \lambda_2 h^2 \f^2 + \lambda_3 \f^4\right),\label{eq:Vtree-SU2}
\end{equation}
where $h$ and $\f$ should be interpreted as the background fields for the SM Higgs doublet and  the new SU(2)$_X$ doublet, respectively.\footnote{Due to the symmetry of the model the potential only depends on the absolute values of the fields and the same is true for the effective potential. Thus, $h$ and $\f$ are the background fields for the radial components of the fields.}
The one-loop correction is given by the general formula of eq.~\eqref{eq:general-one-loop-term}\footnote{As stated before, we used the $\ms$ renormalisation scheme and Landau gauge in our computations.}. In what follows we refer to the one-loop unimproved potential as $V_1=\Vtree+\Vone$. The one-loop RG-improved potential, referred to as $V$, is computed using the tree-level hypersurface method and is defined as in eq.~\eqref{eq:V-improved}, with $t_*$ approximated as $t_*^{(0)}$ given in eq.~\eqref{eq:tstar-0}.

For the sake of this example we fix the values of the scalar couplings and the SU(2)$_X$ coupling $g_X$ at the scale $\mu=246\g$ to the following values: $\lambda_1=0.1754$, $\lambda_2=-0.0049$, $\lambda_3=-0.0038$, $g_X=0.83$. The choice of the couplings is such that the model yields a radiatively generated minimum at $h=246\g$, $\f=2200\g$. Moreover, the model correctly reproduces the masses of the SM particles, and predicts a very small mixing between the two scalar particles, such that it is likely consistent with all current LHC data. 
Furthermore, there are no Landau poles up to the Planck scale and the tree-level stability conditions evaluated at the Planck scale hold.

\begin{figure}[ht]
\center
\includegraphics[height=.31\textwidth]{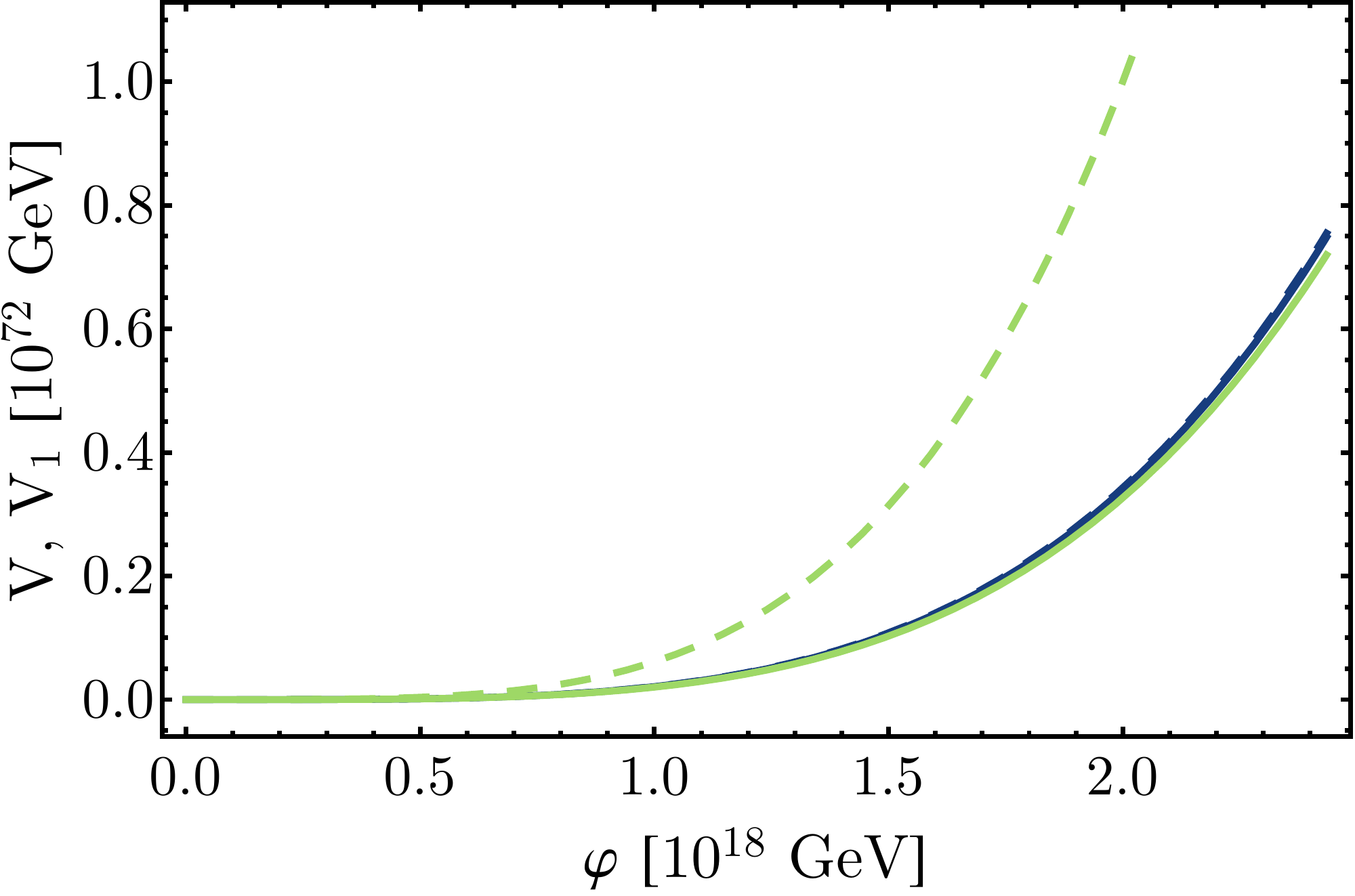}\hspace{.3cm}
\includegraphics[height=.31\textwidth]{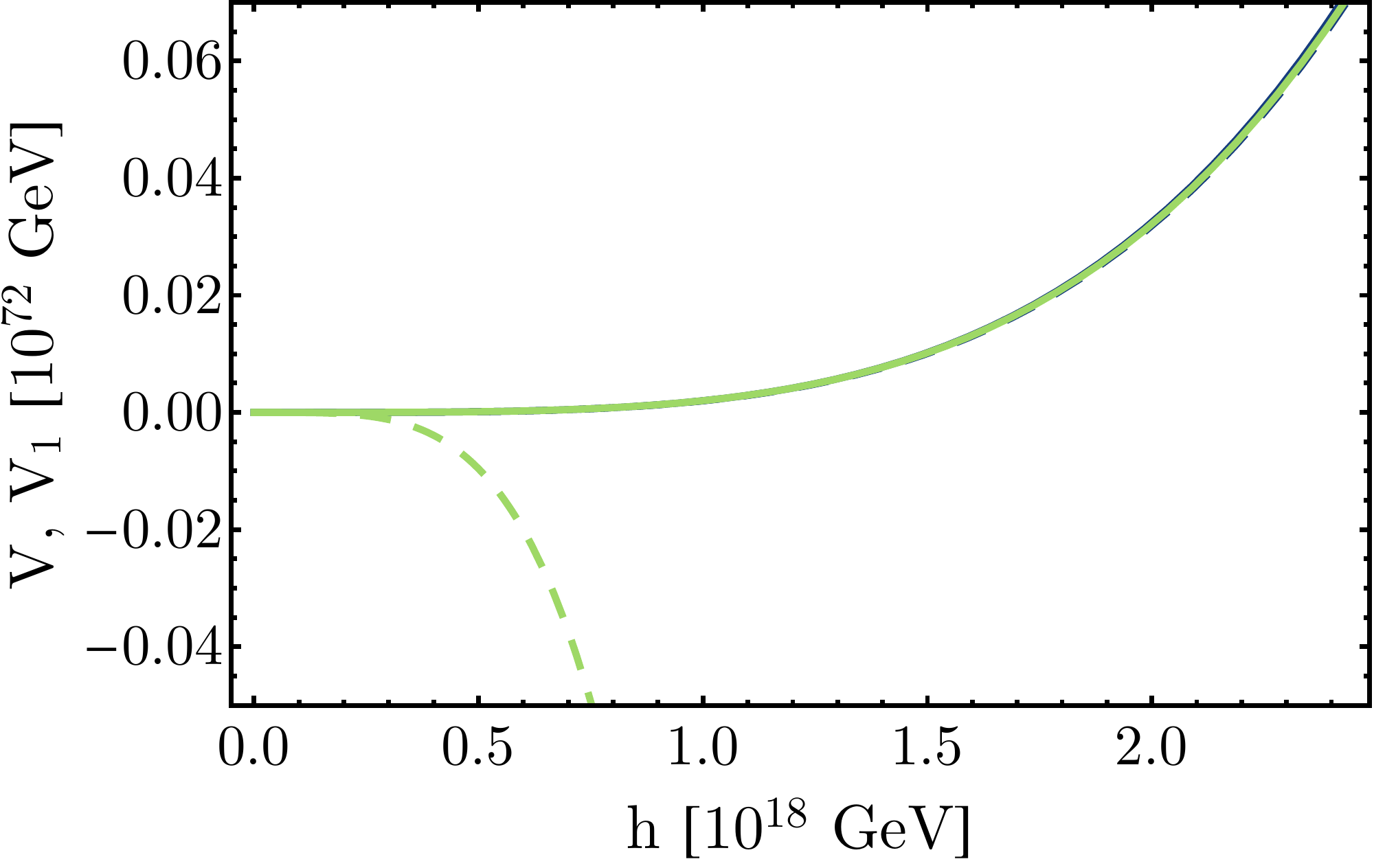}
\caption{The one-loop unimproved effective potential $V_1$ (dashed lines) and the one-loop improved effective potential $V$ (solid lines) for two values of renormalisation scale $\mu$, $\mu=246\g$ (green) and $\mu=M_P$ (dark blue). In the left panel the value of $h$ is fixed to $246\g$ and the potentials are plotted as functions of $\f$. In the right panel $\f=600\g$ and the potentials are plotted along the $h$ direction. Note that three of the curves overlap, that is why it is hard to see the dark blue curves.\label{fig:stability}}
\end{figure}

Let us start from checking the large-field behaviour of the improved and unimproved potential. Since one of the reasons to consider RG improvement is to extend the validity of the effective potential for large fields, we should see a significant difference between $V$ and $V_1$ in this regime. In figure~\ref{fig:stability} one can see $V$ (solid lines) and $V_1$ (dashed lines) plotted for two choices of the renormalisation scale $\mu=246\g,\ M_P$, where $M_P=2.44\cdot10^{18}\g$ is the reduced Planck scale. In the left panel the potentials are displayed along the $\f$ direction with $h$ fixed and in the right panel along the $h$ direction with $\f$ fixed. Of course the improved potential does not depend explicitly on $\mu$ but it has some residual dependence, through the value of $t_*$. This is confirmed by the plots --- the curves corresponding to the improved effective potential defined at different scales differ only slightly. Moreover, it is clear that the improved potential $V$ is positive in the large field limit, in agreement with the fact that the tree-level stability conditions, evaluated at the Planck scale, are satisfied. The behaviour of the unimproved potential is strikingly different. While the potential $V_1$ computed at the Planck scale agrees approximately at large scales with the improved one, the one defined at the electroweak scale diverges significantly from the others. This was to be expected --- since it was defined for a fixed scale, far from the Planck scale, its behaviour for large field values is no longer reliable. In particular, it gives a wrong answer to the question of stability of the effective potential.

This simple example shows clearly that the potential that is RG-improved with the use of the method presented in this article behaves as expected in the large field limit. Moreover, it confirms the usefulness of the improvement procedure for the issue of stability of the potential.

In what follows we study in detail the limitation to our method due to the the ill-posedness of the Cauchy problem with the choice of the tree-level boundary hypersurface for $\mathbb{B}=0$ within the $\tzero$ approximation, expanding on the general discussion of section~\ref{sec:B0}.

\begin{figure}[h!t]
\center
\includegraphics[width=.45\textwidth]{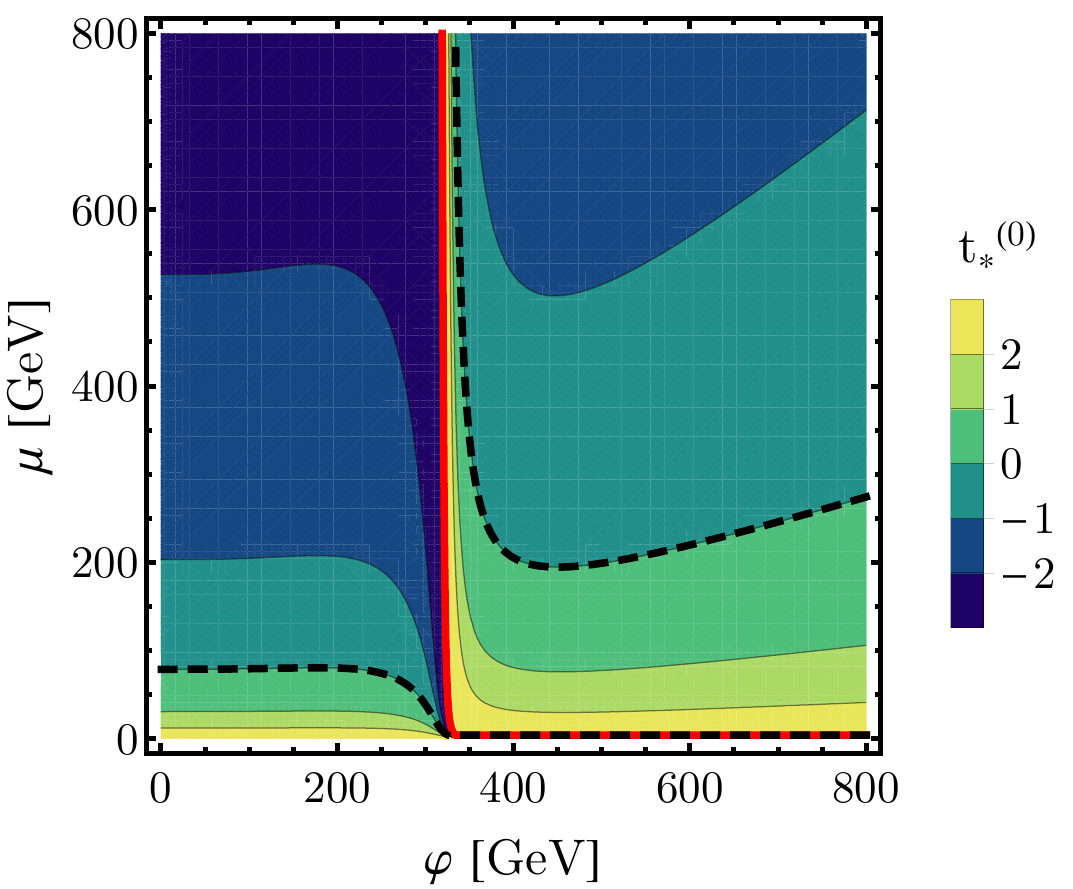}\hspace{.4cm}
\includegraphics[width=.45\textwidth]{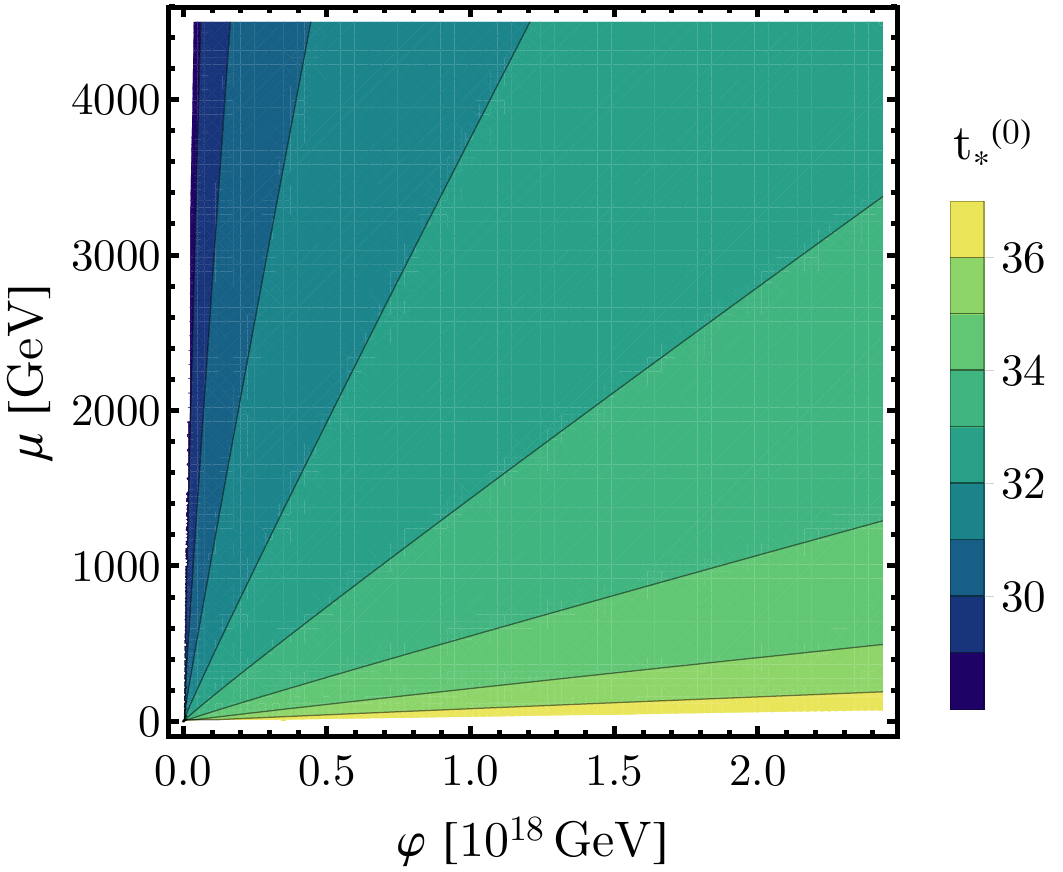}
\caption{Contour plots of $t_*^{(0)}(\mu, h, \f)$ for $h=246\g$ in different ranges of $\f$. In the left panel also the hypersurfaces $\mathbb{B}=0$ (red solid line) and $\Vone=0$ (black dashed line) are displayed. \label{fig:surfaces-t}}
\end{figure}

The tree-level hypersurface given by the equation $\Vone=0$ and the hypersurface of $\mathbb{B}=0$ are defined in the $(\mu, \lambda_1, \lambda_2,\lambda_3,h,\f)$ parameter space. However, since the characteristic equations for the couplings do not depend on the fields we can solve the equations for the couplings and represent the hypersurfaces in a three-dimensional space of $(\mu, h, \f)$. In figure~\ref{fig:surfaces-t} slices of these surfaces for a fixed value of $h=246\g$ are displayed. One can see that the $\Vone=0$ hypersurface is discontinuous across $\mathbb{B}=0$. In the right lower part of the plot the two hypersurfaces converge, $\Vone=0$ being always below $\mathbb{B}=0$. \footnote{One should note, that the region of very small $\mu$ (infrared) should not be trusted, since there the top Yukawa coupling (which contributes to $\mathbb{B}$), as well as $\lambda_1$ develop Landau poles.}

\begin{figure}[h!t]
\center
\includegraphics[width=.48\textwidth]{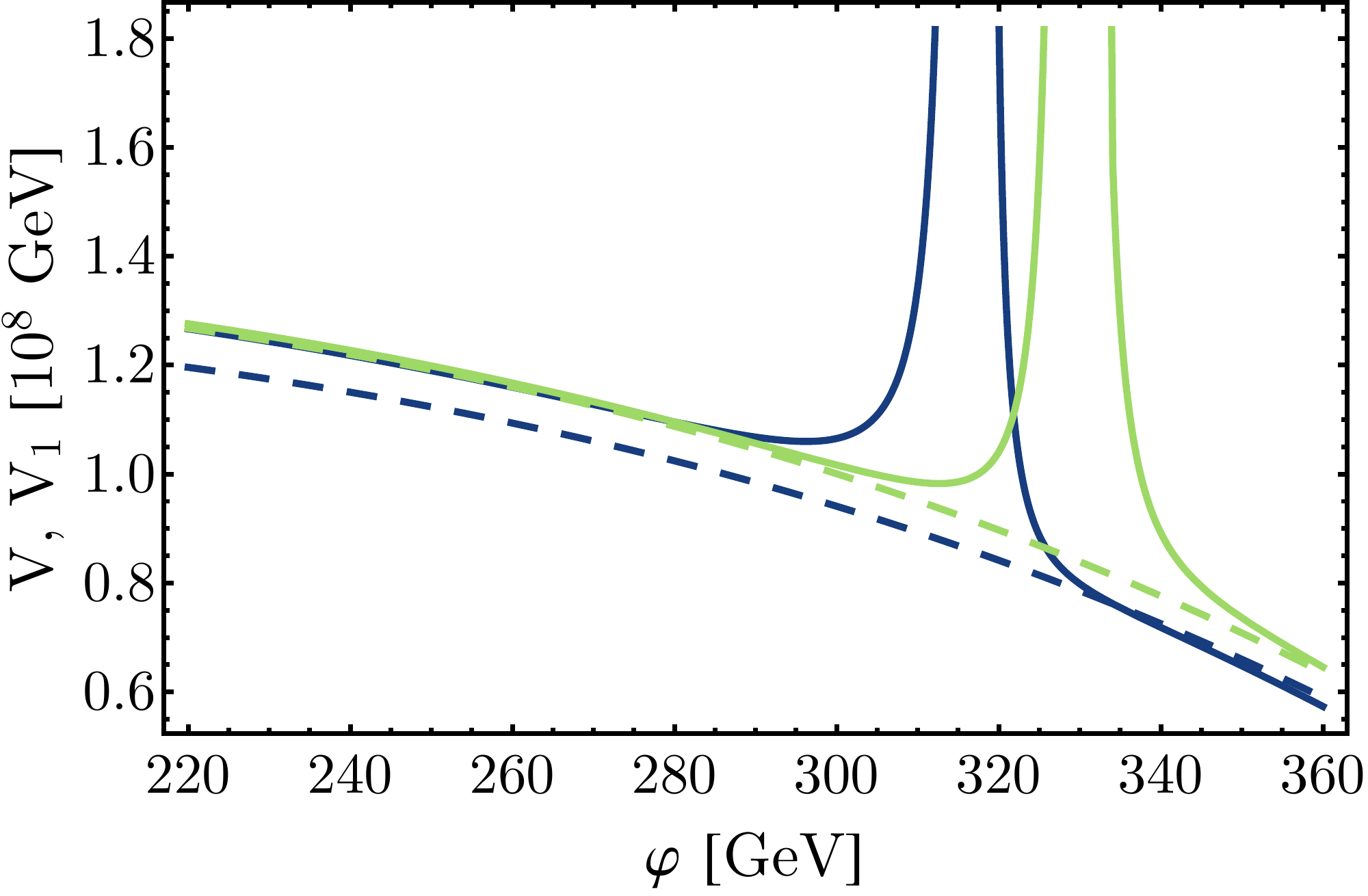}\hspace{.4cm}
\includegraphics[width=.48\textwidth]{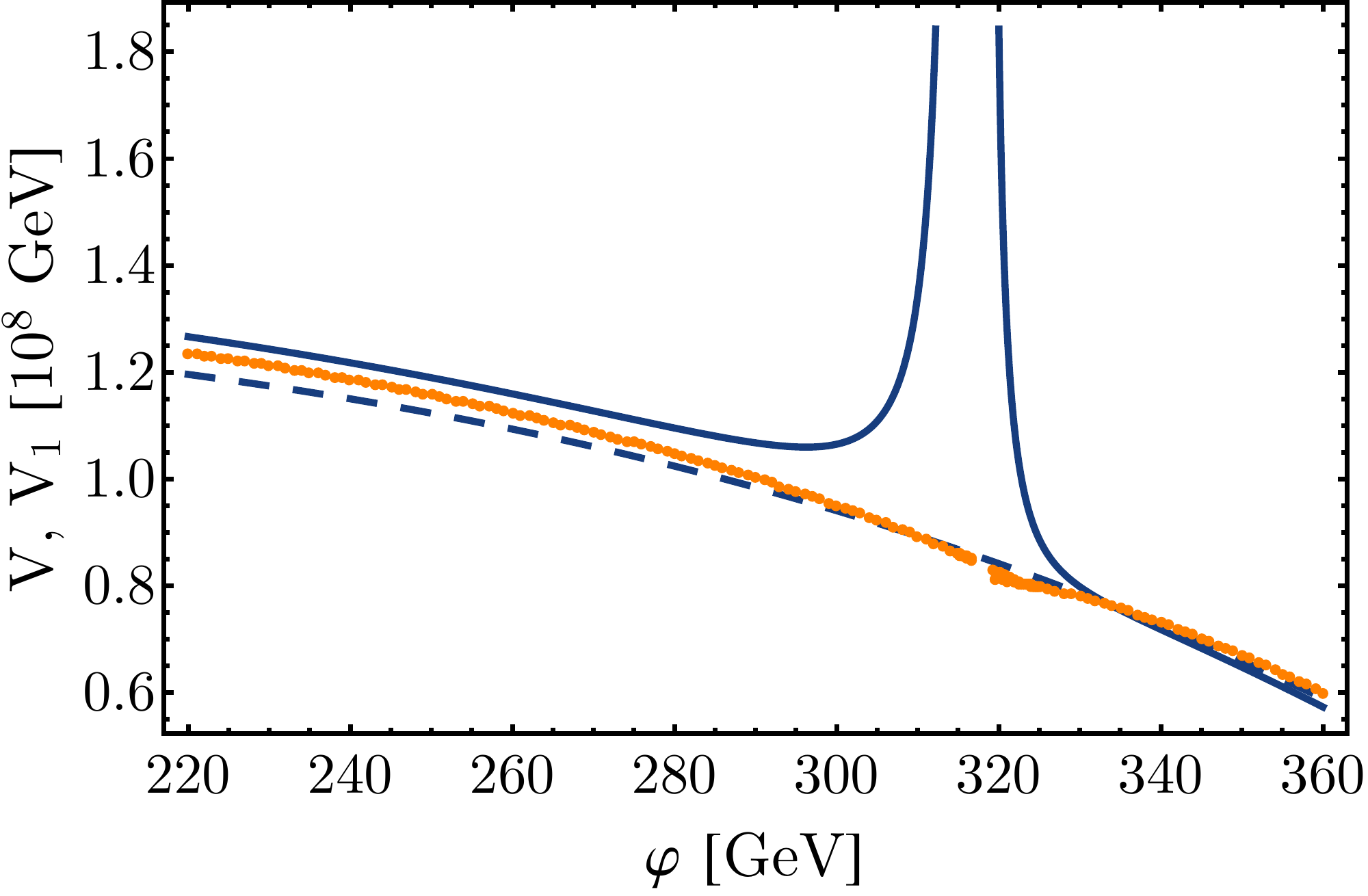}
\caption{Left panel: The one-loop unimproved effective potential $V_1$ (dashed lines) and the one-loop improved effective potential $V$ (solid lines) for two values of renormalisation scale $\mu$, $\mu=246\g$ (green) and $\mu=800\g$ (dark blue). The value of $h$ is fixed to $246\g$ at the scale $\mu=800\g$ (the value at $\mu=246\g$ is obtained as a value of the running field at this scale). Right panel: The one-loop improved (with $\tzero$ approximation) and unimproved potential for $\mu=800\g$ showed in the left panel (dark blue) superimposed with the one-loop improved potential (with full $t_*$) (orange).\label{fig:matching}}
\end{figure}

Figure~\ref{fig:surfaces-t} also shows a contour plot of the $t_*^{(0)}(\mu, h,\f)$ function (with the value of $h$ fixed to  246$\g$). We expect that $t_*^{(0)}$ acquires large (positive ore negative) values as we approach the $\mathbb{B}=0$ surface, since $\Vone$ is characteristic for $\mathbb{B}=0$, and the $\mathbb{B}=0$ surface is characteristic itself. This means that the RG flow becomes tangent to the $\Vone=0$ surface for $\mathbb{B}=0$ and it is impossible to reach the tree-level hypersurface from points for which $\mathbb{B}=0$. This indeed can be seen in figure~\ref{fig:surfaces-t}. It is thus clear that our method does not work for the points for which $\mathbb{B}=0$. It is also not reliable in the vicinity of the $\mathbb{B}=0$ hypersurface, since then $t_*^{(0)}$ is very large and the running couplings enter non-perturbative regime.\footnote{Unavoidably, in typical models under consideration, there are some Landau poles in the deep ultraviolet or infrared.} This can be also seen in the plot of figure~\ref{fig:matching} which shows again the improved and unimproved potentials as functions of $\f$ for two choices of the  renormalisation scale, $\mu=246,\, 800\g$, and fixed $h=246\g$ around the point where $\mathbb{B}=0$. The funnel that can be seen in the improved potential corresponds to the divergence of $t_*^{(0)}$ at $\mathbb{B}=0$. However, as we already pointed out in section~\ref{sec:B0}, when $\mathbb{B}=0$ the one-loop unimproved potential has no explicit $\mu$ dependence and in the vicinity of $\mathbb{B}=0$ it provides a good approximation of the solution to the RG equation. This also holds for small $\mathbb{B}$,\footnote{By small $\mathbb{B}$ we mean $\mathbb{B}$ small in comparison with $\Vone$ such that $t_*^{(0)}$ is very large, see eq.~\eqref{eq:tstar-0}.} and thus we can match the two functions to obtain an (approximate) solution to the RG equation that is valid in the whole parameter space. This can be clearly seen by inspecting the plots for $\mu=246\g$.

The matching of the two functions $V$ and $V_1$ seems more problematic for $\mu=800\g$. One may ask whether applying the full $t_*$ solution (cf.\ eqs.~\eqref{eq:define-implicit-t*} and~\eqref{eq:t*-first}--\eqref{eq:V1/2Brho4}) rather than $t_*^{(0)}$ may improve the situation. If we inspect figure~\ref{fig:surfaces-t} we can see that to the left from $\mathbb{B}=0$ for $\mu=800\g$ the value of the parameter $t_*^{(0)}$ is negative, below -2, which means that this point lies rather close to the Landau poles of the top Yukawa coupling and the $\lambda_1$ coupling. Thus, the running of the couplings may be important in this region, while it is neglected in the $\tzero$ computation. 

We find the solution for $t_*$ numerically, looking for a root of eq.~\eqref{eq:define-implicit-t*} and then apply formula~\eqref{eq:V-improved} to obtain the one-loop improved potential. The result is presented in the right panel of figure~\ref{fig:matching}, together with the approximation obtained with $\tzero$. One can immediately notice that the matching with the one-loop potential is improved when the full solution for $t_*$ is used. The remaining discrepancy may be explained by the proximity of the Landau pole mentioned before, which suggests that in order to obtain a more accurate answer,  higher-loop corrections should be included. Moreover, with the use of $t_*$ we can evaluate the RG improved potential very close to the point where $\mathbb{B}=0$ without encountering divergences. Thus, the results are improved when $t_*$ is used. The only puzzling issue is the behaviour of the improved potential around $\phi=320\g$. In this region two solutions for $t_*$ exist. This possibility has already been anticipated in section~\ref{sec:applicability}. In principle, this is a challenge for our method, however one can see that in the discussed case it is clear which solution is the correct one. We have not found any other value of $\phi$ (for the other values of parameters fixed) for which two solutions for $t_*$ would be possible. Moreover,  as we discussed previously, in the vicinity of the $\mathbb{B}=0$ hypersurface the unimproved one-loop potential constitutes a good approximation so the improvement procedure in this region of parameter space is not needed.

We thus see that the RG improvement procedure applies to models with more than one scalar field and it is useful for the study of stability of the potential. The difficulties associated with the $\mathbb{B}=0$ hypersurface and the uniqueness of the $t_*$ solution that we encountered can be overcome and the most important limitation of our method of RG improvement is the requirement of perturbativity of the couplings evaluated at the $t_*$ (or the approximate $t_*^{(0)}$) scale.

\subsection{Higgs--Yukawa model\label{sec:Casas}}

In this section we apply the method of RG-improvement introduced in the present paper to the Higgs--Yukawa model.  This simple model was discussed in ref.~\cite{Casas} to illustrate the decoupling method of RG improvement and below we use it to display similarities and differences between the two approaches.\footnote{At the final stage of preparation of this paper ref.~\cite{Iso} appeared which discusses the decoupling method in the context of Higgs--Yukawa model and a model with two scalar fields.}

The Higgs-Yukawa model consists of a scalar Higgs field $\phi$ and a Dirac fermion $\psi$, the interactions are defined by the following Lagrangian
\begin{equation}
\mathcal{L}=\frac{1}{2}\partial_{\mu}\phi \partial^{\mu}\phi - \frac{1}{2} m^2 \phi^2 - \frac{1}{4!}\lambda \phi^4 + \bar{\psi}\left(i \slashed{\partial}+g\phi\right)\psi - \Lambda,
\end{equation}
where $\Lambda$ is the cosmological constant term. The one-loop potential is given by formula~\eqref{eq:general-one-loop-term}. The expressions for the field-dependent masses and the $\beta$ and $\gamma$ functions for this model can be found in ref.~\cite{Casas}.\footnote{Due to the decoupling method implemented in ref.~\cite{Casas}, Heaviside theta functions appear in the expressions for $\beta$ and $\gamma$ functions. To obtain the RG functions applicable in our approach one should simply remove the theta functions from these expressions. Moreover, the formula for the $\gamma$ function should be rescaled by a factor of 2 to conform with our convention for the RG equation, see eq.~\eqref{eq:RG-potential2}.}

In short, the approach of ref.~\cite{Casas} consists of implementing the decoupling theorem. In the effective potential, the contributions from respective particle species decouple at a scale $\mu$ equal to their field-dependent mass. Thus, a given particle contributes to the effective potential only at scales above its field-dependent mass. Therefore, if the potential is evaluated at a field-dependent scale $\tilde{\mu}_*$ that is below all the field-dependent masses, the effective potential reduces to the tree-level potential evaluated at $\tilde{\mu}_*$. For more details we refer the reader to ref.~\cite{Casas}.

\begin{figure}[ht]
\center
\includegraphics[width=.48\textwidth]{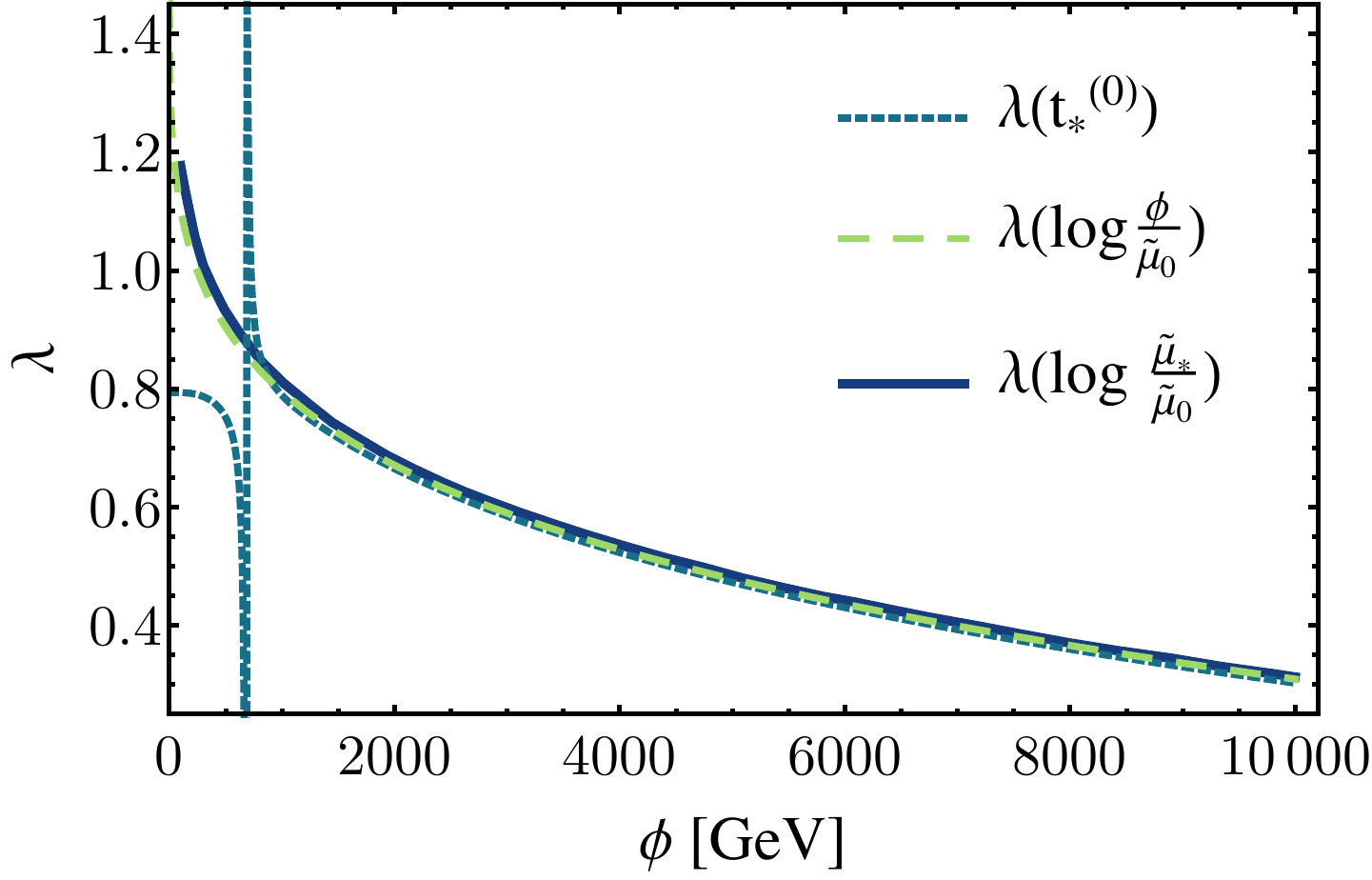}\hspace{.3cm}
\includegraphics[width=.48\textwidth]{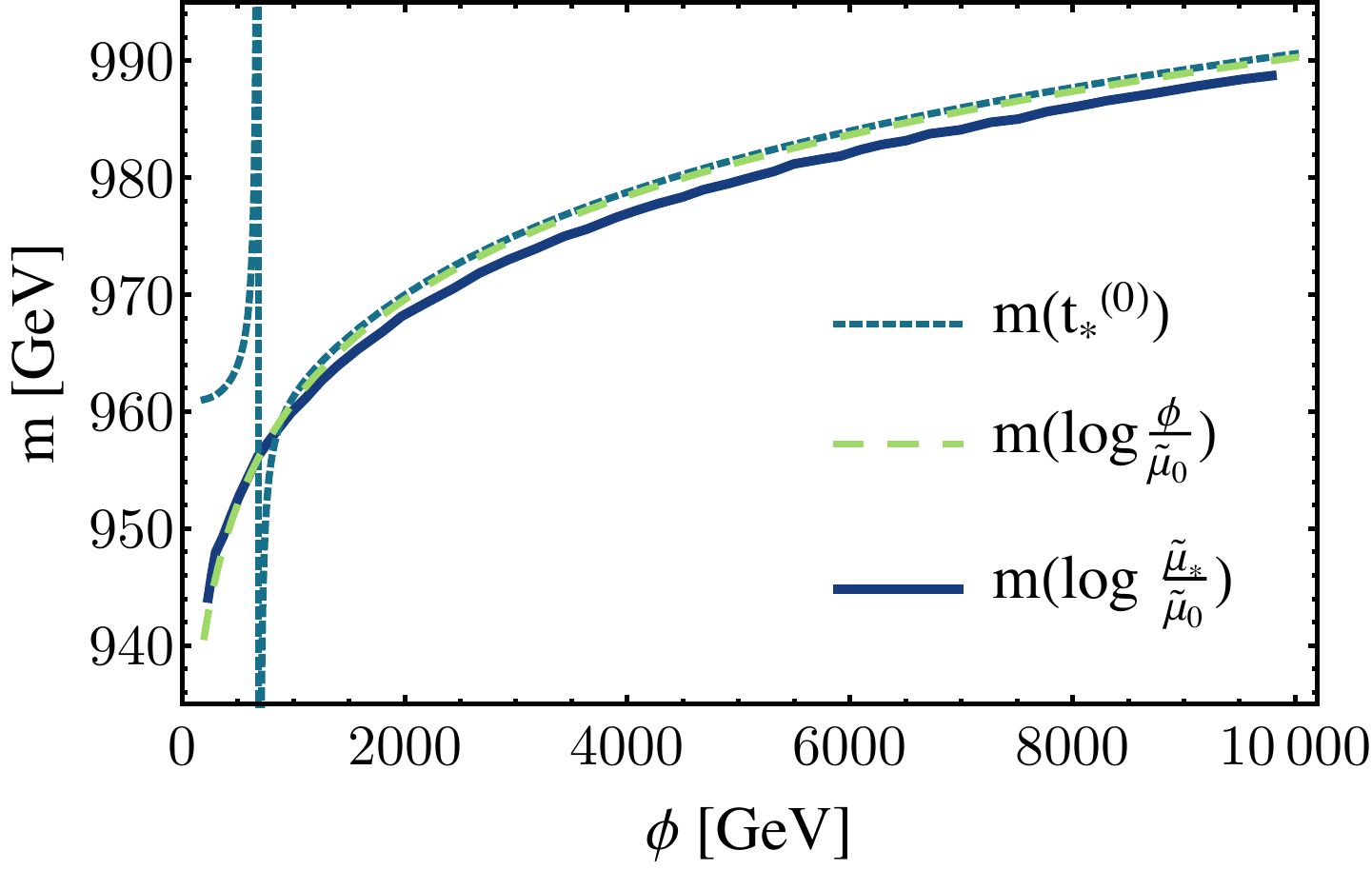}
\caption{The running coupling constant $\lambda$ (left panel) and the running mass term $m$ (right panel) evaluated at different field-dependent scales: $t_*^{(0)}$, $\log \frac{\phi}{\tilde{\mu}_0}$ and $\log\frac{\tilde{\mu}_*}{\tilde{\mu}_0}$, where $\tilde{\mu}_0=\textrm{exp}(3/4)\cdot 10 \textrm{TeV}$. The curves for  $\log\frac{\tilde{\mu}_*}{\tilde{\mu}_0}$ are reproduced from ref.~\cite{Casas}.
\label{fig:Casas-couplings}}
\end{figure}

It is clear that for both the approach of ref.~\cite{Casas} and of the present paper the choice of the correct scale for evaluating the effective potential is crucial. Nonetheless, the details of the approaches are significantly different. It is thus interesting to compare the results obtained with the use of the two methods. In figure~\ref{fig:Casas-couplings} the running coupling $\lambda$ (left panel) and the mass parameter $m$ (right panel) evaluated at $\log \frac{\tilde{\mu}_*}{\tilde{\mu}_0}$ in the approach of ref.~\cite{Casas} and at $t_*^{(0)}$ in our approach, as functions of the scalar field are compared. The definition of $\tilde{\mu}_0$ is such as to incorporate the constant $\chi_a=\frac{3}{2}$ from the one-loop effective potential (eq.~\eqref{eq:general-one-loop-term}), $\tilde{\mu}_0=\exp(3/4)\mu_0$. For comparison we also show the running of the coupling evaluated at the ``traditional'' value $\log \frac{\phi}{\tilde{\mu}_0}$. The initial values for the running are chosen as in ref.\cite{Casas}: $\lambda=0.1$, $g=1$, $m^2=10^6\g^2$ at $\mu_0=10\;\mathrm{TeV}$. 

One can see that going from high energies down the plots agree very well until the point where $\mathbb{B}=0$ and the running coupling evaluated at $t_*^{(0)}$ diverges. This issue has been discussed in the previous section in detail, where we have also explained how this difficulty can be overcome. Interestingly, the running of the couplings with the field is significantly different for the two approaches for the values of $\phi$ to the left of the $\mathbb{B}=0$ funnel. This is only an indicator that the two scales $\tilde{\mu}_*$ and $\mu_*^{(0)}=\mu\exp{t_*^{(0)}}$ indeed differ, however it does not tell us much about the accuracy or consistency of the two methods since the scales $\tilde{\mu}_*$ and $\mu_*^{(0)}$ were chosen such that to give the correct approximation to the effective potential and not the couplings. We note, however, that the behaviour of the couplings evaluated at $\tilde{\mu}_*$ can be well reproduced by choosing the field-dependent scale as is usually done in the one-field case as $\log \frac{\phi}{\tilde{\mu}_0}$.

\begin{figure}[ht]
\center
\includegraphics[height=.31\textwidth]{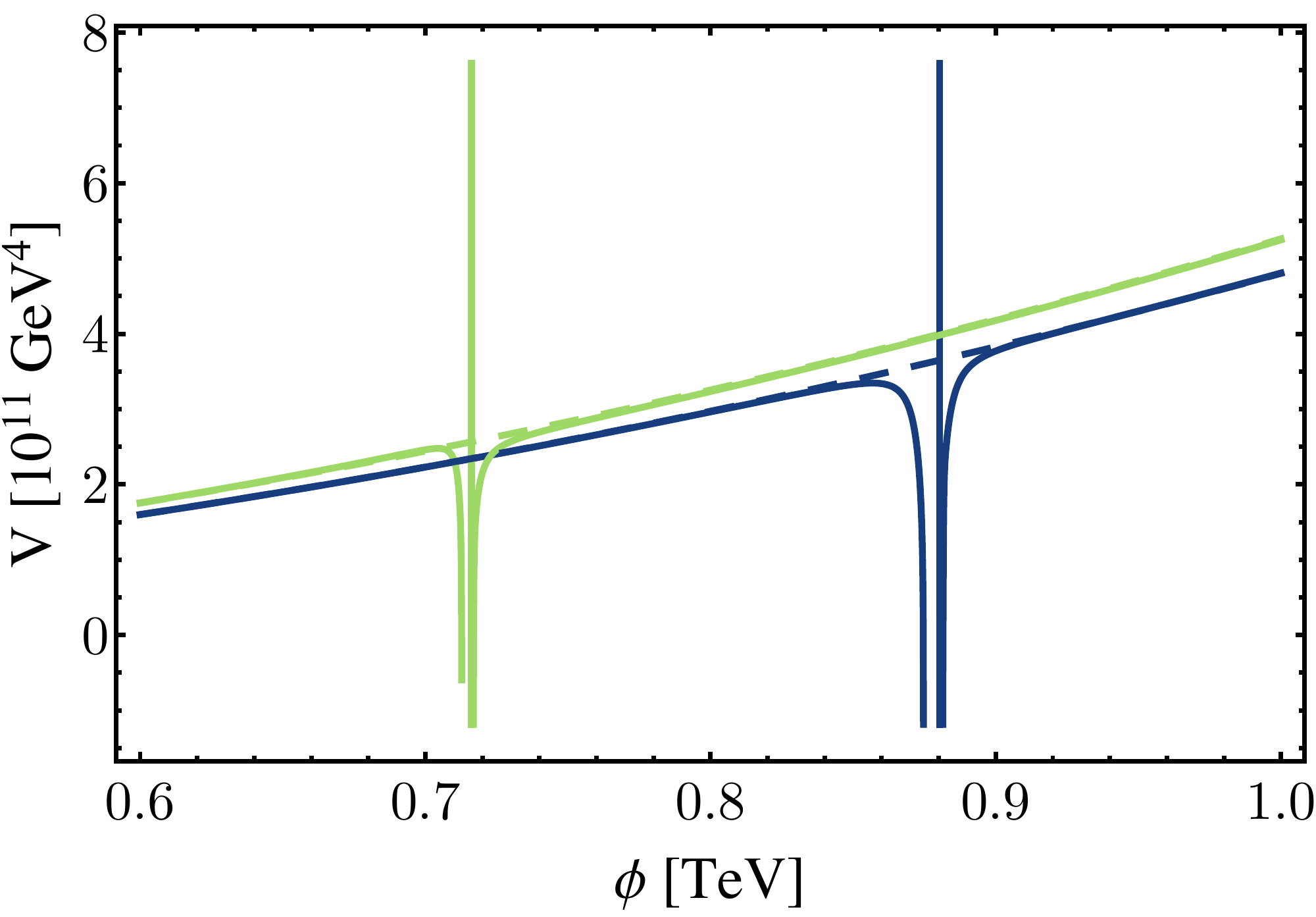}\hspace{.4cm}
\includegraphics[height=.31\textwidth]{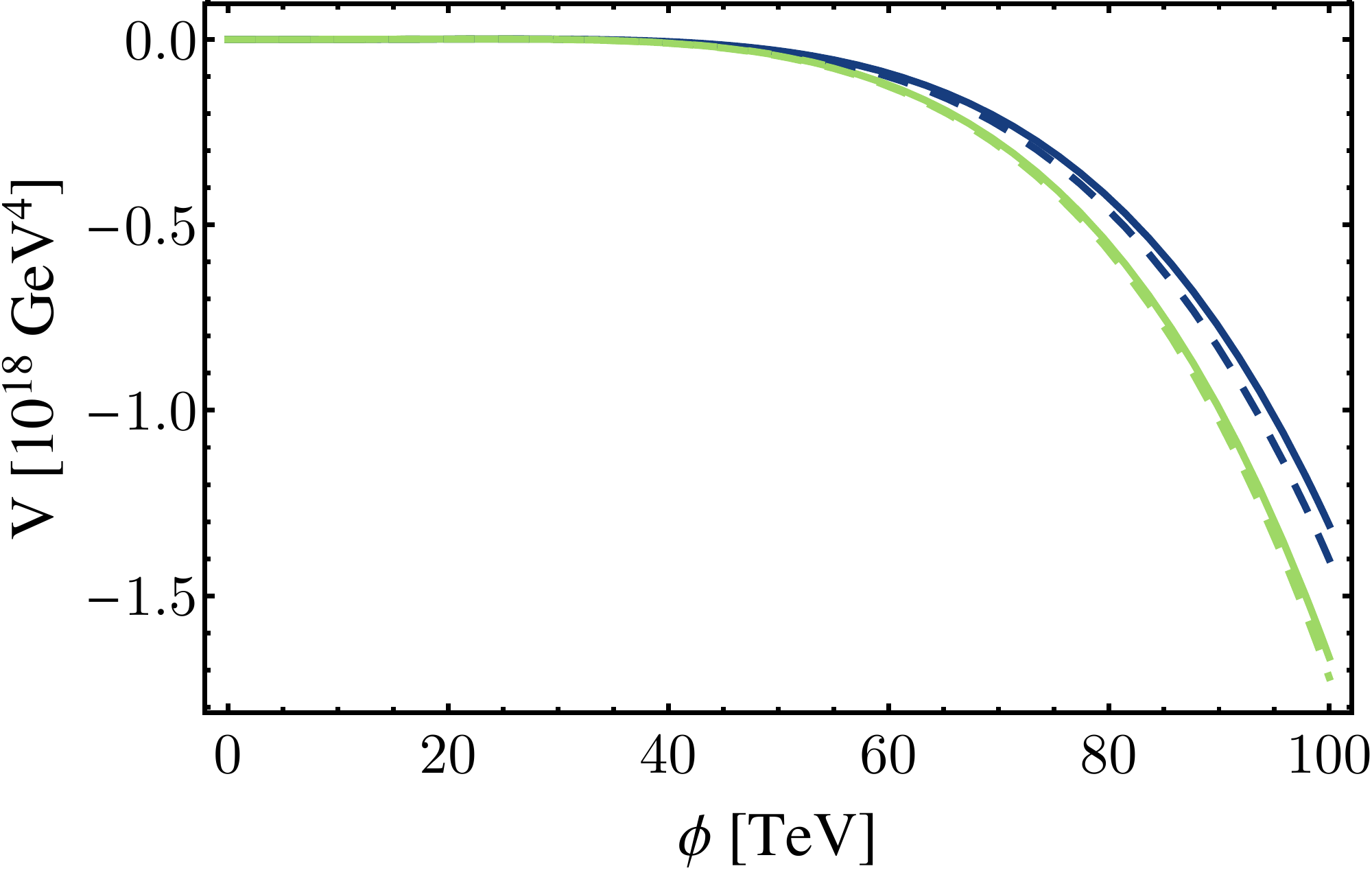}
\caption{Potential as a function of $\phi$ obtained using different methods of RG-improvement for two different reference scales, $\mu_0=246\g$ (dark blue) and $\mu_0=10^4\g$ (green). Solid lines correspond to the tree-level potential evaluated at $t_*^{(0)}$ (see eq.~\eqref{eq:V-improved}), while dashed curves correspond to the one-loop effective potential evaluated at the scale $\log\frac{\phi}{\tilde{\mu}_0}$. \label{fig:Casas-potential}}
\end{figure}

To gain better understanding of these two methods, we should compare the potentials obtained using the two approaches. Unfortunately, in ref.~\cite{Casas} a plot of the RG-improved effective potential as a function of $\phi$ was not given. We will thus compare the RG-improved potential obtained with the use of our method with the one-loop potential evaluated at $\log\frac{\phi}{\tilde{\mu}_0}$ since this choice reproduced well the behaviour of the running couplings presented in ref.~\cite{Casas}. In figure~\ref{fig:Casas-potential} we present the comparison of the RG-improved effective potential defined in eq.~\eqref{eq:V-improved} (with $t_*$ approximated by $t_*^{(0)}$, solid lines) and the one-loop effective potential evaluated at the field-dependent scale $\log\frac{\phi}{\tilde{\mu}_0}$ (dashed lines) for two values of  $\mu_0$, $\mu_0=246\g$ (dark blue) and $\mu_0=10^4\g$ (green). One can see that, apart from the vicinity of the point where $\mathbb{B}=0$, the two potentials agree very well. In particular, they agree very well on both sides of the $\mathbb{B}=0$ hypersurface, which shows that the discrepancies in the running of the couplings visible in fig.~\ref{fig:Casas-couplings} when combined with the running of the field, do not invalidate the RG-improved potential.\footnote{One should note that the field-dependent masses at some point become negative and hence the one-loop potential becomes complex. Therefore to compute $t_*^{(0)}$ we use the real part of the one-loop correction, and similarly fig.~\ref{fig:Casas-potential} shows only the real part of the one-loop potential.}

\begin{figure}[ht]
\center
\includegraphics[height=.31\textwidth]{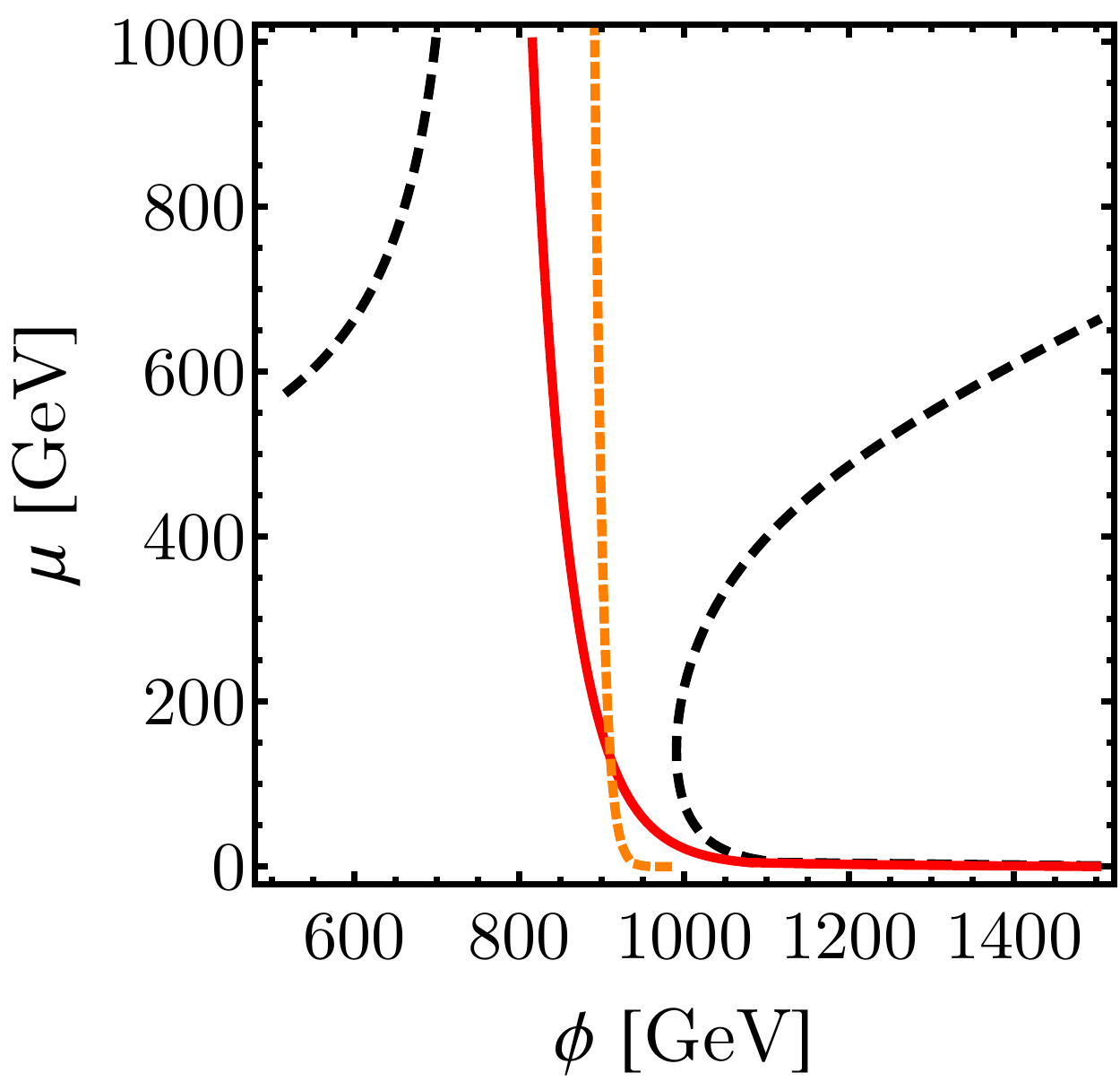}\hspace{1.5cm}
\includegraphics[height=.31\textwidth]{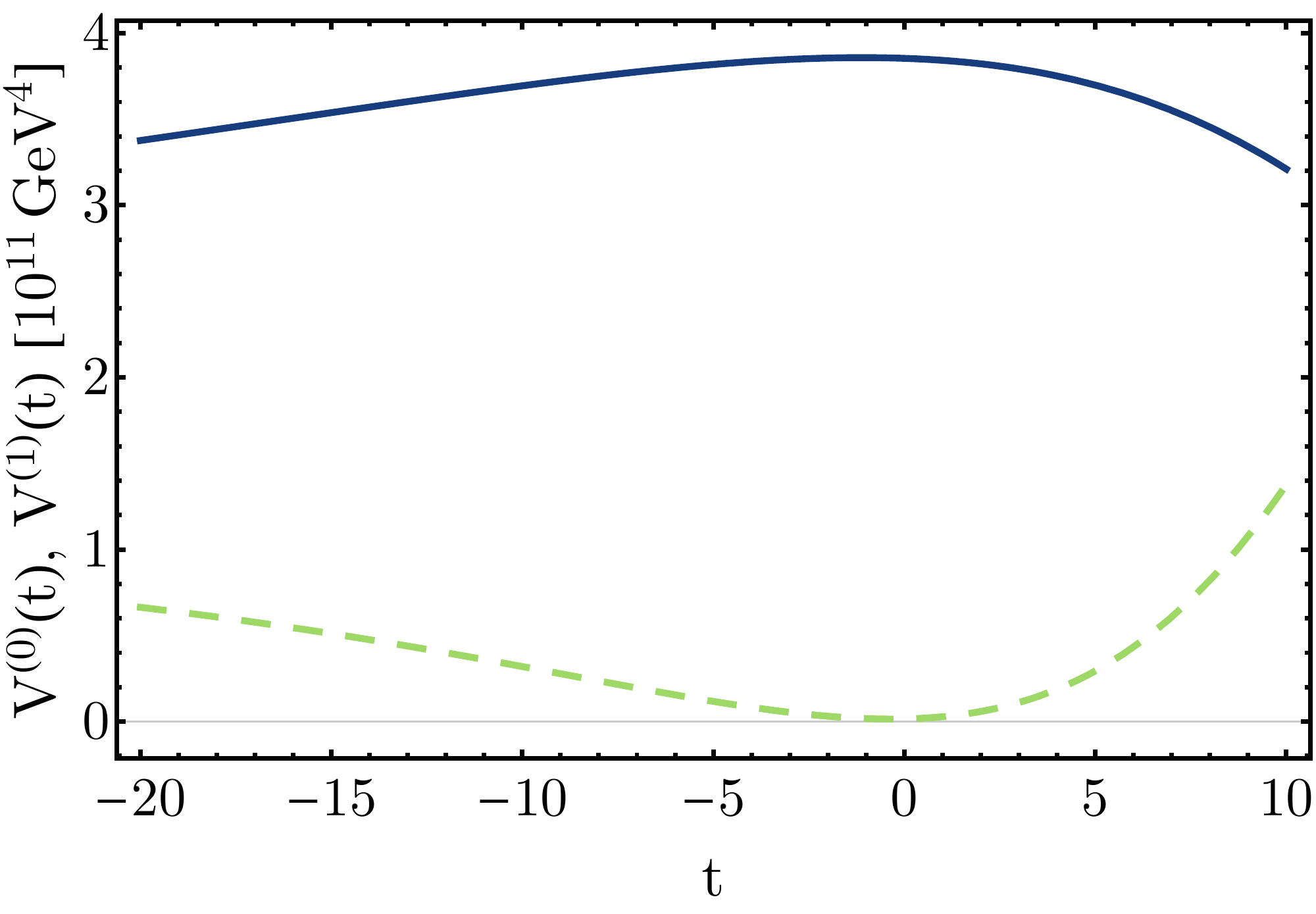}
\caption{Left panel: The $\Vone=0$ (black dashed line) and $\mathbb{B}=0$ (red solid line) hypersurfaces, the short-dashed orange line is a characteristic curve passing throug a point with $\phi=900\g$ and $\mu=400\g$ for $t=0$. Right panel: The tree-level potential (blue solid line) and the one-loop correction (green dashed line) plotted as a function of $t$ with the same initial conditions as the characteristic curve from the left panel. The minimal value of the one-loop correction is $V^{(1)}=1.5\cdot10^9\g^4$ and is attained for $t=-0.35$.\label{fig:no-t*}}
\end{figure}

In the end, let us use the Higgs--Yukawa model to discuss the issue of existence of solutions to eq.~\eqref{eq:define-implicit-t*} for $t_*$.  In the left panel of figure~\ref{fig:no-t*} the $\mathbb{B}=0$ (red solid line) and $\Vone=0$  (black dashed line) hypersurfaces are presented, running of the couplings is taken into account. The left branch of the $\Vone=0$ hypersurface asymptotes the $\mathbb{B}=0$ hypersurface. Looking for a solution for $t_*$ at a given point in the parameter space geometrically can be represented as drawing a characteristic curve passing through this point and searching for its intersection with the $\Vone=0$ hypersurface. The orange short-dashed curve represents a characteristic curve which does not cross the $\Vone=0$ hypersurface. This means that for any point along this curve a solution for $t_*$ does not exist.

This example proves that indeed, as suggested in section~\ref{sec:applicability}, in some small regions of the parameter space it is not possible to nullify the (one-)loop correction to the effective potential by choosing an appropriate scale. However, our example suggests that the occurrence of this problem is limited to the region in the vicinity of the $\mathbb{B}=0$ hypersurface where, as we argued before, the one-loop approximation to the effective potential works well and there is no need to employ RG improvement. Moreover, even if it is not possible to find $t_*$ that would cancel the one-loop correction, we can fix $t$ to a value that minimises its contribution. In the right panel of figure~\ref{fig:no-t*} the tree-level potential is plotted as a function of $t$ for the same initial conditions as that for the characteristic curve in the left panel, together with the one-loop correction. It is clear that if we choose $t$ to minimise the one-loop correction, it is well negligible. We have also checked that for a given scale, while changing the values of $\phi$ we can always find such a value of $t$ that the one-loop correction is negligible in comparison with the tree-level contribution. 

The above considerations show that the issues with finding a unique solution for $t_*$ are rather technical complications than serious impediments to our method of RG improvement.

\subsection{Massless O(N)-symmetric $\phi^4$ theory}\label{chap:O-N-phi4-2}

To illustrate our method of RG-improvement beyond one-loop level we analyse  massless O(N)-symmetric $\phi^4$ theory. In this model there are $N$ scalar fields and a single scalar coupling $\lambda$. The tree-level potential is given by the simple formula
\begin{equation}
V^{(0)}(\lambda,\phi) = \frac{\lambda}{4!}\rho^4, 
\end{equation}
where
\begin{equation}
\rho^2 = \sum_{j = 1}^{N}\phi_j^2. 
\end{equation}
In appendix~\ref{app:ON} we provide formulas for two-loop effective potential and $\beta$ and $\gamma$ functions following refs.~\cite{Kastening, Ford2}. 

In the present section we use the techniques described in appendix~\ref{chap:O-N-phi4} to evaluate the difference between the two-loop and one-loop improved effective potential. The two-loop effective potential is obtained by evaluating the tree-level potential at a field-dependent scale computed at two-loop order, $t_*^{(0)}+t_*^{(1)}$. In general, we expect $t_*^{(1)}$ to be a small correction to $t_*^{(0)}$ as long as the running coupling $\lambda(t)$ remains a small parameter across a large range of scales and no Landau poles occur. We can compute $t_*^{(0)}$ and higher-loop correction $t_*^{(1)}$ using eqs.~(\ref{eq:t0}), (\ref{eq:t1}) and~(\ref{eq:ONphi4-betas}).  It follows that $t_*^{(0)}$ is a sum of a constant term and a term proportional to the radial logarithm $\log\frac{\rho^2}{\mu^2}$, while $t_*^{(1)}$ has the same structure with an additional overall factor of $\lambda$. This confirms that with $t_*^{(0)}$ we can resum the leading, and with $t_*^{(1)}$ the subleading logarithmic contributions, and that $t_*^{(1)}$ is a small correction as long as the coupling constant remains perturbative.

\begin{figure}[ht]
\center
\includegraphics[height=.45\textwidth]{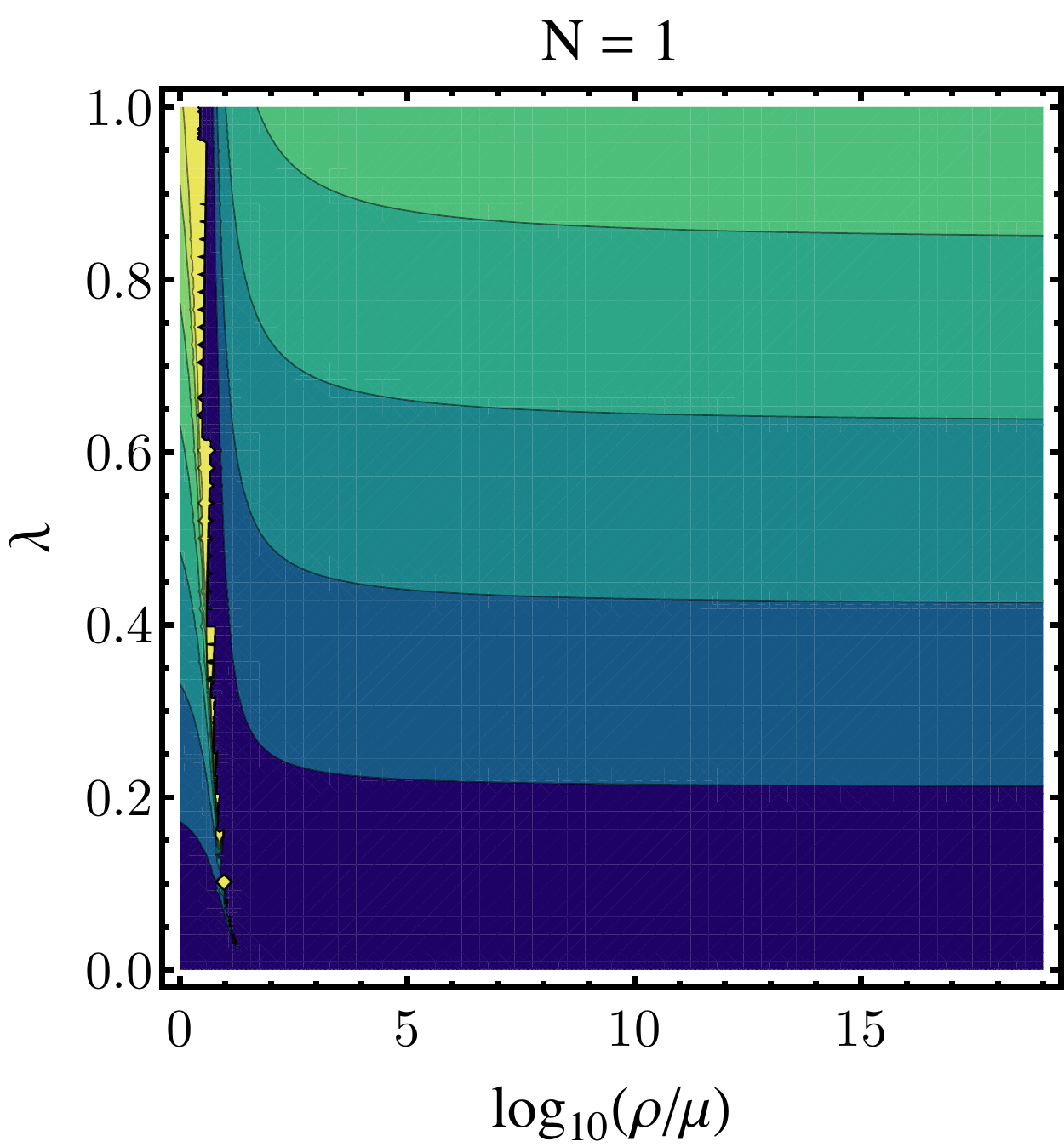}\hspace{.7cm}
\includegraphics[height=.45\textwidth]{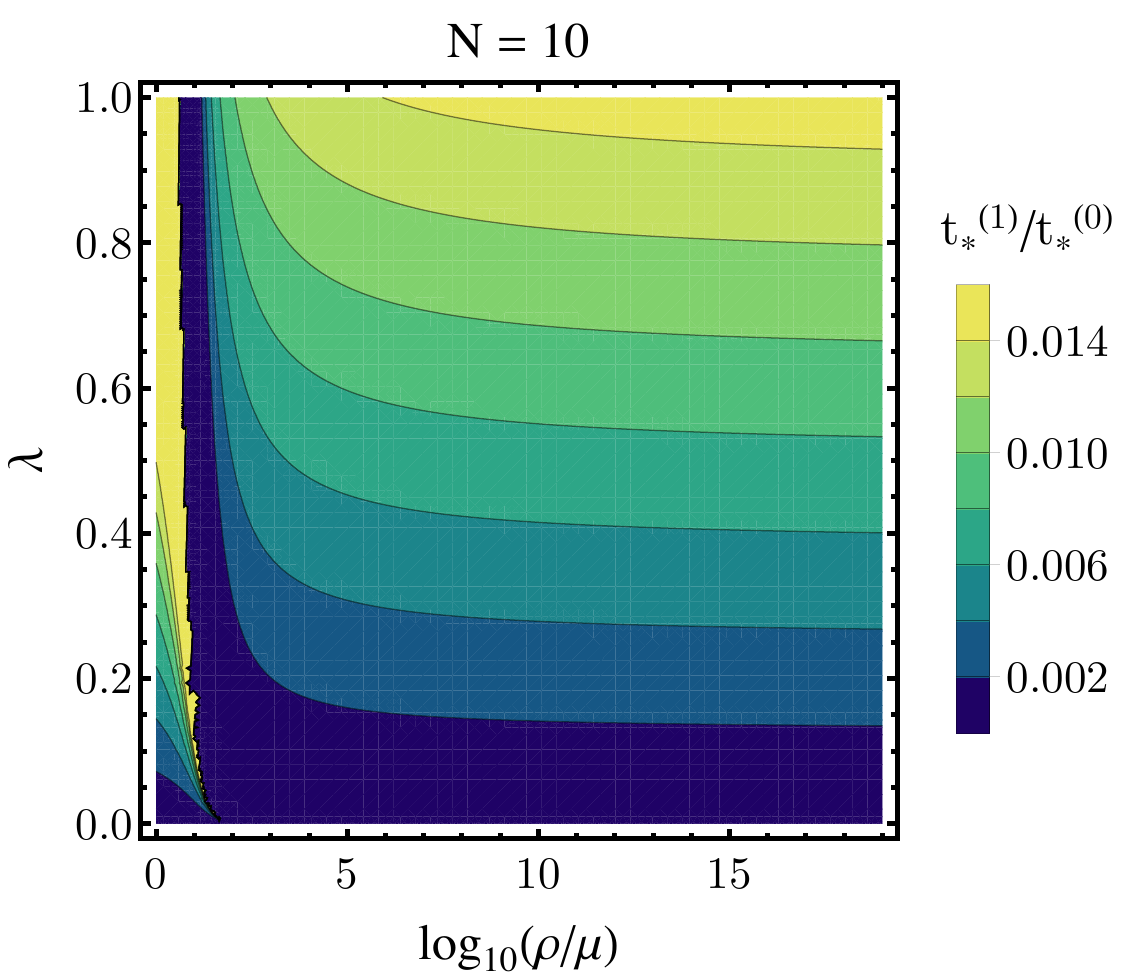}
\caption{Contour plot of the ratio  $t_*^{(1)}/t_*^{(0)}$ for $N=1$ (left) and $N=10$ (right) scalar fields as a function of the coupling $\lambda$ and the decimal logarithm of the ratio $\frac{\rho}{\mu}$.  The colour coding is the same for both of the plots.
\label{fig:ONphi41-10}}
\end{figure}

In figure~\ref{fig:ONphi41-10}, contour plots of the ratio $\frac{t_*^{(1)}}{t_*^{(0)}}$ are shown for two values of the number $N$ of scalar fields as functions of the coupling $\lambda$ and the ratio $\frac{\rho}{\mu}$.
We note that, as expected, the ratio $t_*^{(1)}/t_*^{(0)}$ remains small across a large region of parameter space. The discontinuous behaviour of this ratio around $\frac{\rho}{\mu}=10^2$ is due to $t_*^{(0)}$ changing sign. For bigger $N$ the contribution from higher loop orders becomes more important, since each of the scalar fields contributes to the loop corrections with the same coupling. Nonetheless, also in the $N=10$ case the higher-loop corrections are very modest.

\begin{figure}[h!]
\center
\includegraphics[width=.48\textwidth]{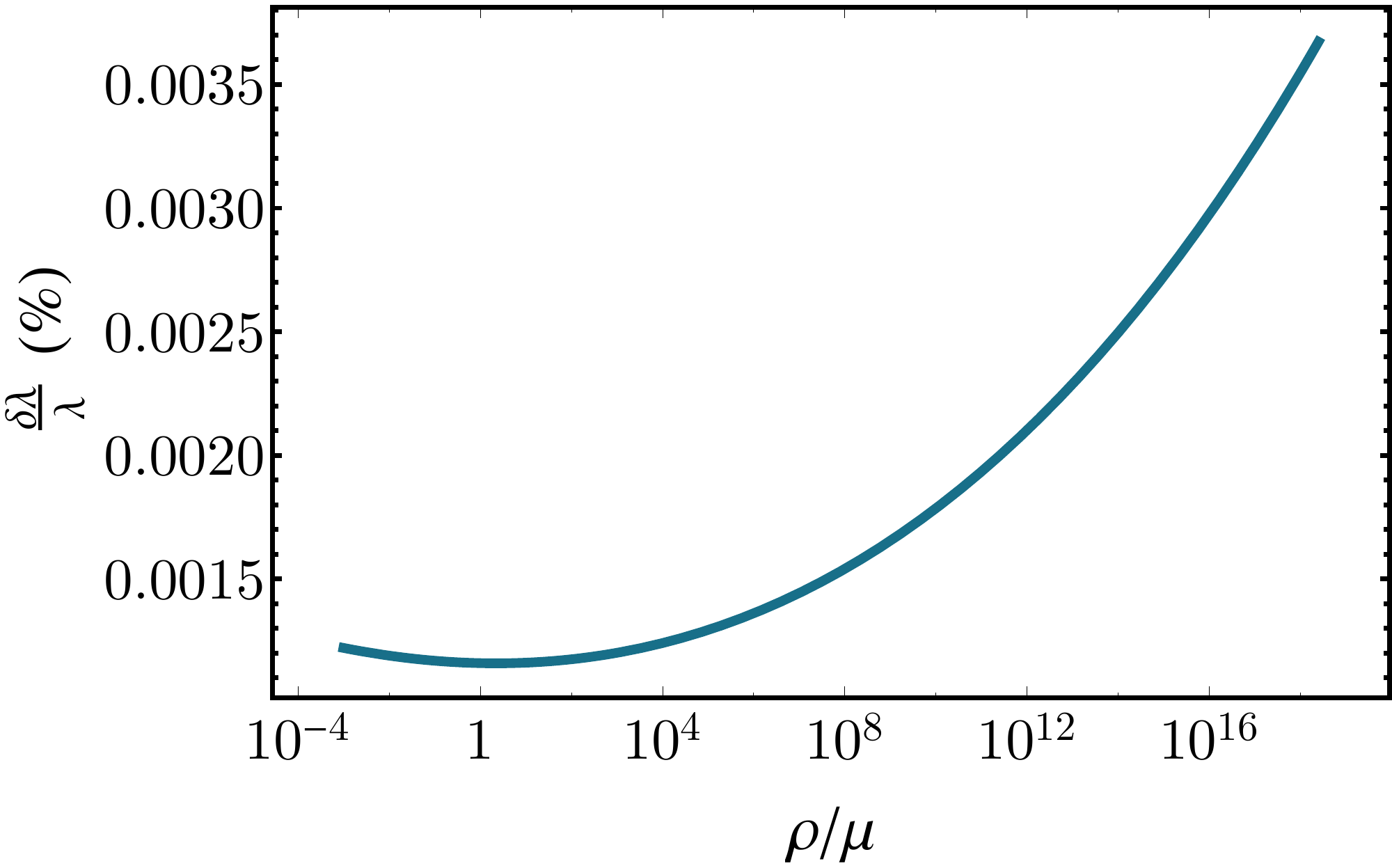}\hspace{.4cm}
\includegraphics[width=.48\textwidth]{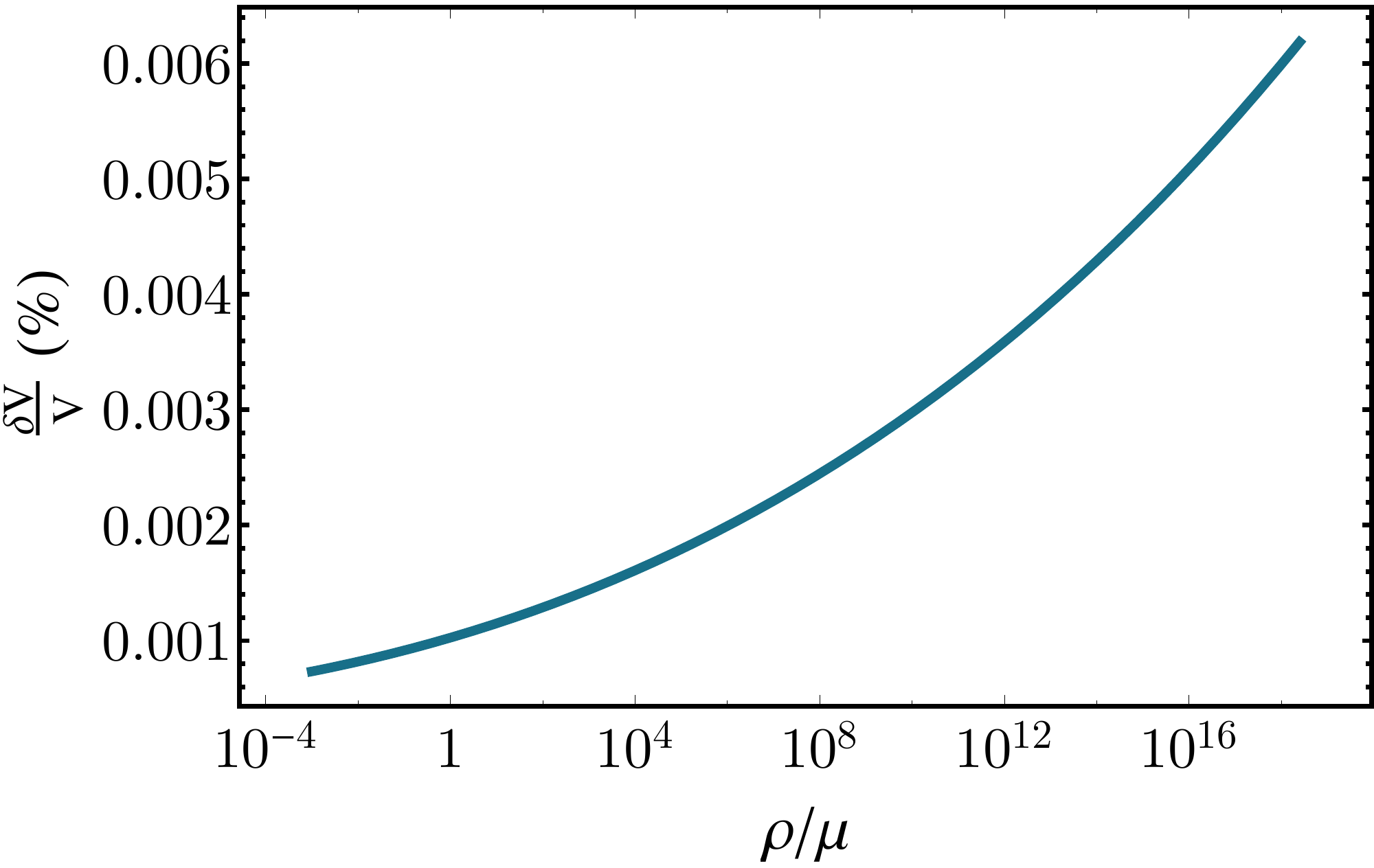}
\caption{The relative differences between the one-loop and two-loop running coupling ~\eqref{eq:ONphi4-deltalambda} (left panel) and improved potentials~\eqref{eq:ONphi4-deltaV}  (right panel) for $N=10$ scalar fields.}
\label{fig:ONphi4-deltas10}
\end{figure}

 If we understand $\lambda(t_*^{(0)})$ to be the one-loop running coupling evaluated at the field-dependent scale $t_*^{(0)}$ and, analogously, the quantity $\lambda(t_*^{(0)}+t_*^{(1)})$ to be the two-loop running coupling evaluated at the field-dependent scale $t_*^{(0)}+t_*^{(1)}$, we can define the relative difference
\begin{equation}\label{eq:ONphi4-deltalambda}
\frac{\delta\lambda}{\lambda} :=\frac{\lambda(t_*^{(0)}+t_*^{(1)})-\lambda(t_*^{(0)})}{\lambda(t_*^{(0)})} \ .
\end{equation}
In the same way, we define the one-loop and two-loop improved potentials and their relative difference as
\begin{align}
V_{\text{one-loop improved}} & := \frac{\lambda(t_*^{(0)})}{4!}\rho^4 \ , \notag\\
V_{\text{two-loop improved}} &:= \frac{\lambda(t_*^{(0)}+t_*^{(1)})}{4!}\rho^4(t_*^{(0)}+t_*^{(1)}) \ , \notag\\
\frac{\delta V}{V}&:=\frac{V_{\text{two-loop improved}}-V_{\text{one-loop improved}}}{V_{\text{one-loop improved}}} \ .\label{eq:ONphi4-deltaV}
\end{align}
In Figure~\ref{fig:ONphi4-deltas10} we show the relative differences of eqs.~(\ref{eq:ONphi4-deltalambda}) and~(\ref{eq:ONphi4-deltaV}) across a large range of field values. As expected, the differences are small. This implies that truncating the RG functions at one-loop order provides a reliable approximation of the full effective potential.

\section{Conclusions\label{sec:conclusions}}

We have presented a new method to RG improve the effective potential in the general case in which multiple scalar fields are present. We use the freedom to choose the single renormalisation scale $\mu$ to evaluate the effective potential of the theory on a hypersurface in the parameter space in which quantum corrections vanish. This is equivalent to solving the RG equation for the effective potential with a suitably chosen boundary conditions. The resulting potential has the form of the tree-level potential evaluated at a field-dependent scale. Since it contains no explicit logarithmic terms it is valid as long as the running couplings remain perturbative. Moreover, higher loop corrections are not expected to be big since they can only come from the running of the couplings and fields between slightly different scales. We discusses the validity and accuracy of our method in section~\ref{sec:applicability}.

While this method does not resum the leading logarithms as they are defined in the literature, we have shown that it reproduces the result of a resummation of the dominant logarithmic corrections, in the regions where a dominant (pivot) logarithm can be defined, without the need to explicitly specify the pivot scale. This is possible by understanding that one may work with different sets of logarithms (cf.\,eq.~(\ref{eq:sets-of-logs})) and that evaluating the tree-level potential at suitably chosen scale $t_*$ (cf. eqs.~\eqref{eq:V1/2Brho4}, \eqref{eq:t*-first}) automatically resums the dominant logarithmic contributions to the effective potential. Nonetheless, we underline that our method can be applied regardless of the existence of a pivot scale fulfilling inequality~\eqref{eq:dominant-pivot}.

Moreover, once the $\beta$-functions and anomalous dimensions are known and the boundary values for the running parameters are given, this method can be numerically implemented in a straightforward way, as we have shown on several examples in the present paper and we will also show in an accompanying paper~\cite{project2}. Indeed, given the (approximate) definition of $t_*$ in eqs.~\eqref{eq:V1/2Brho4}, (\ref{eq:tstar-0}), it is a simple matter to numerically evaluate the running couplings  $\bar{\lambda}(t_*)$ at given field values. We therefore conclude that the method presented is a numerically simple alternative to  other available techniques, such as multi-scale or decoupling methods.

It is also worth noting that
the results of this paper show that the one-loop RG-improved effective potential is given by the tree-level potential evaluated at $t_*$ also for large field values, as long as the running couplings remain perturbative. This implies that, to study the stability of the improved potential, it is enough to study the tree-level stability conditions with running couplings evaluated at large scales.
Even though expected, this result is not trivial, for quantum corrections could lead to other conditions on stability.

There are several further applications of the method presented in this paper. One of them is the study of radiative symmetry breaking in classically conformal models (for recent studies see e.g.\ refs.~\cite{Hambye, Carone, Khoze,Helmboldt,Khoze-Ro,Khoze-Plascencia, Lindner, Latosinski,Karam2} and references therein). In such models the symmetry breaking is entirely due to loop effects and thus inclusion of higher-loop terms in the improved potential may be important. With the above technique of RG improvement, we will be able to study the effective potential of these models across a large range of scales (field values), which will aid the numerical search for a global and dynamically generated minimum. We will apply our method  to evaluate the accuracy of commonly applied perturbative methods in an accompanying paper~\cite{project2}. The applications of the method presented herein are of course not limited to conformal models. It can be applied to any model with extended scalar sector, e.g.\ to models accommodating scalar dark matter, neutrino masses, accounting for strongly first order phase transition and baryogenesis, new scalar resonances, etc.

\acknowledgments
T.P. and B.{\'S}. acknowledge funding from the D-ITP consortium, a program of the NWO that is funded by the Dutch Ministry of Education, Culture and Science (OCW). This work is part of the research programme of the Foundation for Fundamental Research on Matter (FOM), which is part of the Netherlands Organisation for Scientific Research (NWO). B.{\'S}. acknowledges the support from the National Science Centre, Poland, the HARMONIA project under contract UMO- 2015/18/M/ST2/00518 (2016-2019) and from the Foundation for Polish Science (FNP). L.C. acknowledges the support of Utrecht University (UU), the Netherlands, in the form of the Utrecht Excellence Scholarship (UES) received from September 2015 to June 2017.


\appendix
\section{The Tree-Level Hypersurface: A General Approach}\label{chap:O-N-phi4}

In this appendix we generalise the method of RG-improvement presented in section~\ref{chap:RG} to all loop orders. This method can be used to RG-improve the effective potential in a general theory with $N_{\phi}$ scalar fields and $N_{\lambda}$ couplings. The key step is to evaluate the effective potential on a hypersurface in parameter space in which all quantum corrections vanish. 
Naturally, to fully determine the field-dependent value of the renormalisation scale on this surface, knowledge of all loop-orders is necessary. However, this method can be made practical by noting that this field-dependent scale can be computed as a power series in $\hbar$. The truncations of the RG functions at any given loop order can produce reliable approximations of the full effective potential if the running couplings are sufficiently small. In particular, truncation of the $\beta$-functions and anomalous dimensions to one-loop order yields the results of section~\ref{chap:RG}. In this appendix we discuss the subdominant terms originating from higher loop-orders that were not contemplated in section~\ref{chap:RG}. We will conclude that this general method, which can be easily implemented numerically, is a viable alternative to other currently available techniques of RG improvement. 


\subsection{Vanishing Loop Corrections: General Formulas}\label{sec:improve-curve-app}
As before, we   consider a theory with $N_{\phi}$ scalar fields, $N_{\lambda}$ couplings and $N_m$ mass eigenvalues. The couplings (possibly including mass terms) are denoted by $\lambda = (\lambda_1, ..., \lambda_{N_{\lambda}})$, the classical scalar fields by $\phi = (\phi_1,...,\phi_{N_{\phi}})$ and the mass eigenvalues by $m = (m_1, ..., m_{N_m})$. The mass logarithms are defined as
\begin{equation}
L_a = \log\frac{m_a^2(\lambda,\phi)}{\mu^2} \ , \ a =1,...,N_m.
\end{equation}
We  denote the effective potential by $V(\mu;\lambda,\phi)$, which is a function defined on a domain of the parameter space spanned by $(\mu;\lambda,\phi)$.

To generalise the method of RG-improvement presented in section~\ref{chap:RG} beyond one-loop order, let us now search for a field-dependent scale at which all loop-corrections vanish, as opposed to just the one-loop term. For this purpose, we write the effective potential of the theory as
\begin{equation}\label{eq:eff2}
V(\mu;\lambda,\phi) = V^{(0)}(\lambda,\phi)+q(\mu,\lambda,\phi) \ ,
\end{equation}
where we defined the variable
\begin{equation}\label{eq:CW-charac}
q \equiv q(\mu,\lambda,\phi) = \sum_{l = 1}^{\infty}\hbar^lV^{(l)}(\mu,\lambda,\phi) \ ,
\end{equation}
which encodes the quantum corrections. We note that the solution to the equation
\begin{equation}\label{eq:q}
q(\bar{\mu}(t),\bar{\lambda}(t),\bar{\phi}(t)) = 0 \ ,
\end{equation}
which we   denote by
\begin{equation}\label{eq:charac-displacement}
t_{*} \equiv t_{*}(\mu,\lambda,\phi) \ ,
\end{equation}
defines a displacement along the characteristic curve to a point of vanishing quantum corrections. For brevity, let us denote\footnote{For simplicity, as was done before in section~\ref{sec:hyper-one-loop}, we suppress the arguments which encode the initial conditions for the running, i.e. we denote $\bar{\lambda}_i(t_*,\mu,\lambda,\phi)\equiv \bar{\lambda}_i(t_*)$.}
\begin{equation}
\begin{aligned}
\mu_* &\equiv \bar{\mu}(t_*)  = \mu e^{t_{*}} \ ,\\
\lambda_{i*} &\equiv \bar{\lambda}_i(t_*) \ , \ \bar{\lambda}_i(0) = \lambda_i \ ,\\
\phi_{a*} &\equiv \bar{\phi}_a(t_*) \ , \ \bar{\phi}_a(0) = \phi_a \ ,
\end{aligned}
\end{equation}
Due to the scale-invariance of the effective potential, in analogy with eq.~(\ref{eq:t-invariance}), we obtain
\begin{equation}\label{eq:invariant-V}
V^{(0)}(\lambda_*,\phi_*) = V^{(0)}(\bar{\lambda}(t_*),\bar{\phi}(t_*))+q(\bar{\mu}(t_*),\bar{\lambda}(t_*),\bar{\phi}(t_*)) = V(\mu;\lambda,\phi) \ .
\end{equation}
Moreover, quantum corrections   satisfy
\begin{equation}\label{eq:vanishing-q}
0 = q(\bar{\mu}(t_*),\bar{\lambda}(t_*),\bar{\phi}(t_*)) = q(\mu,\lambda,\phi) +\sum_{n = 1}^{\infty}\frac{1}{n!}\left.\frac{\mathrm{d}^n q}{\mathrm{d}t^n}\right|_{t = 0}t_{*}^{n}  = q(\mu,\lambda,\phi) - \sum_{n = 1}^{\infty}\frac{1}{n!}\left.\frac{\mathrm{d}^n V^{(0)}}{\mathrm{d}t^n}\right|_{t = 0}t_{*}^{n} \ ,
\end{equation}
where we used the fact that $V$ is scale independent and thus  $\frac{\mathrm{d}^nV}{\mathrm{d}t^n} \equiv 0$ and therefore,
\begin{equation}
\frac{\mathrm{d}^nV^{(0)}}{\mathrm{d}t^n} = -\frac{\mathrm{d}^n q}{\mathrm{d}t^n} \ , \ n>0 \ .
\end{equation}
Using eq.~(\ref{eq:vanishing-q}), we find
\begin{equation}\label{eq:series-tree}
V^{(0)}(\lambda_*,\phi_*) = V^{(0)}(\lambda,\phi)+\sum_{n = 1}^{\infty}\frac{1}{n!}\left.\frac{\mathrm{d}^n V^{(0)}}{\mathrm{d}t^n}\right|_{t = 0}t_{*}^{n}  = V^{(0)}(\lambda,\phi)+q(\mu,\lambda,\phi) \ ,
\end{equation}
which is merely a rewriting of eq.~(\ref{eq:invariant-V}).

On the hypersurface $q = 0$, the potential can be written as the tree-level form $V^{(0)}(\lambda_*,\phi_*)$ and only the running couplings $\lambda_* \equiv \bar{\lambda}(t_*)$ and fields $\phi_* \equiv \bar{\phi}(t_*)$ appear. It is worth emphasizing that in this way, any logarithmic dependence is implicit. This amounts to a resummation of all logarithms. A perturbative treatment is valid if the running couplings are small and, therefore, running towards this tree-level hypersurface   minimises the effect of radiative corrections in truncations of the effective potential at a given loop order.\footnote{Since standard perturbation theory is an asymptotic series in the coupling constants, our RG improvement method applies, as standard perturbation theory, up to some fixed order in the couplings.}

We can solve for $t_{*}$ in perturbation theory using 
eq.~(\ref{eq:vanishing-q}).  Employing a shorthand notation for a multi-index, $\{\alpha\}:= (\alpha_1,...,\alpha_{N-1})$ (see also appendix~\ref{sec:pert-struct-2} for more details regarding notation), we write
\begin{equation}
t_{*} = \sum_{l = 0}^{\infty}\hbar^l t_{*}^{(l)} \ ,
\end{equation}
which we insert in
\begin{equation}\notag
q(\mu,\lambda,\phi) = \sum_{n=1}^{\infty}\frac{1}{n!}\left.\frac{\mathrm{d}^nV^{(0)}}{\mathrm{d}t^n}\right|_{t = 0}t_{*}^n \ ,
\end{equation}
to obtain
\begin{align*}
\sum_{s = 1}^{\infty}\hbar^sV^{(s)}(\mu,\lambda,\phi) &= \sum_{n=1}^{\infty}\frac{1}{n!}\sum_{l_1,...,l_n = 1}^{\infty}\hbar^{l_1+\cdots+l_n}\prod_{a = 1}^{n}\left[\mathrm{d}^{(l_a)}\right]_{t=0}V^{(0)}\sum_{k_1,...,k_n=0}^{\infty}\hbar^{k_1+\cdots+k_n}\prod_{b=1}^{n}t_{*}^{(k_b)}  \\
& = \sum_{n=1}^{\infty}\sum_{l = 0}^{\infty}\sum_{k = 0}^{\infty}\frac{\hbar^{l+n+k}}{n!}\left[\sum_{\{l\}=1}^{l+n}\prod_{a=1}^n\mathrm{d}^{(l_a)}V^{(0)}\right]_{t=0}\sum_{\{k\}=0}^k\prod_{b=1}^nt_{*}^{(k_b)} \\
& = \sum_{s=1}^{\infty}\hbar^s\sum_{n=1}^{s}\sum_{l=0}^{s-n}\frac{1}{n!}\left[\sum_{\{l\}=1}^{l+n}\prod_{a=1}^n\mathrm{d}^{(l_a)}V^{(0)}\right]_{t=0}\sum_{\{k\}=0}^{s-n-l}\prod_{b=1}^nt_{*}^{(k_b)} \ .
\end{align*}
Note that we have employed changes of summation variables to write the right-hand side of the above equation in a similar form to the left-hand side. We are lead to the formula
\begin{equation}\label{eq:loop-orders-t}
\sum_{n=1}^{s}\sum_{l=0}^{s-n}\frac{1}{n!}\left[\sum_{\{l\}=1}^{l+n}\prod_{a=1}^n\mathrm{d}^{(l_a)}V^{(0)}\right]_{t=0}\sum_{\{k\}=0}^{s-n-l}\prod_{b=1}^nt_{*}^{(k_b)} = V^{(s)}(\mu,\lambda,\phi) \ ,\ (s\geq1) \ .
\end{equation}
To lowest order $(s = 1)$, formula~(\ref{eq:loop-orders-t}) gives 
\begin{equation}\label{eq:t0}
\begin{aligned}
\left[\mathrm{d}^{(1)}V^{(0)}\right]_{t=0} t_{*}^{(0)} = V^{(1)}(\mu,\lambda,\phi) \ , \\
t_{*}^{(0)} = \frac{V^{(1)}(\mu,\lambda,\phi)}{\left[\mathrm{d}^{(1)}V^{(0)}\right]_{t=0}} \ ,
\end{aligned}\end{equation}
which agrees with eq.~(\ref{eq:tstar-0}), once one notes that we have $\left[\mathrm{d}^{(1)}V^{(0)}\right]_{t=0} = 2\mathbb{B}$ from eq.~(\ref{eq:recursive-w00}).
The next order $(s = 2)$ is obtained with the equation
\begin{align*}
&\sum_{n=1}^{2}\sum_{l=0}^{2-n}\frac{1}{n!}\left[\sum_{\{l\}=1}^{l+n}\prod_{a=1}^n\mathrm{d}^{(l_a)}V^{(0)}\right]_{t=0}\sum_{\{k\}=0}^{2-n-l}\prod_{b=1}^nt_{*}^{(k_b)} = V^{(2)}(\mu,\lambda,\phi) \ , \\*
& \left[\mathrm{d}^{(1)}V^{(0)}\right]_{t=0}t_{*}^{(1)}+\left[\mathrm{d}^{(2)}V^{(0)}\right]_{t=0}t_{*}^{(0)}+\frac{1}{2}\left[\left(\mathrm{d}^{(1)}\right)^2V^{(0)}\right]_{t=0}\left(t_{*}^{(0)}\right)^2= V^{(2)}(\mu,\lambda,\phi) \ ,
\end{align*}
which yields
\begin{equation}\label{eq:t1}
t_{*}^{(1)} = \frac{V^{(2)}(\mu,\lambda,\phi)-\left[\mathrm{d}^{(2)}V^{(0)}\right]_{t=0}t_{*}^{(0)}-\frac{1}{2}\left[\left(\mathrm{d}^{(1)}\right)^2V^{(0)}\right]_{t=0}\left(t_{*}^{(0)}\right)^2}{\left[\mathrm{d}^{(1)}V^{(0)}\right]_{t=0}} \ .
\end{equation}
One can continue in this way to determine $t_{*} \equiv t_{*}(\mu,\lambda,\phi)$ to an arbitrary loop order using formula~(\ref{eq:loop-orders-t}). In particular, we note that $t_*$ inherits from the effective potential the invariance under redefinitions of the pivot scale at each order in perturbation theory. This can be explicitly verified in eqs.~(\ref{eq:t0}) and~(\ref{eq:t1}). 

In the next subsections we discuss the issues of resummation of logarithmic corrections in more detail. In particular, we define the leading functions of the $t_*$ expansion. Moreover, we discuss the pivot expansion beyond leading order. We show that, as at one-loop order, the method of tree-level hypersurface reproduces the results of the pivot logarithm resummation, once a dominant pivot logarithm can be determined.

\subsection{The Perturbative Structure of the Effective Potential}\label{sec:pert-struct-2}

In what follows  we  use a modified version of the multi-index notation adopted from ref.~\cite{FritzJohn}. A point in $\mathbb{Z}^n$ with non-negative components is called a multi-index and is denoted by Greek letters, for example $\alpha = (\alpha_1,...,\alpha_n)$. Given a multi-index $\alpha$, we denote
\begin{align*}
|\alpha| &:=\sum_{i = 1}^{N}\alpha_i \ ,\\
x^{\boldsymbol{\alpha}} &:= \prod_{i = 1}^n x_i^{\alpha_i} = x_1^{\alpha_1}\cdots x_n^{\alpha_n}\ , \\
A_{\boldsymbol{\alpha}} &:= A_{\alpha_1...\alpha_N} \ ,
\end{align*}
where $x$ is a vector field and $A$ is a tensor field. 
Given some function $f(\alpha) = f(\alpha_1,...,\alpha_N)$, we will also make use of the change of variables
\begin{align*}
&\sum_{\alpha_1 = 0}^{\infty}\cdots\sum_{\alpha_N = 0}^{\infty}f(\alpha) =\\
&= \sum_{a= 0}^{\infty}\sum_{\alpha_1 = 0}^{a}\sum_{\alpha_2 = 0}^{a-\alpha_1}\cdots\sum_{\alpha_{N-1} = 0}^{a-\alpha_1-...-\alpha_{N-2}}f(\alpha_1,\alpha_2,...,\alpha_{N-1},a-\alpha_1-...-\alpha_{N-1}) \equiv\\
&\equiv \sum_{a= 0}^{\infty}\sum_{\{\alpha\} = 0}^{a}f(\alpha_1,\alpha_2,...,\alpha_{N-1},a-\alpha_1-...-\alpha_{N-1}) \ ,
\end{align*}
where in the last line we defined a convenient short-hand notation. For example, we may write
\begin{align*}
\sum_{\{n\}=0}^{n}\phi^{\mathbf{n}} = \sum_{n_1=0}^{n}\sum_{n_2 = 0}^{n-n_1}\cdots\sum_{n_{N_{\phi}-1}=0}^{n-n_1-...-n_{N_{\phi}-2}}\phi_1^{n_1}\cdots\phi_{N_{\phi}}^{n_{N_{\phi}}}  \ .
\end{align*}

In perturbation theory, the effective potential is written as the loop expansion in eq.~(\ref{eq:eff}). The general structure of the renormalised $l$-th loop order term in the $\ms$ scheme is~\cite{Kastening2,Bando2,Ford}
\begin{equation}\label{eq:loops}
V^{(l)}(\mu;\lambda,\phi) = \sum_{n=0}^l\sum_{\{n\}=0}^{n}v^{(l)}_{\mathbf{n}}\prod_{a=1}^{N_{m}}L_{a}^{n_a}\equiv\sum_{n=0}^l\sum_{\{n\}=0}^{n}v^{(l)}_{\mathbf{n}}L^{\mathbf{n}} \ ,
\end{equation}
where the coefficients $v^{(l)}_{\mathbf{n}} = v^{(l)}_{n_1...n_{N_\phi}}$ are functions of the couplings and the fields. The potential depends logarithmically on the renormalisation scale $\mu$ through the powers of the mass logarithms, which originate from the regularisation of momentum integrals.


In certain regions of the parameter space, in particular for large field values, 
it is
necessary to reorganise the perturbative expansion in eq.~(\ref{eq:eff}) in order to resum the large logarithms that appear in eq.~(\ref{eq:loops}).
To achieve this, we make use of the renormalisation group (RG), which we studied in section~\ref{chap:RG}. A typical reorganisation or improvement of the effective potential can be written as
\begin{equation}\label{eq:leading-functions}
V(\mu;\lambda,\phi) = \sum_{l = 0}^{\infty}\hbar^l V^{(l)}(\mu,\lambda,\phi) = \sum_{l = 0}^{\infty}\hbar^l f_l(\hbar;\mu,\lambda,\phi) \ ,
\end{equation}
where the functions $f_l$ may contain all powers of $\hbar$. We will call $f_l$ the $l$-th-to-leading functions~\cite{Kastening2, Bando2}. If such functions can be found in closed form and satisfy
\begin{equation}\notag
\left|\frac{f_{l+1}}{f_l}\right| < 1 \ ,
\end{equation}
than one can truncate the right hand side of eq.~(\ref{eq:leading-functions}) to a given order in $\hbar$. For suitable choices of $f_l$, the region of parameter space for which the right hand side of eq.~(\ref{eq:leading-functions}) is perturbative can thus be larger than the corresponding region for which the left hand side is perturbative~\cite{Ford}.

In the case of large logarithms, we can define the $l$-th-to-leading logarithms in analogy to the one-field case studied in section~\ref{sec:pre-RG} (cf. eq.~(\ref{eq:n-to-leading-logs-phi4})). This can be done by changing the summation variables in eqs.~(\ref{eq:eff}) and (\ref{eq:loops}) to obtain
\begin{equation}\label{eq:pot-multi-leading}
V(\mu;\lambda,\phi) = \sum_{l = 0}^{\infty}\hbar^{l}\sum_{n=0}^l\sum_{\{n\}=0}^{n}v^l_{\mathbf{n}}L^{\mathbf{n}}  = \sum_{l = 0}^{\infty}\sum_{n=0}^{\infty}\hbar^{l+n}\sum_{\{n\}=0}^{n}v^{l+n}_{\mathbf{n}}L^{\mathbf{n}}  =  \sum_{l = 0}^{\infty}\hbar^l f_l(\hbar;\mu,\lambda,\phi) \ ,
\end{equation}
where
\begin{equation}\label{eq:multi-leading}
f_l(\hbar;\mu,\lambda,\phi) = \sum_{n=0}^{\infty}\hbar^n\sum_{\{n\}=0}^{n}v^{l+n}_{\mathbf{n}}L^{\mathbf{n}}.
\end{equation}
 We   refer to eq.~(\ref{eq:multi-leading}) as the $l$-th-to-leading logarithms, since this is how they are usually defined in the literature~\cite{Ford}.  



Let us now derive the $l$-th-to-leading functions of the $t_*$ expansion of the result obtained in the previous subsection.

In the same way we derived eq.~(\ref{eq:loop-orders-t}), we can write
\begin{align*}
V(\mu;\lambda,\phi) &= V^{(0)}(\lambda_*,\phi_*) = V^{(0)}(\lambda,\phi)+\sum_{n = 1}^{\infty}\frac{1}{n!}\left.\frac{\mathrm{d}^nV^{(0)}}{\mathrm{d}t^n}\right|_{t = 0}t_{*}^n  \\
& = V^{(0)}(\lambda,\phi)+\sum_{n = 1}^{\infty}\sum_{l = 0}^{\infty}\sum_{k = 0}^{\infty}\frac{\hbar^{l+k+n}}{n!}\sum_{\{l\}=1}^{l+n}\sum_{\{k\}=0}^k \prod_{a=1}^{n}\left[\mathrm{d}^{(l_a)}\right]_{t=0}V^{(0)}\prod_{b=1}^n t_{*}^{(k_b)}  \\
& = V^{(0)}(\lambda,\phi)+\sum_{s=0}^{\infty}\hbar^s\sum_{n=1}^{\infty}\frac{\hbar^n}{n!}\sum_{l = 0}^s\sum_{\{l\}=1}^{l+n}\sum_{\{k\}=0}^{s-l} \prod_{a=1}^{n}\left[\mathrm{d}^{(l_a)}\right]_{t=0}V^{(0)}\prod_{b=1}^n t_{*}^{(k_b)} \ .
\end{align*}
Defining
\begin{equation}
T_{n}^{(s+n)} := \frac{1}{n!}\sum_{l = 0}^s\sum_{\{l\}=1}^{l+n}\sum_{\{k\}=0}^{s-l} \prod_{a=1}^{n}\left[\mathrm{d}^{(l_a)}\right]_{t=0}V^{(0)}\prod_{b=1}^n t_{*}^{(k_b)} \ ,
\end{equation}
we obtain
\begin{align*}
V(\mu;\lambda,\phi) &=  V^{(0)}(\lambda,\phi)+\sum_{s=0}^{\infty}\hbar^s\sum_{n=1}^{\infty} \hbar^n T_{n}^{(s+n)} = \sum_{s = 0}^{\infty}\hbar^s f_s(\hbar;\mu,\lambda,\phi) \ ,
\end{align*}
where the $s$-th-to-leading function is defined as
\begin{equation}\label{eq:leading-t}
f_s(\hbar;\mu,\lambda,\phi) = \delta_{s,0}V^{(0)}(\lambda,\phi)+ \sum_{n = 1}^{\infty}\hbar^n T_{n}^{(s+n)} \ .
\end{equation}
In particular, the leading function reads
\begin{equation}\label{eq:first-leading-t}
\begin{aligned}
f_0(\hbar;\mu,\lambda,\phi) &= V^{(0)}(\lambda,\phi)+\sum_{n =1}^{\infty}\hbar^n T_{n}^{(n)} = V^{(0)}(\lambda,\phi)+\sum_{n =1}^{\infty}\frac{1}{n!}\left[\left(\hbar\mathrm{d}^{(1)}\right)^nV^{(0)}\right]_{t=0}\left(t_{*}^{(0)}\right)^n \\
&= V^{(0)}\left(\bar{\lambda}\left(t_{*}^{(0)}\right), \bar{\phi}\left(t_{*}^{(0)}\right)\right) \ ,
\end{aligned}\end{equation}
where $\bar{\lambda}\left(t_{*}^{(0)}\right)$ and $\bar{\phi}\left(t_{*}^{(0)}\right)$ are understood here as one-loop running parameters.
Note that to compute the leading function, one needs only knowledge of the one-loop RG functions. The objects $T_n^{(s+n)}$ were defined such that they are formally of order $\hbar^{s+n}$, in analogy to the usual definition of the $s$-th-to-leading logarithms.

In what follows we derive the leading functions of the pivot logarithm expansion and we show that these functions are automatically included in the leading functions of the $t_*$ expansion. This will prove that with the $t_*$ method we resum the powers of the pivot logarithm, without the need to specify it.

\subsection{Subleading Contributions in the Pivot Logarithm Expansion}\label{sec:pivot-log-app}
In section~\ref{sec:pivot-log}, we showed that knowledge of the one-loop $\beta$-functions and anomalous dimensions was sufficient to resum the leading function of the pivot logarithm expansion (cf. eq.~(\ref{eq:leading-logs-pivot})). We will now follow the work of B. Kastening in \cite{Kastening2}, in which the pivot logarithm method was applied to $O(N)$-symmetric $\phi^4$-theory, and establish a general way of computing the $k$-th-to-leading functions in the pivot logarithm expansion.

Instead of using eqs.~(\ref{eq:recursive-wnn}) and~(\ref{eq:recursive-wns}) to resum (sub)leading functions, it is more convenient to follow \cite{Kastening, Kastening2} and use recursive relations for the $f_k$ functions.
In order to obtain these relations, we define
\begin{equation}
L_{\mathcal{M}} = \frac{\hbar}{2}\log\frac{\mathcal{M}^2}{\mu^2} \ 
\end{equation}
and write $f_k$ as function of the pivot logarithm,
\begin{equation}\notag
f_k(\hbar;\mu,\lambda,\phi)\equiv f_k(L_{\mathcal{M}},\lambda,\phi) = \sum_{n=0}^{\infty}2^nw_{n}^{(n+k)}\left(\lambda,\phi\right) L_{\mathcal{M}}^n \ , 
\end{equation}
such that the insertion of eq.~(\ref{eq:resum-pivot}) into eq.~(\ref{eq:RG-potential2}) now yields
\begin{align*}
0&=\sum_{k= 0}^{\infty}\hbar^k\mu\frac{\mathrm{d}f_k}{\mathrm{d}\mu}  \\
& = \sum_{k = 0}^{\infty}\hbar^k\left(-\hbar\frac{\partial f_k}{\partial L_{\mathcal{M}}}+\sum_{i=1}^{N_{\lambda}} \beta_i \frac{\partial f_k}{\partial \lambda_i}-\frac{1}{2}\sum_{a=1}^{N_{\phi}}\gamma_a\phi_a\frac{\partial f_k}{\partial\phi_a}\right)  \\
& = \sum_{k=0}^{\infty}\hbar^{k+1}\left(-\frac{\partial f_{k}}{\partial L_{\mathcal{M}}} +\sum_{l=1}^{k+1}\sum_{i=1}^{N_{\lambda}} \beta_i^{(l)} \frac{\partial f_{k-l+1}}{\partial \lambda_i}-\frac{1}{2}\sum_{l=1}^{k+1}\sum_{a=1}^{N_{\phi}}\gamma_a^{(l)}\phi_a\frac{\partial f_{k-l+1}}{\partial\phi_a}\right) \ .
\end{align*}
We thus obtain the recursive equations
\begin{equation}\label{eq:recursive-fk-pivot}
\frac{\partial f_{k}}{\partial L_{\mathcal{M}}} -\sum_{l=1}^{k+1}\sum_{i=1}^{N_{\lambda}} \beta_i^{(l)} \frac{\partial f_{k-l+1}}{\partial \lambda_i}+\frac{1}{2}\sum_{l=1}^{k+1}\sum_{a=1}^{N_{\phi}}\gamma_a^{(l)}\phi_a\frac{\partial f_{k-l+1}}{\partial\phi_a} = 0 \ ,
\end{equation}
supplemented by the boundary conditions
\begin{equation}\label{eq:boundary-fk-pivot}
f_k(0,\lambda,\phi) = w_0^{(k)} \ .
\end{equation}
Let us solve eq.~(\ref{eq:recursive-fk-pivot}) for the first leading function. The Cauchy problem for $f_0$ is
\begin{align*}
&\frac{\partial f_{0}}{\partial L_{\mathcal{M}}} -\sum_{i=1}^{N_{\lambda}} \beta_i^{(1)} \frac{\partial f_{0}}{\partial \lambda_i}+\frac{1}{2}\sum_{a=1}^{N_{\phi}}\gamma_a^{(1)}\phi_a\frac{\partial f_{0}}{\partial\phi_a} = 0 \ , \\
&f_0(0,\lambda,\phi) = w_0^{(0)} \equiv V^{(0)} \ ,
\end{align*}
which can be solved with the method of characteristics. The boundary hyperplane is chosen to be $L_{\mathcal{M}} = 0$, which is a regular and noncharacteristic hypersurface. 

We can proceed in full analogy to the procedure outlined in sections~\ref{sec:pre-RG} and~\ref{sec:hyper-one-loop} and obtain
$$
f_0(L_{\mathcal{M}},\lambda,\phi) = V^{(0)}(\bar{\lambda}_i(L_{\mathcal{M}},\lambda, \phi),\bar{\phi}_a(L_{\mathcal{M}},\lambda,\phi)),
$$
which agrees with eq.~(\ref{eq:leading-logs-pivot}). In general, we can use the method of characteristics to solve eq.~(\ref{eq:recursive-fk-pivot}) for $f_k$ with $f_{s} \ (0 \leq s < k)$ as sources. Due to the boundary conditions (\ref{eq:boundary-fk-pivot}), knowledge of the RG functions up to $(k+1)$-th loop order is necessary to compute the $k$-th-to-leading function. In particular, one needs only the one-loop order RG functions to compute the leading function $f_0$, as we saw in section~\ref{chap:RG}.

As was underlined before, the reliability of the resummation of the powers of the pivot logarithm depends crucially on the right choice of the pivot logarithm. If it is not the dominant one, in the sense of inequality~\eqref{eq:dominant-pivot}, the result of the resummation is not a reliable approximation of the effective potential. Evidently, the difficulty in determining the (dominant) pivot logarithm for each region in parameter space remains beyond one-loop order. Nonetheless, from eq.~(\ref{eq:t0}), we see that the leading function of the $t_*$ expansion in eq.~(\ref{eq:first-leading-t}) automatically includes the leading function  of the pivot expansion given in eq.~(\ref{eq:pivot-f0}) and, therefore, the expansion in powers of $t_*$ resums the leading powers of a pivot logarithm and also includes terms that are subleading, containing the logarithms of the ratios $\log \frac{m_a^2}{\mathcal{M}^2}$. As we showed in section~\ref{chap:RG}, the invariance under redefinitions of $\mathcal{M}$ guarantees that the dominant logarithms are captured in this resummation. The subleading terms  are resummed with higher orders in $t_*$ (cf.\,eq.~\eqref{eq:loop-orders-t}) which yield the subleading functions as defined in eq.~(\ref{eq:leading-t}).

\section{Massless O(N)-symmetric $\phi^4$ theory\label{app:ON}}
Massless $\phi^4$-theory is an example of a classically conformal theory. This implies that there is no vacuum energy and no mass parameters at tree-level. We will consider $N_{\phi} = N$ scalar fields and $N_{\lambda} = 1$ coupling, such that the theory has $O(N)$ symmetry. For convenience, we set $\hbar = 1$. In the $\ms$ scheme, the renormalised effective potential up to two-loop order is \cite{Kastening, Ford2}
\begin{align}
V(\mu,\lambda,\phi) &= V^{(0)}(\lambda,\phi) + V^{(1)}(\mu,\lambda,\phi)+V^{(2)}(\mu,\lambda,\phi) \ , \notag\\
V^{(0)}(\lambda,\phi) &= \frac{\lambda}{4!}\rho^4 \ , \notag\\
V^{(1)}(\mu,\lambda,\phi) &=\frac{1}{64\pi^2}\left[m_H^4\left(\log\frac{m_H^2}{\mu^2}-\frac{3}{2}\right)+(N-1)m_G^4\left(\log\frac{m_G^2}{\mu^2}-\frac{3}{2}\right)\right] \ , \notag\\
V^{(2)}(\mu,\lambda,\phi) &=\frac{1}{8(4\pi)^4}\lambda^2\rho^2m_H^2\left(\log^2\frac{m_H^2}{\mu^2}-4\log\frac{m_H^2}{\mu^2}+8\Omega(1)+5\right)+\frac{1}{8(4\pi)^4}\lambda m_H^4\left(\log\frac{m_H^2}{\mu^2}-1\right)^2\notag\\
&+\frac{N-1}{(4\pi)^4}\left\{\frac{1}{72}\lambda^2\rho^2\left[(m_H^2+2m_G^2)\left(\log^2\frac{m_G^2}{\mu^2}-4\log\frac{m_G^2}{\mu^2}+8\Omega\left(\frac{m_H^2}{m_G^2}\right)+5\right)\right.\right.\notag\\
&\left.\left.+ 2m_H^2\log\frac{m_H^2}{m_G^2}\left(\log\frac{m_G^2}{\mu^2}-4\right)\right]+\frac{1}{12}\lambda m_H^2m_G^2\left[\log\frac{m_H^2}{\mu^2}\log\frac{m_G^2}{\mu^2}-\log\frac{m_H^2}{\mu^2}-\log\frac{m_G^2}{\mu^2}+1\right]\right\}\notag\\
&+\frac{N^2-1}{(4\pi)^4}\frac{\lambda}{24}m_G^4\left(\log\frac{m_G^2}{\mu^2}-1\right)^2 \ , \label{eq:phi4-two-loop}
\end{align}
where we defined
\begin{equation}\label{eq:masses-etc-O-N}
\begin{aligned}
\rho^2 &= \sum_{a = 1}^{N}\phi_j^2 \ , \\
m_H^2 &= \frac{\lambda}{2}\rho^2\ , \ m_G^2 =\frac{\lambda}{6}\rho^2 \ ,\\
\Omega(x)&=\left\{\begin{matrix}\frac{\sqrt{x(4-x)}}{x+2}\int_0^{\arcsin\left(\frac{\sqrt{x}}{2}\right)}\log(2\sin t)\mathrm{d}t & \text{for} \ x \leq 4\\
\frac{\sqrt{x(x-4)}}{x+2}\int_0^{\arccosh\left(\frac{\sqrt{x}}{2}\right)}\log(2\cosh t)\mathrm{d}t & \text{for} \ x > 4\end{matrix}\right. \ .
\end{aligned}
\end{equation}
In terms of the pivot mass $\mathcal{M} = \rho$, we can rewrite (\ref{eq:phi4-two-loop}) as follows.
\begin{align*}
V(\mu,\lambda,\phi)&= w_0(\lambda)+w_1(\lambda)\log\frac{\rho^2}{\mu^2}+w_2(\lambda)\log^2\frac{\rho^2}{\mu^2} \ , \\
w_0(\lambda) &= w_0^{(0)}(\lambda)+w_0^{(1)}(\lambda)+w_0^{(2)}(\lambda) \ , \\
w_1(\lambda) &= w_1^{(1)}(\lambda)+w_1^{(2)}(\lambda) \ , \\
w_2(\lambda) &= w_2^{(2)}(\lambda) \ , \\
w_0^{(0)}(\lambda) &= \frac{\lambda}{4!}\rho^4 \ , \\
w_0^{(1)}(\lambda) &= \frac{1}{64\pi^2}\left[m_H^4\left(\log\frac{m_H^2}{\rho^2}-\frac{3}{2}\right)+(N-1)m_G^4\left(\log\frac{m_G^2}{\rho^2}-\frac{3}{2}\right)\right] \ , \\
w_0^{(2)}(\lambda) &=V^{(2)}(\mu=\rho,\lambda,\phi)\ ,\\
w_1^{(1)}(\lambda) &= \frac{1}{64\pi^2}\left[m_H^4+(N-1)m_G^4\right] \ , \\
w_1^{(2)}(\lambda) &=\frac{1}{8(4\pi)^4}\lambda^2\rho^2 m_H^2\left(2\log\frac{m_H^2}{\rho^2}-4\right)+\frac{1}{4(4\pi)^4}\lambda m_H^4\left(\log\frac{m_H^2}{\rho^2}-1\right)\\
&+\frac{N-1}{(4\pi)^4}\left\{\frac{1}{72}\lambda^2\rho^2\left[(m_H^2+2m_G^2)\left(2\log\frac{m_G^2}{\rho^2}-4\right)+2 m_H^2\log\frac{m_H^2}{m_G^2}\right]\right.\\
&+\left.\frac{1}{12}\lambda m_H^2m_G^2\left[\log\frac{m_H^2}{\rho^2}+\log\frac{m_G^2}{\rho^2}-2\right]\right\}+\frac{N^2-1}{(4\pi)^4}\frac{\lambda}{12} m_G^4 \left(\log\frac{m_G^2}{\rho^2}-1\right) \ ,\\
w_2^{(2)}(\lambda) &=\frac{1}{8(4\pi)^4}\lambda^2\rho^2 m_H^2+\frac{1}{8(4\pi)^4}\lambda m_H^4+\frac{N-1}{(4\pi)^4}\left\{\frac{1}{72}\lambda^2\rho^2\left[m_H^2+2m_G^2\right]+\frac{1}{12}\lambda m_H^2m_G^2\right\}\\
&+\frac{N^2-1}{(4\pi)^4}\frac{\lambda}{24} m_G^4 \ .
\end{align*}
The $\beta$-function and anomalous dimension can be computed with standard techniques and, up to two-loop order, read~\cite{Sher, Ford, Ford2}
\begin{equation}\label{eq:ONphi4-betas}
\begin{aligned}
\beta^{(1)} &= \frac{N+8}{3(4\pi)^2}\lambda^2 \ , \\
\beta^{(2)} &= -\frac{3N+14}{3(4\pi)^4}\lambda^3 \ ,\\
\gamma^{(1)} &= 0 \ , \\
\gamma^{(2)}&=\frac{N+2}{18(4\pi)^4}\lambda^2 \ .
\end{aligned}
\end{equation}
Let us now verify the solutions of eq.~(\ref{eq:recursive-w00}). Taking a first derivative, we obtain
\begin{align*}
\mathrm{d}^{(1)}w_0^{(0)} = \beta^{(1)}\frac{\partial w_0^{(0)}}{\partial\lambda} = \frac{N+8}{3(4\pi)^2}\lambda^2\frac{\rho^4}{4!} \ .
\end{align*}
With the definitions of $m_H$ and $m_G$ given in eq.~(\ref{eq:masses-etc-O-N}) and the formulas given on the previous page, one may easily verify that the right hand side of the above equation is indeed equal to $2w_1^{(1)}$. Then, taking a second derivative, we find
\begin{align*}
\left[\mathrm{d}^{(1)}\right]^2w_0^{(0)} = 2\beta^{(1)}\frac{\partial w_1^{(1)}}{\partial\lambda} = \frac{\lambda^3}{36}\frac{(N+8)^2}{3(4\pi)^4}\rho^4 \ .
\end{align*}
Again, it is straightforward to verify that the right hand side of the above equation is equal to $8w_2^{(2)}$, with the formulas previously given.

\bibliography{RG-biblio}
\end{document}